%% file: main.tex
\documentclass{article}

\usepackage[script]{changdefs}
\usepackage[numbers,compress]{natbib}
\usepackage{isomath} 
\usepackage{stackrel}

\usepackage{graphicx}
\usepackage{wrapfig}
\graphicspath{{./figures/}}

\usepackage{booktabs}
\usepackage{footmisc} 

\numberwithin{equation}{section}

\usepackage{pifont}
\newcommand{\cmark}{\ding{51}}
\newcommand{\xmark}{\ding{55}}

\usepackage{footnote}
\makesavenoteenv{tabular}
\makesavenoteenv{table}

\newcommand{\bigsquare}{\,\fbox{$\phantom{o}$}}

\newcommand{\rrvd}{\dot{\rrv}}
\newcommand{\rrvdd}{\ddot{\rrv}}

\newcommand{\ppvh}{\hat{\ppv}}
\newcommand{\vvvh}{\hat{\vvv}}

\newcommand{\ppvb}{\bar{\ppv}}
\newcommand{\rrvb}{\bar{\rrv}}

\newcommand{\Svh}{\hat{\Sv}}
\newcommand{\Jth}{\hat{\Jt}}
\newcommand{\Kth}{\hat{\Kt}}
\newcommand{\Kbh}{\hat{\Kb}}

\newcommand{\bmrrv}{\vec{\bm{r}}}
\newcommand{\bmrrvd}{\dot{\vec{\bm{r}}}}
\newcommand{\bmkkv}{\vec{\bm{k}}}
\newcommand{\bmppv}{\vec{\bm{p}}}
\newcommand{\bmppvd}{\dot{\vec{\bm{p}}}}
\newcommand{\bmvvv}{\vec{\bm{v}}}
\newcommand{\bmgammav}{\vec{\bm{\gamma}}}
\newcommand{\bmgammavd}{\dot{\vec{\bm{\gamma}}}}

\newcommand{\bmclE}{\bm{\clE}}

\newcommand{\bfAt}{\tilde{\bfA}}
\newcommand{\bfGammat}{\tilde{\bfGamma}}
\newcommand{\bfAb}{\bar{\bfA}}
\newcommand{\bfBb}{\bar{\bfB}}
\newcommand{\bfLb}{\bar{\bfL}}
\newcommand{\bfPb}{\bar{\bfP}}
\newcommand{\bfSb}{\bar{\bfS}}
\newcommand{\bfTb}{\bar{\bfT}}
\newcommand{\bfVb}{\bar{\bfV}}
\newcommand{\bfffb}{\bar{\bfff}}
\newcommand{\bfLambdab}{\bar{\bfLambda}}

\newcommand{\sfbJb}{\bar{\sfbJ}}
\newcommand{\sfbKb}{\bar{\sfbK}}

\newcommand{\XC}{\textnormal{XC}}
\newcommand{\XCb}{{\overline{\XC}}}
\newcommand{\ee}{\textnormal{ee}}
\newcommand{\Ne}{\textnormal{Ne}}
\newcommand{\ext}{\textnormal{ext}}
\newcommand{\TF}{\textnormal{TF}}
\newcommand{\TFD}{\textnormal{TFD}}
\newcommand{\TFW}{\textnormal{TFW}}
\newcommand{\vW}{\textnormal{vW}}
\newcommand{\HF}{\textnormal{HF}}
\newcommand{\HK}{\textnormal{HK}}
\newcommand{\KS}{\textnormal{KS}}
\newcommand{\Nsp}{N_\textnormal{sp}}
\newcommand{\Npr}{N_\textnormal{pr}}
\newcommand{\Iupr}{\clI_\textnormal{upr}}

\newcommand{\LDA}{\textnormal{LDA}}
\newcommand{\tnCb}{{\bar{\textnormal{C}}}}
\newcommand{\WS}{\textnormal{WS}}
\newcommand{\GGA}{\textnormal{GGA}}
\newcommand{\GEA}{\textnormal{GEA}}

\newcommand{\rmDeltab}{\bar{\rmDelta}}
\newcommand{\CE}{\mathrm{CE}}
\newcommand{\DM}{\textnormal{DM}}
\newcommand{\Levy}{\textnormal{Lv}}
\newcommand{\Lieb}{\textnormal{Lb}}

\begin{document}
\abovedisplayskip=3pt
\belowdisplayskip=2pt
\abovedisplayshortskip=3pt
\belowdisplayshortskip=2pt

\title{A Verbose Note on Density Functional Theory}
\author{Chang Liu \\ Microsoft Research AI for Science \\
\texttt{changliu@microsoft.com}
}
\date{}
\maketitle

\abstract{
  This note is intended for expanding the details on the derivation and properties of density functional theory, in hope to make them more systematic, better motivated, and step-by-step for readers new to the domain. The note starts with basic concepts in quantum mechanics, then takes the step towards many-body systems using the tools of second quantization and Fock space, with some highlights on properties of the Coulomb system. Given these general technical preparations, the Hartree-Fock method is naturally unrolled, with expressions for various cases and quantities. Density functional theory is then presented, with a motivating reasoning of the Kohn-Sham formulation, and a connection and comparison of the expressions with the Hartree-Fock method. Lieb's celebrated 1983 paper on functional analysis on density functionals is also summarized.
}

\setcounter{tocdepth}{3}
\tableofcontents

\setcounter{section}{-1}

\include{preamble}

\include{quantum-basics}

\include{quantum-many-body}

\include{hf}

\include{dft}

\subsection*{Acknowledgement}
The note is composed with helpful discussions and checks by Jiacheng You, He Zhang, Siyuan Liu, and Feidiao Yang.

\bibliographystyle{abbrvnat}
\bibliography{main}

\end{document}

%% file: preamble.tex
\section{Preamble} \label{sec:pre}

\subsection{Notation}

\begin{itemize}
  \item Let $\rrv = (x,y,z) \in \bbR^3$ denote a general spacial position.
    For multiple particles (electrons), let $\rrv_i$ denote the position of the $i$-th, and denote $\bmrrv := (\rrv_1, \cdots, \rrv_N)$.
    The same goes for $\ppv$, $\nabla$, and spin $s$ (except that it takes discrete values).

    A molecular conformation is defined by the charges and positions of the $N_\tnA$ nuclei $\{(Z_A, \rrv_A)\}_{A \in [N_\tnA]}$.

    Let $\rmxx := (\rrv, s)$, and similarly denote $\bfxx := (\rmxx_1, \cdots, \rmxx_N)$.
    Define $\int \ud \rmxx \, f(\rmxx) := \sum_s \int \ud \rrv \, f(\rrv, s)$.

  \item Upper (resp. lower) Greek letters $\Psi$, $\Phi$ denote time-dependent (resp. time-independent) wavefunctions.
    $\psi$ denotes a general wavefunction or a many-body wavefunction.
    Others like $\phi, \eta, \chi$ denote one-body wavefunctions.

    $\psi_{(n)}$ denotes an eigenvector of the $n$-th energy eigenstate with eigenvalue $E_{(n)}$.
    $\phi_i$ denotes the $i$-th one-body wavefunction (orbital, basis, \ldots).

    $\psi^s$ or $\phi^s$ is used for emphasizing the spacial wavefunction of a given spin $s$.
    For specific value, use $\psi^{s = \frac{1}{2}}$ or $\psi^\uparrow$.
    Other cases denote the usual exponent.

    $\psi_\tnS$ denotes a Slater-determinant wavefunction (often with orthonormal orbitals).

  \item $\rho$ / $\Gamma$ represents a general density / density matrix, and $\rho^{(1)}$, $\rho^{(2)}$ / $P^{(1)}$, $P^{(2)}$ represent reduced densities / density matrices.
    $\bfGamma$, $\bfP$ are their matrix representations under a basis.

  \item Function symbols with tilde (\eg, $\psit$) represent the Fourier-transformed functions of the respective functions.

  \item $\Hh$, $\pph$, $\Th$, $\Vh$, $\Wh$ denote physical operators on wavefunctions (Hermite/self-adjoint operators on Hilbert space).
    $\Vh$ and $\Wh$ are devoted for one-particle and two-particle operators, respectively.
    The same symbols denote the respective extensions to apply on a many-body wavefunction, and in that case $\Vh_i$ or $\Wh_{ij}$ are used for the one-particle or two-particle operator on the specified particle(s)
    ($\Vh_\cdot$ or $\Wh_{\cdot\cdot}$ for general one-particle or two-particle operator without specifying the operand).

    $\bfV$, $\bfW$ are the corresponding matrix representations under a basis.
    Superscript denotes matrix row index.
    Sans-serif bolds, \eg, $\sfbJ$, $\sfbK$, denote tensor representations.

  \item $E_\textnormal{subs}^\textnormal{supers}$, $T_\textnormal{subs}^\textnormal{supers}$, $V_\textnormal{subs}^\textnormal{supers}$:
    for the corresponding value or functional of wavefunction/density/density-matrix.
    ``$\textnormal{subs}$'' specifies what energy/potential it is (\eg, $E_\ee$, $E_\ext$; for $T$, only $T_\tnS$ is available, which denotes the kinetic energy of a non-interacting system; $V_\tnS$ represents the effective potential for a non-interacting system), and ``$\textnormal{supers}$'' specifies the version (often name initials).

  \item $F[f]$ represents a functional that maps $f$ to $\bbR$ or $\bbC$,
    while $g_{[f]}$ represents a function (of $\rrv$, $\bmrrv$, $\bfxx$, \etc) or a functional determined by function $f$.

  \item For $N \in \bbN^*$, $[N] := \{1, \cdots, N\}$. \quad
    $\sum_{i, i' \in [N]} \cdots := \sum_{i \in [N]} \sum_{i' \in [N]} \cdots$. \quad
    $\sum_{i \ne j} \cdots := \sum_{i} \sum_{j: \ne i} \cdots$. \quad
    $\sum_{i < j} \cdots := \sum_{i} \sum_{j: > i} \cdots$. \quad
    $\int \ud\bfxx \cdots := \int \ud\rmxx_1 \cdots \ud\rmxx_N \cdots$. \quad
    $\int \ud\bfxx_{i:j} \cdots := \int \ud\rmxx_i \ud\rmxx_{i+1} \cdots \ud\rmxx_j \cdots$.

\end{itemize}

\subsection{Functions of Complex Variable} \label{sec:pre-complex}

A complex function $f: \bbC \to \bbC, z = (x + \ii y) \mapsto u(x,y) + \ii v(x,y)$ is \textbf{complex-differentiable (analytic)}, \emph{iff.} $u$ and $v$ are real-differentiable and satisfy \textbf{Cauchy-Riemann equations}:
\begin{align}
  \partial_x u = \partial_y v, \quad
  \partial_y u = -\partial_x v.
  \label{eqn:cauchy-riemann}
\end{align}
To see the necessity, $\lim_{\veps \to 0} \frac{f(x_0 + \ii y_0 + \veps) - f_0}{\veps}
= \lim_{\veps \to 0} \lrparen{ \frac{u(x_0 + \veps, y_0) - u_0}{\veps} + \ii \frac{v(x_0 + \veps, y_0) - v_0}{\veps} }
= \partial_x u + \ii \partial_x v$
and $\lim_{\veps \to 0} \frac{f(x_0 + \ii y_0 + \ii \veps) - f_0}{\ii \veps}
= \lim_{\veps \to 0} \lrparen{ \frac{u(x_0, y_0 + \veps) - u_0}{\ii \veps} + \ii \frac{v(x_0, y_0 + \veps) - v_0}{\ii \veps} }
= -\ii \partial_y u + \partial_y v$ should coincide: both are $\partial_z f(z = x_0 + \ii y_0)$.
Note \textbf{none} of $z / \lrvert{z}^2$, $z^2 / \lrvert{z}^2$, $\lrvert{z}^2 z$ and even $\lrvert{z}^2$, is analytic.

In this note we use a more general notion of derivative of $f$:
\begin{align}
  \partial_z f(z = x + \ii y) := \partial_x u(x,y) - \ii \partial_y u(x,y), \qquad u(x,y) := \Re f(x + \ii y).
  \label{eqn:general-complex-deriv}
\end{align}
When $f$ is analytic, \eqnref{cauchy-riemann} gives $-\partial_y u = \partial_x v$ so this $\partial_z f(z = x + \ii y) = \partial_x u + \ii \partial_x v$ coincides with the standard derivative.
Even if $f(z)$ is not analytic, it can be defined (as long as each $\partial_x u$ and $\partial_y u$ can),
and could already serve for variation calculation if $f = u$ is itself real, since each of $x$ and $y$ is an individual/independent variable, and $\partial_x f = \partial_y f = 0$ \emph{iff.} $\partial_x u - \ii \partial_y u = 0$.
If the function form is the same for both $x$ and $y$, \ie, $f(z) = f(x) + f(y)$ (\eg, the case for deriving \eqnref{schrodinger-stationary-common}), then $\partial_{z^*} f$ has the same form as treating $z$ as real.

Under this notion, $\partial_{z_1^*} z_1^* z_2 = z_2$, $\partial_{z_1} z_1^* z_2 = z_2^*$ hence also $\partial_{z_1^*} z_1 z_2 = z_2^*$, and $\partial_{z^*} \lrvert{z}^2 = 2 z$, $\partial_z \lrvert{z}^2 = 2 z^*$.

\subsection{Gradient w.r.t Complex Tensors} \label{sec:pre-complex-grad}

\begin{align}
  (\nabla_\bfvv f)^i := \partial_{(\bfvv^\dagger)_i} f = \partial_{(\bfvv^i)^*} f,
  \label{eqn:complex-grad}
  \quad \text{and similarly} \quad
  (\nabla_\bfA f)^i_j := \partial_{(\bfA^\dagger)^j_i} f = \partial_{(\bfA^i_j)^*} f.
\end{align}
When adapting the above generalized complex derivative \eqnref{general-complex-deriv},
\begin{align}
  \partial_{(\bfvv^i)^*} (\bfvv^j)^* z^k_l
  ={} & (\partial_{\Re \bfvv^i} + \ii \partial_{\Im \bfvv^i}) (\Re \bfvv^j \Re z^k_l + \Im \bfvv^j \Im z^k_l)
  = \delta^i_j \Re z^k_l + \ii \delta^i_j \Im z^k_l
  = \delta^i_j z^k_l, \\
  \partial_{(\bfvv^i)^*} \bfvv^j z^k_l
  ={} & (\partial_{\Re \bfvv^i} + \ii \partial_{\Im \bfvv^i}) (\Re \bfvv^j \Re z^k_l - \Im \bfvv^j \Im z^k_l)
  = \delta^{ij} \Re z^k_l - \ii \delta^{ij} \Im z^k_l
  = \delta^{ij} (z^k_l)^*
  = \delta^{ij} (z^*)^{l'}_{k'} \delta^{kk'} \delta_{ll'}.
\end{align}
Hence,
\begin{align}
  (\nabla_\bfvv \bfvv^\dagger \bfuu)^k = \partial_{(\bfvv^k)^*} (\bfvv^i)^* \bfuu^i = \delta^k_i \bfuu^i = \bfuu^k
  & \, \Longrightarrow \,
  \nabla_\bfvv \bfvv^\dagger \bfuu = \bfuu, \\
  (\nabla_\bfuu \bfvv^\dagger \bfuu)^k = \partial_{(\bfuu^k)^*} (\bfvv^i)^* \bfuu^i = \delta^{ki} \bfvv^{i'} \delta_{ii'} = \delta^k_{i'} \bfvv^{i'} = \bfvv^k
  & \, \Longrightarrow \,
  \nabla_\bfuu \bfvv^\dagger \bfuu = \bfvv,
  \label{eqn:complex-grad-inprod}
\end{align}
and
\begin{align}
  & (\nabla_\bfvv \bfvv^\dagger \bfA \bfvv)^k = \partial_{(\bfvv^k)^*} (\bfvv^i)^* \bfA^i_j \bfvv^j
  = \delta^k_i \bfA^i_j \bfvv^j + \delta^{kj} \bfvv^i (\bfA^i_{j'})^* \delta_{jj'}
  = \bfA^k_j \bfvv^j + \delta^k_{j'} \bfvv^i (\bfA^\dagger)^{j'}_i
  = \bfA^k_j \bfvv^j + \bfvv^i (\bfA^\dagger)^k_i \\
  \Longrightarrow {} \, &
  \nabla_\bfvv \bfvv^\dagger \bfA \bfvv = (\bfA + \bfA^\dagger) \bfvv
  \, \stackrel{\text{if $\bfA$ is Hermitian}}{\Longrightarrow} \,
  \nabla_\bfvv \bfvv^\dagger \bfA \bfvv = 2 \bfA \bfvv.
  \label{eqn:complex-grad-quadratic}
\end{align}
Also note $\nabla_{\bfvv^\dagger} f = (\nabla_\bfvv f)^\dagger$.

%% file: quantum-basics.tex
\section{Quantum Mechanics Basics} \label{sec:qm}

\begin{table}[h]
  \centering
  \caption{Timetable of quantum physics and quantum chemistry.}
  \label{tab:timetable}
  \begin{tabular}{c|c}
    \toprule
    \begin{tabular}[t]{rl}
      1900 & Planck's quantum hypothesis \\
      1913 & Bohr's atom model \\
      1924 & de Broglie wave \\
      1925 & Einstein's intro of de Broglie wave \\
      1926 & Schr\"odinger eq, \\
           & Pauli exclusion principle, \\
           & antisymm wavefn (Heisenberg, Dirac) \\
           & Born's probabilistic interp of wavefn \\
      1927 & Born-Oppenheimer approx, \\
           & Thomas's DFT (uniform electron gas) \\
      1928 & Hartree method, \\
           & Thomas-Fermi model \\
      \end{tabular}
      &
      \begin{tabular}[t]{rl}
      1929 & Slater determinant, \\
           & LCAO-MO (Lennard-Jones) \\
      1930 & Hartree-Fock method (and Slater), \\
           & Thomas-Fermi-Dirac method (w/ X fn'al) \\
      1935 & von Weizs\"acker's kinetic energy fn'al \\
      1950 & Roothaan method (also Hall) \\
      1964 & Hohenberg-Kohn theorem \\
      1965 & Kohn-Sham method \\
      1979 & Univ. fn'al (Levy constraint search) \\
      1975 & N-repr. of KS density (Gilbert) \\
      1983 & Levy-Lieb fn'al ($\exists$ lowest eng., solves V-repr.) \\
    \end{tabular}
    \\
    \bottomrule
  \end{tabular}
\end{table}

\subsection{Origin and Interpretation of Schr\"odinger Equation} \label{sec:qm-schr}

\subsubsection{History}
\begin{itemize}
  \item Originates from Planck’s explanation to black-body radiation.
  \item Bohr explored the idea of quantization and successfully explained the electron distribution of hydrogen atom.
  \item The foundation/principle of quantization is then developed in two ways: matrix mechanics (Heisenberg, Born, Pauli) and wave mechanics (Schr\"odinger).
    They are found equivalent (Dirac) and are two pictures of quantum mechanics.
  \item The Schr\"odinger picture is commonly considered.
\end{itemize}

\subsubsection{Wavefunction: quantum description of physical state}

De Broglie wave (1924): generalize the wave-particle duality from photons to matter particles.
\begin{itemize}
  \item A free particle with $(E, \ppv)$ $\Longleftrightarrow$ a plane wave
    \begin{align}
      \Psi(\rrv, t) = A \exp(-\ii (\omega t - \kkv \cdot \rrv)) = A \exp(-\ii (E t - \ppv \cdot \rrv) / \hbar),
      \quad \text{where }
      E = h \nu = \hbar \omega, \;
      \ppv = \text{`` } h / \lambdav \text{ ''} = \hbar \kkv.
    \end{align}
  \item A general particle $\Longleftrightarrow$ Superposition of multiple plane waves:
    \begin{align}
      \Psi(\rrv, t) = (2\pi \hbar)^{-3/2} \int_{\bbR^3} \ud \ppv \, \Phi(\ppv, t) \exp(\ii \ppv \cdot \rrv / \hbar),
    \end{align}
    where $\Phi(\ppv, t)$ is the ``superposition coefficient'' of different plane waves (different $\ppv$) from Fourier transform:\footnote{
      \label{ftn:fourier-transform}
      Why it holds over the unbounded space (while commonly it is over a period):
      Let $\int_\bbR \ud x \, \exp(\ii kx)$ denote the limit in a symmetric way:
      $\lim_{B \to \infty} \int_{-B}^B \ud x \, \exp(\ii kx)
      = \lim_{B \to \infty} \frac{1}{k} (\sin kx - \ii \cos kx) |_{-B}^B
      = 2 \lim_{B \to \infty} \frac{1}{k} \sin kB
      =: 2 g(k)$,
      which is $2\pi \delta(k)$ in the sense that:
      $\int_\bbR \ud k \, f(k) g(k) = \int_\bbR \ud k \, \lim_{B \to \infty} f(k) \frac{\sin kB}{k}
      = \int_\bbR \ud (kB) \, \lim_{B \to \infty} f(kB/B) \frac{\sin kB}{kB}
      \stackrel{\cdots}{=} \lim_{B \to \infty} \int_\bbR \ud \alpha \, f(\alpha/B) \frac{\sin \alpha}{\alpha}
      \stackrel{\cdots}{=} \int_\bbR \ud \alpha \, f(0) \frac{\sin \alpha}{\alpha}
      = \pi f(0)$.
      For the last equality, $\int_\bbR \ud \alpha \, \frac{\sin \alpha}{\alpha}
      = 2 \int_0^\infty \ud \alpha \, \frac{\sin \alpha}{\alpha}
      = 2 \int_0^\infty \ud \alpha \, \frac{\sin \alpha}{\alpha} \exp(-\nu \alpha) \big|_{\nu = 0}
      = 2 \int_0^\infty \ud \alpha \, \sin \alpha \int_0^\infty \ud v \, \exp(-v \alpha)
      = 2 \int_0^\infty \ud v \, \int_0^\infty \ud \alpha \, \sin \alpha \exp(-v \alpha)
      = 2 \int_0^\infty \ud v \, \frac{1}{1 + v^2}
      = 2 \arctan v \, |_0^\infty
      = \pi$, where we have used the Laplace transform of $\sin \alpha$.
      This leads to:
      \begin{align}
        \int_\bbR \ud x \, \exp(\ii kx) = 2\pi \delta(k),
        \quad \text{and} \quad
        \int_{\bbR^3} \ud \rrv \, \exp(\ii (\ppv - \ppv') \cdot \rrv / \hbar) = (2 \pi \hbar)^3 \delta(\ppv - \ppv').
      \end{align}
    }
    \begin{align}
      \Phi(\ppv, t) = (2\pi \hbar)^{-3/2} \int_{\bbR^3} \ud \rrv \, \Psi(\rrv, t) \exp(-\ii \ppv \cdot \rrv / \hbar).
    \end{align}
  \item Supported by the electron interference experiment (Davisson \& Germer, 1927).
\end{itemize}

Born's probabilistic interpretation of wavefunction (1926):
\begin{itemize}
  \item Particle density $\rho(\rrv, t) = \lrvert{\Psi(\rrv, t)}^2$.
  \item Mean position $\rrvb = \int \ud \rrv \, \psi^*(\rrv) \rrv \psi(\rrv)$ at some time $t_0$ ($\psi(\rrv) := \Psi(\rrv, t_0)$).
  \item Mean momentum $\ppvb = \int \ud \ppv \, \phi^*(\ppv) \ppv \phi(\ppv)$.
    Using Fourier transform for $\phi$ from $\psi$, we have:\footnote{\vspace{-12pt}\begin{align}
        \ppvb ={} &  (2\pi \hbar)^{-3/2} \int \ud \ppv \Big( \int \ud \rrv \, \psi(\rrv) \exp(-\ii \ppv \cdot \rrv / \hbar) \Big)^* \ppv \phi(\ppv)
        =     (2\pi \hbar)^{-3/2} \int \ud \ppv \int \ud \rrv \, \psi^*(\rrv) \exp(\ii \ppv \cdot \rrv / \hbar) \ppv \phi(\ppv) \\
        ={} & (2\pi \hbar)^{-3/2} \int \ud \rrv \, \psi^*(\rrv) \int \ud \ppv \, \exp(\ii \ppv \cdot \rrv / \hbar) \ppv \phi(\ppv)
        =     (2\pi \hbar)^{-3/2} \int \ud \rrv \, \psi^*(\rrv) \int \ud \ppv \, (-\ii \hbar) \nabla \exp(\ii \ppv \cdot \rrv / \hbar) \phi(\ppv) \\
        ={} & (2\pi \hbar)^{-3/2} \int \ud \rrv \, \psi^*(\rrv) (-\ii \hbar) \nabla \int \ud \ppv \, \exp(\ii \ppv \cdot \rrv / \hbar) \phi(\ppv)
        =     \int \ud \rrv \, \psi^*(\rrv) (-\ii \hbar) \nabla \psi(\rrv).
    \end{align}}
    \begin{align}
      \ppvb = \int \ud \rrv \, \psi^*(\rrv) \ppvh \psi(\rrv), \quad
      \text{where $\ppvh := -\ii \hbar \nabla$ is the momentum operator}.
    \end{align}
  \item Mean kinetic energy $\Tb = \int \ud \ppv \, \phi^*(\ppv) \frac{\ppv^2}{2m} \phi(\ppv)$, or in terms of $\psi(\rrv)$,
    $\Tb = \int \ud \rrv \, \psi^*(\rrv) \Th \psi(\rrv)$, where $\Th := \frac{\ppvh^2}{2m} = -\frac{\hbar^2}{2m} \nabla^2$.
\end{itemize}

Quantum mechanics probability vs. statistical mechanics probability:
\begin{itemize}
  \item In QM, the distribution $\rho_{(n)}(\rrv) = \lrvert{\psi_{(n)}(\rrv)}^2$ describes the \emph{spatial uncertainty in one state}. It is one state.
  \item In SM, the distribution is over \emph{states}: $p_{(n)}$ for discrete energy states or $p(E)$ for continuous energy states.
\end{itemize}

\subsubsection{Schr\"odinger equation: quantum description of physical dynamics}

\begin{align}
  \ii \hbar \fracpartial{}{t} \Psi(\rrv, t) = \Hh \Psi(\rrv, t), \quad
  \text{where $\Hh := \Th + V(\rrv)$ is the Hamiltonian operator}.
  \label{eqn:schrodinger}
\end{align}

\paragraph{Intuition 1:}
For a free particle described by a plane wave function $\Psi(\rrv, t) = A \exp(-\ii (E t - \ppv \cdot \rrv) / \hbar)$,
l.h.s $= -\frac{\hbar^2}{2m} \nabla^2 \Psi(\rrv, t) = \frac{\ppv^2}{2m} \Psi(\rrv, t) = E \Psi(\rrv, t) =$ r.h.s.
For a general particle, since the equation is linear, it is similar by superpositioning plane waves.

\paragraph{Intuition 2:} Conservation of particle number/probability (when the potential $V$ is real).
By Schr\"odinger equation,
\begin{align}
  \fracpartial{\rho(\rrv, t)}{t} \equiv{} & \Psi^* \fracpartial{\Psi}{t} + \fracpartial{\Psi^*}{t} \Psi
  = \ii \frac{\hbar}{2m} (\Psi^* \nabla^2 \Psi - \Psi \nabla^2 \Psi^*) = -\nabla \cdot \Jv, \\
  \text{where } \Jv :={} & -\ii \frac{\hbar}{2m} (\Psi^* \nabla \Psi - \Psi \nabla \Psi^*)
  = \frac{1}{2m} (\Psi^* \ppvh \Psi - \Psi \ppvh \Psi^*)
  = \frac{1}{2m} (\Psi^* \ppvh \Psi + \Psi \ppvh^* \Psi^*) \\
  ={} & \Re(\Psi^* \vvvh \Psi), \quad \vvvh := \ppvh / m, \\
  ={} & \rho(\rrv, t) \frac{1}{2} \lrparen*[\Big]{ \frac{\vvvh \Psi}{\Psi} + \frac{\vvvh^* \Psi^*}{\Psi^*} }
  = \rho(\rrv, t) \Re \lrparen*[\big]{ \frac{\vvvh \Psi}{\Psi} }.
\end{align}
This is in the form of \textbf{continuity equation}.
Particularly, for a free particle, $\Jv = \frac{1}{2m} (\Psi^* \ppv \Psi + \Psi \ppv \Psi^*) = \Psi^* \Psi \frac{\ppv}{m} = \rho \vvv$.

\paragraph{Intuition 3:} Schr\"odinger's explanation to quantization as the discrete spectrum of $\Hh$, based on an analytical mechanics formulation. See \secref{analytical-mechanics} below.

\paragraph{Stationary Schr\"odinger equation}
When $\Hh$ does not explicitly depend on time, $\Psi(\rrv, t) = \psi(\rrv) \chi(t)$ can be separated (not necessarily),
which turns the equation to $\psi(\rrv) \ii \hbar \chi'(t) = (\Hh \psi(\rrv)) \chi(t)$, or $\frac{1}{\chi(t)} \ii \hbar \chi'(t) = \frac{1}{\psi(\rrv)} \Hh \psi(\rrv)$.
The l.h.s is constant of $\rrv$ and the r.h.s is constant of $t$, so both sides must be a constant $E$ of both $\rrv$ and $t$.
This leads to the original/stationary Schr\"odinger equation:
\begin{align}
  \Hh \psi(\rrv) = E \psi(\rrv),
  \label{eqn:schrodinger-stationary}
\end{align}
and $\ii \hbar \chi'(t) = E \chi(t)$ whose solution is $\chi(t) = A \exp(-\ii Et/\hbar)$.
A solution $(E_{(n)}, \psi_{(n)})$ to the stationary equation is called a \textbf{stationary state} or \textbf{eigenstate}.
Its total wavefunction
\begin{align}
  \Psi_{(n)}(\rrv, t) = A_{(n)} \psi_{(n)}(\rrv) \exp(-\ii E_{(n)} t / \hbar)
\end{align}
makes mechanical quantities stationary (time-independent):
$\Ob := \int \ud \rrv \, \Psi_{(n)}^*(\rrv, t) \Oh \Psi_{(n)}(\rrv, t) = \lrvert{A_{(n)}}^2 \int \ud \rrv \, \psi_{(n)}^*(\rrv) \Oh \psi_{(n)}(\rrv)$.

General solution in the time-independent Hamiltonian case is in the form
\begin{align}
  \Psi(\rrv, t) = \sum_n c_{(n)} \psi_{(n)}(\rrv) \exp(-\ii E_{(n)} t / \hbar).
\end{align}
\emph{Each physical measurement only observes one single stationary state (wavefunction collapse)}.
Let $\{\psi_{(n)}\}_n$ be orthonormalized, which is achievable.
Then the energy $\Eb = \sum_n \lrvert{c_{(n)}}^2 E_{(n)}$,\footnote{$
  \Eb = \int \ud \rrv \, \Psi^*(\rrv, t) \Hh \Psi^*(\rrv, t)
  = \int \ud \rrv \big( \sum_n c_{(n)}^* \psi_{(n)}^*(\rrv) \exp(\ii E_{(n)} t / \hbar) \big) \big( \sum_{n'} E_{(n')} c_{(n')}^* \psi_{(n')}(\rrv) \exp(-\ii E_{(n')} t / \hbar) \big)
  = \sum_{n,n'} c_{(n)}^* c_{(n')} E_{(n')} \exp(\ii (E_{(n)} - E_{(n')}) t / \hbar) \int \ud \rrv \, \psi_{(n)}^*(\rrv) \psi_{(n')}(\rrv)
  = \sum_{n,n'} c_{(n)}^* c_{(n')} E_{(n')} \exp(\ii (E_{(n)} - E_{(n')}) t / \hbar) \delta_{n,n'}
$.}
so $\lrvert{c_{(n)}}^2$ is the probability to be in state $n$.

\subsubsection{Analytical mechanics origin: Schr\"odinger's intuition} \label{sec:analytical-mechanics}

\paragraph{Lagrangian mechanics and the least action principle}
For $N$ particles with general coordinates $\bmrrv := (\rrv_1, \cdots, \rrv_N)$ moving in potential $V(\bmrrv, t)$ with kinetic energy $T(\bmrrvd, t)$, %
define their \textbf{Lagrangian}, and the \textbf{action} along a curve $(\bmrrv_t)_{t \in [t_1, t_2]}$ as:
\begin{align}
  L(\bmrrv, \bmrrvd, t) := T(\bmrrvd, t) - V(\bmrrv, t),
  \qquad
  \clS[(\bmrrv_t)_{t \in [t_1, t_2]}] := \int_{t_1}^{t_2} \ud t \, L(\bmrrv_t, \bmrrvd_t, t).
\end{align}
Fixing the ending points of all curves, the dynamics between the two points, \ie, the curve $(\bmrrv_t)_{t \in [t_1, t_2]}$ that the particles move along, is given by the \textbf{least action principle} as the \textbf{extremal curve}.
Explicitly, it is given by the \textbf{Euler-Lagrange equation (E-L equation)} (Euler, 1753; Lagrange, 1754; Euler, 1766):\footnote{
  For \eqnref{action-variation}, $
  \rmdelta \clS = \int_{t_1}^{t_2} \ud t \, \lrparen{ \nabla_{\bmrrv} L \cdot \rmdelta \bmrrv_t + \nabla_{\bmrrvd} L \cdot \fracdiff{}{t} \rmdelta \bmrrv_t }
  = \int_{t_1}^{t_2} \ud t \, \nabla_{\bmrrv} L \cdot \rmdelta \bmrrv_t
  + \int_{t_1}^{t_2} \ud t \, \fracdiff{}{t} \lrparen{ \nabla_{\bmrrvd} L \cdot \rmdelta \bmrrv_t }
  - \int_{t_1}^{t_2} \ud t \, \lrparen{ \fracdiff{}{t} \nabla_{\bmrrvd} L } \cdot \rmdelta \bmrrv_t
  = \left. \nabla_{\bmrrvd} L \cdot \rmdelta \bmrrv_t \right|_{t_1}^{t_2}
    + \int_{t_1}^{t_2} \ud t \, \lrparen{ \nabla_{\bmrrv} L - \fracdiff{}{t} \nabla_{\bmrrvd} L } \cdot \rmdelta \bmrrv_t
  $.
}
\begin{gather}
  \rmdelta \clS
  = \left. \nabla_{\bmrrvd} L \cdot \rmdelta \bmrrv_t \right|_{t_1}^{t_2}
    + \int_{t_1}^{t_2} \ud t \, \lrparen{ \nabla_{\bmrrv} L - \fracdiff{}{t} \nabla_{\bmrrvd} L } \cdot \rmdelta \bmrrv_t
  = 0, \quad
  \forall \rmdelta \bmrrv_t \st \rmdelta \bmrrv_{t_1} = \rmdelta \bmrrv_{t_2} = 0, \label{eqn:action-variation} \\
  \Longrightarrow \quad
  \nabla_{\bmrrv} L - \fracdiff{}{t} \nabla_{\bmrrvd} L = 0, \forall \text{ a.e. } t \in [t_1, t_2]. \label{eqn:euler-lagrange}
\end{gather}
For classical particles and Euclidean coordinates, $T(\bmrrvd, t) = \sum_i \frac{1}{2} m_i \rrvd_i^2$, which recovers Newton's law, $m_i \rrvdd_i = -\nabla_{\rrv_i} V$.

\paragraph{Noether's theorem} Uniformities lead to conservatives (1918).
\begin{itemize}
  \item Uniformity in space: the replacement $\bmrrv \mapsto \bmrrv + \rmDelta \rrv, \forall \rmDelta \rrv$ gives the same Lagrangian. So $
    \rmDelta L = \sum_i \nabla_{\rrv_i} L \cdot \rmDelta \rrv = 0,
    \Longrightarrow
    \sum_i \nabla_{\rrv_i} L = 0$ so $V$ is uniform, and $
    \stackrel{\text{\eqnref{euler-lagrange}}}{\Longrightarrow}
    \fracdiff{}{t} \sum_i \nabla_{\rrvd_i} L = 0$,
    so the (total) \textbf{momentum} $\Pv := \sum_i \nabla_{\rrvd_i} L$ is a conservative.
  \item Uniformity in time: the replacement $t \mapsto t + \rmDelta t, \forall \rmDelta t$ gives the same Lagrangian. So $
    \rmDelta L = \partial_t L \rmDelta t = 0,
    \Longrightarrow
    \partial_t L = 0$ so $L$ thus $V$ and $T$ does not explicitly depend on time, so $
    \fracdiff{L(\bmrrv, \bmrrvd)}{t} = \sum_i \nabla_{\rrv_i} L \cdot \rrvd_i + \sum_i \nabla_{\rrvd_i} L \cdot \rrvdd_i,
    \stackrel{\text{\eqnref{euler-lagrange}}}{\Longrightarrow}
    \fracdiff{L}{t}
    = \sum_i \fracdiff{}{t} \left( \nabla_{\rrvd_i} L \cdot \rrvd_i \right),
    \Longrightarrow
    \fracdiff{}{t} \left( \sum_i \nabla_{\rrvd_i} L \cdot \rrvd_i - L \right) = 0$,
    so the \textbf{energy} $E := \sum_i \nabla_{\rrvd_i} L \cdot \rrvd_i - L$ is a conservative.
\end{itemize}

\paragraph{Hamiltonian mechanics}
Formulation based on energy (conservative) thus superior to Lagrangian, excels in applicability to mechanics problems.
Define the (general) \textbf{momentum} $\ppv_i := \nabla_{\rrvd_i} L(\bmrrv, \bmrrvd, t)$,
and the \textbf{Hamiltonian} as the Legendre transform of the Lagrangian $L$ in argument $\bmrrvd$:
\begin{align}
  H(\bmrrv, \bmppv, t) := \sup_{\bmrrvd} \Big( \sum_i \ppv_i \cdot \rrvd_i - L(\bmrrv, \bmrrvd, t) \Big)
  = \bmppv \cdot \bmrrvd - L(\bmrrv, \bmrrvd, t) \Big|_{\bmrrvd = \bmvvv(\bmrrv, \bmppv, t) \text{ solved from } \bmppv = \nabla_{\bmrrvd} L(\bmrrv, \bmrrvd, t)},
  \label{eqn:am-hamiltonian}
\end{align}
which takes $\bmppv$ as the basic/free variable in place of $\bmrrvd$.
Then we have:
\begin{align}
  \begin{cases}
    \nabla_{\bmrrv} H
    = \bmppv \cdot \nabla_{\bmrrv} \bmvvv - \nabla_{\bmrrv} L - \nabla_{\bmrrvd} L \cdot \nabla_{\bmrrv} \bmvvv
    = -\nabla_{\bmrrv} L, \\
    \nabla_{\bmppv} H
    = \bmvvv + \bmppv \cdot \nabla_{\bmppv} \bmvvv - \nabla_{\bmrrvd} L \cdot \nabla_{\bmppv} \bmvvv
    = \bmvvv, \\
    \partial_t H = \bmppv \cdot \partial_t \bmvvv - \partial_t L - \nabla_{\bmrrvd} L \cdot \partial_t \bmvvv
    = -\partial_t L.
  \end{cases}
\end{align}
The E-L \eqnref{euler-lagrange} becomes $\nabla_{\bmrrv} L - \fracdiff{}{t} \bmppv = 0$, \ie, $\nabla_{\bmrrv} L = \bmppvd$.
This gives the \textbf{canonical equation} / \textbf{Hamilton equation}:
\begin{align}
  \bmrrvd = \nabla_{\bmppv} H, \quad \bmppvd = -\nabla_{\bmrrv} H.
  \label{eqn:ham-canonical}
\end{align}
Alternatively, when $L$ does not explicitly depend on time, the total differential
$\ud H = \bmppv \cdot \ud \bmrrvd + \bmrrvd \cdot \ud \bmppv - \nabla_{\bmrrv} L \cdot \ud \bmrrv - \nabla_{\bmrrvd} L \cdot \ud \bmrrvd
= \bmppv \cdot \ud \bmrrvd + \bmrrvd \cdot \ud \bmppv - \bmppvd \cdot \ud \bmrrv - \bmppv \cdot \ud \bmrrvd
= \bmrrvd \cdot \ud \bmppv - \bmppvd \cdot \ud \bmrrv$
also gives \eqnref{ham-canonical}.
In this case, $H(\bmrrv, \bmppv) \equiv E$ also does not explicitly depend on time and is conserved.

\paragraph{Hamilton-Jacobi equation}
Consider the action in Hamilton's formulation. Define \textbf{Hamilton's principal function}:
\begin{align}
  S(\bmrrv, t; \bmrrv_0) := \min_{(\bmgammav_\tau)_{\tau \in [0,t]}: \bmgammav_0 = \bmrrv_0, \bmgammav_t = \bmrrv} \clS[(\bmgammav_\tau)_\tau]
  = \int_0^t \ud \tau \, L(\bmgammav^\star_\tau, \bmgammavd^\star_\tau, \tau),
\end{align}
\ie, the least action between $\bmrrv$ at time $t$ and some given $\bmrrv_0$ at time $0$, where $(\bmgammav^\star_\tau)_\tau$ is the extremal curve between the two sets of points.
\itemone From \eqnref{action-variation}, $\nabla_{\bmrrv} S
  = \left. \nabla_{\bmrrvd} L \cdot \nabla_{\bmrrv} \bmgammav^\star_\tau \right|_0^t
    + \int_0^t \ud \tau \, \lrparen{ \nabla_{\bmrrv} L - \fracdiff{}{\tau} \nabla_{\bmrrvd} L } \cdot \nabla_{\bmrrv} \bmgammav^\star_\tau
  $, where the derivatives of $L$ are evaluated at $(\bmgammav^\star_\tau, \bmgammavd^\star_\tau, \tau)$ at each $\tau$.
  Since $(\bmgammav^\star_\tau)_\tau$ is the extremal curve, $\nabla_{\bmrrv} L - \fracdiff{}{\tau} \nabla_{\bmrrvd} L \equiv 0$,
  and $\nabla_{\bmrrv} \bmgammav^\star_0 = 0$,
  so $\nabla_{\bmrrv} S
  = \nabla_{\bmrrvd} L(\bmgammav^\star_t, \bmgammavd^\star_t, t) \cdot \nabla_{\bmrrv} \bmgammav_t^\star
  = \nabla_{\bmrrvd} L(\bmgammav^\star_t, \bmgammavd^\star_t, t)$,
  which is $\bmppv$ on the moving curve:
  \begin{align}
    \nabla_{\bmrrv} S = \bmppv.
  \end{align}
\itemtwo Consider a physical moving curve $\bmrrv_t$.
  Then we have $\fracdiff{S(\bmrrv_t, t; \bmrrv_0)}{t} = L(\bmrrv, \bmrrvd, t)$.\footnote{
    The change of $\bmrrv_t$ by $t$ yields the same extremal curve since $\bmrrv_t$ is a physical moving curve:
    $(\bmgammav^{\star (\bmrrv_{t + s}, t + s)}_\tau)_{\tau \in [0, t]} = (\bmgammav^{\star (\bmrrv_t, t)}_\tau)_{\tau \in [0, t]}$,
    so the integrand is a function only of $t$.
    Alternatively, $\fracdiff{S(\bmrrv_t, t; \bmrrv_0)}{t}
    = L(\bmrrv, \bmrrvd, t) + \int_0^t \ud \tau \, \lrparen{ \nabla_{\bmrrv} L - \fracdiff{}{\tau} \nabla_{\bmrrvd} L } \cdot \partial_t \bmgammav^\star_\tau
    = L(\bmrrv, \bmrrvd, t)$.
  }
  On the other hand, formally, $\fracdiff{S(\bmrrv_t, t; \bmrrv_0)}{t}
  = \nabla_{\bmrrv} S \cdot \bmrrvd + \partial_t S = \bmppv \cdot \bmrrvd + \partial_t S$,
  which is $H(\bmrrv, \bmppv, t) + L(\bmrrv, \bmrrvd, t) + \partial_t S$ by definition \eqnref{am-hamiltonian}.
  Then we arrive at the (time-dependent) \textbf{Hamilton-Jacobi equation (HJE)}:
  \begin{align}
    -\partial_t S = H \Big( \bmrrv, \nabla_{\bmrrv} S, t \Big).
    \label{eqn:hamilton-jacobi}
  \end{align}
  When the Hamiltonian does not explicitly depend on time $t$, it is a constant $E$, which leads to $\partial_t S = -E$, $\Longrightarrow S(\bmrrv, t) = S_0(\bmrrv) - E t$, and the time-independent Hamilton-Jacobi equation:
  \begin{align}
    H \Big( \bmrrv, \nabla_{\bmrrv} S_0 \Big) = E.
    \label{eqn:hamilton-jacobi-time-indep}
  \end{align}
  \begin{itemize}
    \item HJE: a single first-order PDE for a function of $N$ coordinates and time. \\
      E-L Eq.: time evolution of $N$ coordinates; $N$ equations; second-order. \\
      Hamilton Eq.: time evolution of $2N$ coordinates; $2N$ equations; first-order.
    \item HJE ``is particularly useful in identifying conserved quantities for mechanical systems''.
    \item HJE ``is also the only formulation of mechanics in which the motion of a particle can be represented as a wave'', ``analogy between the propagation of light and the motion of a particle''.
  \end{itemize}

\paragraph{Connection to Schr\"odinger equation}
HJE generalizes the duality between trajectories and wave fronts by the variational principle in geometrical optics to mechanical systems:
\begin{align}
  \text{wave-front (isosurface of $S(\bmrrv, t)$) } \underset{\text{HJE}}{\stackrel{\text{E-L}}{\rightleftharpoons}} \text{ trajectory}.
\end{align}
So we can treat $S$ as the phase of a wave: $\Psi(\bmrrv, t) = A \exp(\ii S(\bmrrv, t) / \hbar)$ (let $S$ be complex to allow change of amplitude),
which leads to: $S(\bmrrv, t) = -\ii \hbar \log \Psi(\bmrrv, t) + \const$. So:
\begin{align}
  \bmppv \Psi = \nabla_{\bmrrv} S \Psi = -\ii \hbar \nabla_{\bmrrv} \Psi,
\end{align}
which explains the momentum operator $\ppvh = -\ii \hbar \nabla$.
HJE~\eqref{eqn:hamilton-jacobi} then yields:
\begin{align}
  \ii \hbar \partial_t  \Psi = H \Big( \bmrrv, -\ii \hbar \frac{1}{\Psi} \nabla_{\bmrrv} \Psi, t \Big) \Psi,
  \label{eqn:hje-schrodinger}
\end{align}
which explains Schr\"odinger \eqnref{schrodinger} $\ii \hbar \partial_t \Psi = \Hh \Psi$.
For the time-independent case, we first have:
$\Psi(\bmrrv, t) = A \exp(\ii S_0(\bmrrv) / \hbar) \exp(-\ii E t / \hbar) =: \psi(\bmrrv) \exp(-\ii E t / \hbar)$ is separable.
By the time-independent HJE \eqnref{hamilton-jacobi-time-indep}, we have:
\begin{align}
  H \Big( \bmrrv, -\ii \hbar \frac{1}{\psi} \nabla_{\bmrrv} \psi \Big) \psi = E \psi,
  \label{eqn:hje-schrodinger-stationary}
\end{align}
which explains the stationary Schr\"odinger \eqnref{schrodinger-stationary} $\Hh \psi = E \psi$.
For $H(\bmrrv, \bmppv) = \sum_i \frac{\ppv_i^2}{2 m_i} + V(\bmrrv)$, \eqnref{hje-schrodinger-stationary} becomes $0 =
\sum_i \frac{1}{2 m_i} \lrparen{-\ii \hbar \frac{1}{\psi} \nabla_{\rrv_i} \psi}^* \cdot \lrparen{-\ii \hbar \frac{1}{\psi} \nabla_{\rrv_i} \psi} \psi + (V - E) \psi
= \sum_i \frac{\hbar^2}{2 m_i} \frac{\psi}{\lrvert{\psi}^2} \lrVert{\nabla_{\rrv_i} \psi}^2 + (V - E) \psi$.
Multiplied by $\psi^*$, it becomes $\sum_i \frac{\hbar^2}{2 m_i} \lrVert{\nabla_{\rrv_i} \psi}^2 + (V - E) \lrvert{\psi}^2
= \sum_i \frac{\hbar^2}{2 m_i} \lrparen{ \lrVert{\nabla_{\rrv_i} u}^2 + \lrVert{\nabla_{\rrv_i} v}^2 } + (V - E) (u^2 + v^2)
= 0$, where $u$, $v$ denote the real, imaginary parts of $\psi$, which are independent $\bbR^{3N} \to \bbR$ functions.
So the variational principle of its integral is applied for $u$ and $v$ separately, yielding $
2 (V - E) u - \sum_i \nabla_{\rrv_i}  \cdot \lrparen{ 2 \frac{\hbar^2}{2 m_i} \nabla_{\rrv_i} u } = 0
$ and similarly for $v$.
Written combined, the result is:
\begin{align}
  -\sum_i \frac{\hbar^2}{2 m_i} \nabla^2_{\rrv_i} \psi + V \psi = E \psi,
  \label{eqn:schrodinger-stationary-common}
\end{align}
which explains the Hamilton operator $\Hh = -\sum_i \frac{\hbar^2}{2 m_i} \nabla^2_i + V$.\footnote{
  Another intuition: $\int_{\bbR^3} \ud \rrv_i \, \lrVert*[\big]{ \nabla_{\rrv_i} \psi }^2
  = \int_{\bbR^3} \ud \rrv_i \, \nabla_i \psi^* \cdot \nabla_i \psi
  = \int_{\bbR^3} \ud \rrv_i \, \nabla_i \cdot (\psi^* \nabla_i \psi) - \int_{\bbR^3} \ud \rrv_i \, \psi^* \nabla^2_i \psi
  = -\int_{\bbR^3} \ud \rrv_i \, \psi^* \nabla_{\rrv_i}  \cdot \nabla_{\rrv_i}  \psi$,
  if $\psi$ diminishes sufficiently fast to make $\oint_{\partial \bbR^3} \ud \Sv_i \cdot (\psi^* \nabla_i \psi) = 0$.
}
Adopting the general complex derivative \eqnref{general-complex-deriv}, \eqnref{schrodinger-stationary-common} can also be seen as derived from:
\begin{align}
  \fracdelta{}{\psi^*} \Big( \sum_i \frac{\hbar^2}{2 m_i} \lrVert{\nabla_{\rrv_i} \psi}^2 + (V - E) \lrvert{\psi}^2 \big) = 0.
  \label{eqn:schrodinger-variation-hje}
\end{align}

Note that the eigen-energy as the solution to \eqnref{schrodinger-stationary} must be real,
otherwise the wavefunction $\chi(t) = A \exp(-\ii E t / \hbar)$ would go to infinity for large $t$.
Other physical quantities also must be real, since physical measurement is real. So,
\begin{align}
  & \text{Any mechanical quantity must be real} \\
  \Longleftrightarrow{} & \text{All eigenvalues of the corresponding operator is real} \\
  \Longleftrightarrow{} & \text{The operator is \textbf{Hermitian (self-adjoint)}: } \braket*{\phi}{\Oh \psi} = \braket*{\Oh \phi}{\psi}.
\end{align}

\subsection{Basic Conclusions} \label{sec:qm-basic}

Locally integrable functions have weak derivative.
Note the test functions in defining the weak derivative is compactly supported.

The $N$-particle wavefunction $\psi(\rmxx_1, \cdots, \rmxx_N)$ has zero boundary integral to make the kinetic energy operator $\Th$ Hermitian:
$\braket*{\psi}{\Th \phi} = \obraket*{\psi}{\Th}{\phi} = -\sum_i \int \ud \rmxx_i \, \psi^* \nabla_i \cdot (\nabla_i \phi)
= -\sum_i \int \ud \rmxx_i \, \nabla_i \cdot (\psi^* \nabla_i \phi) + \sum_i \int \ud \rmxx_i \, \nabla_i \psi^* \cdot \nabla_i \phi$,
and symmetrically
$\braket*{\Th \psi}{\phi} = \obraket*{\phi}{\Th}{\psi}^* = -\sum_i \int \ud \rmxx_i \, \nabla_i \cdot (\phi \nabla_i \psi^*) + \sum_i \int \ud \rmxx_i \, \nabla_i \phi \cdot \nabla_i \psi^*$.
To make them equal, $0 = \sum_i \int \ud \rmxx_i \, \big( \nabla_i \cdot (\psi^* \nabla_i \phi) - \nabla_i \cdot (\phi \nabla_i \psi^*) \big)
= \sum_i \sum_{s_i} \lim_{r_i \to \infty} \oint_{\bbS(r_i)} \ud \Sv \cdot \big( \psi^* \nabla_i \phi - \phi \nabla_i \psi^* \big)$ for any $\psi$ and $\phi$, so each term has to be zero.
We then have $\obraket*{\psi}{\Th}{\psi} = \sum_i \int \ud \rmxx_i \, \lrVert{\nabla_i \psi}^2$.

$\psi(\rmxx_1, \cdots, \rmxx_N)$ is in the $(1,2)$-Sobolev space $\bbH^1$ ($\psi$ and all its 1st-order weak derivatives are in $\bbL^2$; it is a Hilbert space)~\citep{lieb1983density}, in order to make kinetic energy finite.

The ground state is spacially non-degenerate. If the Hamiltonian contains spin, the ground state seems non-degenerate ([\href{https://thesis.library.caltech.edu/1007/1/Cohen_m_1956.pdf}{ref. thesis}, Appendix~A]).

DFT initialization methods:
[\href{https://manual.q-chem.com/4.3/sect-initialguess.html}{ref. webpage}]

Natural orbitals: eigenvectors the 1-RDM of a N-wavefunction.
Natural atomic orbital: atomic orbitals of an atom in a molecular environment%
. Can be constructed from free-atom natural atomic orbitals.
[\href{https://nbo7.chem.wisc.edu/webnbo_css.htm}{ref. webpage}].

H\"uckel method: assume $\pi$-bond MOs are the linear combinations of (unhybridized) p-atomic-orbitals of involved atoms.
[\href{https://chem.libretexts.org/Bookshelves/Inorganic_Chemistry/Map\%3A_Inorganic_Chemistry_(Housecroft)/04\%3A_Experimental_techniques/4.13\%3A_Computational_Methods/4.13C\%3A_Huckel_MO_Theory}{ref. webpage}].
Extended H\"uckel method: also consider $\sigma$-bonds [\href{https://en.wikipedia.org/wiki/Extended_H\%C3\%BCckel_method}{Wikipedia}][Hoffmann 1963],
which leverages the Wolfsberg-Helmholz approximation [1952] for off-diagonal Hamiltonian elements, and uses approximate valence state ionization potentials (IPs) for diagonal elements.

Differential virial theorem and implication to $E_\XCb$: \citep{holas1995exact}.

\paragraph{Mixed state vs. superposition of eigenstates}
``The density matrix was first introduced by von Neumann (von Neumann 1927) and Landau (Landau 1927) independently to describe the quantum mechanical natures of statistical systems.''~\citep{tsuneda2014density}.

([\href{https://physics.stackexchange.com/questions/80434/how-is-a-quantum-superposition-different-from-a-mixed-state}{ref. post}],
[\href{https://chem.libretexts.org/Bookshelves/Physical_and_Theoretical_Chemistry_Textbook_Maps/Supplemental_Modules_(Physical_and_Theoretical_Chemistry)/Quantum_Tutorials_(Rioux)/Quantum_Fundamentals/83\%3A_Visualizing_the_Difference_Between_a_Superposition_and_a_Mixture}{ref. webpage}])

Let $\{\ket{\psi_{(1)}}, \ket{\psi_{(2)}}\}$ be orthonormal eigenstates of observable $\Ah$.
Then the equally-weighted \emph{superposition state}, $\ket{\psi} = \frac{1}{\sqrt{2}} (\ket{\psi_{(1)}} + \ket{\psi_{(2)}})$, is a pure state, meaning that
``there is not a 50\% chance the system is in state $\ket{\psi_{(1)}}$ and 50\% in state $\ket{\psi_{(2)}}$, but there is a 0\% chance that the system is in either state, and a 100\% chance the system is in state $\ket{\psi}$;
the point is that these statements are all made before making any measurements.''

Although both the superposition state, whose density matrix is $\Gamma_{[\psi]} = \ket{\psi}\bra{\psi} = \frac{1}{2} \begin{psmallmatrix} 1 & 1 \\ 1 & 1 \end{psmallmatrix}$,
and the equally-weighted \emph{mixed state} of the two pure states $\ket{\psi_{(1)}}$ and $\ket{\psi_{(2)}}$, whose density matrix is $\Gamma_\text{mix} = \frac{1}{2} \begin{psmallmatrix} 1 & 0 \\ 0 & 1 \end{psmallmatrix}$,
give the same $\Ah$ measurement result of 50\% $A_{(1)}$ and 50\% $A_{(2)}$,
they differ when measuring a second observable $\Bh$ s.t. $[\Ah, \Bh] \ne 0$.
Let $\{\ket{\phi_{(1)}}, \ket{\phi_{(2)}}\}$ be orthonormal eigenstates of $\Bh$ and suppose $\ket{\psi_{(1)}} = \frac{1}{\sqrt{2}} (\ket{\phi_{(1)}} + \ket{\phi_{(2)}})$ and $\ket{\psi_{(2)}} = \frac{1}{\sqrt{2}} (\ket{\phi_{(1)}} - \ket{\phi_{(2)}})$ (the relation must not be diagonal due to the non-commutability).
The superposition state $\ket{\psi} = \ket{\phi_{(1)}}$ so it gives 100\% $B_{(1)}$ and 0\% $B_{(2)}$ measurement result,
while the mixed state $\Gamma_\text{mix}$ takes the same matrix form under the $\{\phi_{(1)}, \phi_{(2)}\}$ basis so it still gives 50\% $B_{(1)}$ and 50\% $B_{(2)}$ measurement result.

A general \emph{density operator} is a trace-one linear kernel (positive semi-definite Hermitian operator) on the Hilbert space of quantum states, and satisfies proper symmetry in many-body cases.
The spectral theorem of kernels gives it a universal expression:
\begin{align}
  \exists \text{ orthonormal } \{\ket{\kappa}\}_\kappa, \text{ real non-negative } \{\lambda_\kappa\}_\kappa \text{ with } \sum_\kappa \lambda_\kappa = 1,
  \st \Gamma = \sum_\kappa \lambda_\kappa \ket{\kappa} \bra{\kappa}.
  \label{eqn:kernel-spectral}
\end{align}

For a pure state, $\Ab = \obraket*{\psi}{\Ah}{\psi} = \tr\big( \Ah \ket{\psi} \bra{\psi} \big) = \tr( \Ah \Gamma_{[\psi]} )$,
so we can extend the measurement to be taken under a general density matrix state:
\begin{align}
  \Ab = \tr(\Ah \Gamma).
  \label{eqn:mean-dm}
\end{align}
For a complete basis $\{\ket{\alpha}\}_\alpha$, let $\bfS^{\alpha'}_\alpha := \braket{\alpha'}{\alpha}$ be the overlap matrix (or, Gram matrix).
Then for any vector $\ket{v}$, its expansion is $\ket{v} = \sum_\alpha \bfvv^\alpha \ket{\alpha}$, where $\bfvv^\alpha = \sum_\beta (\bfS^{-1})^\alpha_\beta \braket{\beta}{v}$, \ie, the projection operator is $\sum_{\alpha \beta} \ket{\alpha} (\bfS^{-1})^\alpha_\beta \bra{\beta}$.
An observable operator has a common matrix expression:
\begin{align}
  \Ah = \sum_{\alpha,\alpha'} \bfA^\alpha_{\alpha'} \ket{\alpha} \bra{\alpha'}, \quad
  \text{where } \bfA = \bfS^{-1} \bfAt \bfS^{-1}, \bfAt^\alpha_{\alpha'} := \obraket*{\alpha}{\Ah}{\alpha'},
  \label{eqn:opr-matrix}
\end{align}
and so does the density matrix:
\begin{align}
  \Gamma = \sum_{\beta,\beta'} \bfGamma^\beta_{\beta'} \ket{\beta} \bra{\beta'}, \quad
  \text{where } \bfGamma = \bfS^{-1} \bfGammat \bfS^{-1}, \bfGammat^\alpha_{\alpha'} := \obraket*{\alpha}{\Gammah}{\alpha'}.
  \label{eqn:dm-matrix}
\end{align}
This gives:
\begin{align}
  \Ab ={} & \tr(\Ah \Gamma) = \sum_{\alpha,\alpha',\beta,\beta'} \bfA^\alpha_{\alpha'} \bfS^{\alpha'}_\beta \bfGamma^\beta_{\beta'} \bfS^{\beta'}_\alpha
  = \tr(\bfA \bfS \bfGamma \bfS) = \tr(\bfS^{-1} \bfAt \bfS^{-1} \bfGammat) = \tr(\bfA \bfGammat) = \tr(\bfAt \bfGamma).
  \label{eqn:mean-dm-matrix}
\end{align}
If the basis is orthonormal, then $\tr(\Ah \Gamma) = \tr(\bfA \bfGamma) = \tr(\bfAt \bfGammat)$.
If using the spectral basis $\{\ket{\kappa}\}_\kappa$ of $\Gamma$ in which case $\bfGamma = \Diag(\bmlambda)$, then $\tr(\Ah \Gamma) = \bmlambda\trs \diag(\bfA) = \sum_\kappa \lambda_\kappa \obraket*{\kappa}{\Ah}{\kappa}$.

That a matrix cannot be written as a vector outer-product means (1) it is an entangled state if the matrix represents the state of multiple particles/qubits under the product basis, or (2) it is a mixed state if the matrix is a density matrix.

\paragraph{Misc}

If an operator $\Ah$ does not contain spins (or does not operates on spins), then its spin-basis density matrix is a scalar matrix: $\bfA^{s,\alpha}_{s',\alpha'} = \delta^s_{s'} \bfA^\alpha_{\alpha'}$.

Any unitary operator $\Uh$ can be expressed by an Hermitian operator $\Hh$ in the way $\Uh = e^{\ii \Hh}$. Alternatively, $\{\ii \Hh\}$ is the Lie algebra of the Lie group $\{\Uh\}$.

Tiling theorem~\citep{ceperley1991fermion, foulkes2001quantum}: For $N$ fermions, all the ground-state nodal pockets belong to the same class (can be produced by permuting one nodal pocket).
Roughly means besides the boundary and the case $\rmxx_i = \rmxx_j$, $\psi_{(0)} \ne 0$ a.s.

%% file: quantum-many-body.tex
\section{Quantum Many-Body Systems} \label{sec:qmb}

\subsection{General Setup and Conclusions}

\subsubsection{Wavefunction, density and density matrix}

For $N$ electrons each with coordinate $\rmxx_i := (\rrv_i, s_i)$ where $s_i \in \lrbrace{ -\frac{1}{2}, +\frac{1}{2} }$ or $s_i \in \{\uparrow, \downarrow\}$ is its spin,
denote their wavefunction as $\psi(\bfxx) = \psi(\rmxx_1, \cdots, \rmxx_N)$ (may also denoted as $\braket*{\bfxx}{\psi}$).
\begin{itemize}
  \item Antisymmetry: $\psi(\cdots, \rmxx_i, \cdots, \rmxx_j, \cdots) = -\psi(\cdots, \rmxx_j, \cdots, \rmxx_i, \cdots)$.

    This indicates $\psi(\cdots, \rmxx, \cdots, \rmxx, \cdots) = -\psi(\cdots, \rmxx, \cdots, \rmxx, \cdots) = 0$.

  \item Inner product: $\braket{\phi}{\psi} = \int \ud\bfxx \, \phi^*(\bfxx) \psi(\bfxx)
    = \int \ud\rmxx_1 \cdots \ud \rmxx_N \, \phi^*(\rmxx_1, \cdots, \rmxx_N) \psi(\rmxx_1, \cdots, \rmxx_N)$.

  \item Normalization condition: $\braket{\psi}{\psi} = \int \ud\bfxx \, \lrvert{\psi(\bfxx)}^2 = 1$.
\end{itemize}

\begin{itemize}
  \item \textbf{Joint density}: $\lrvert{\psi(\bfxx)}^2$.

    \itemn The joint density normalizes to $1$, according to the normalization condition of the wavefunction.

    Due to the wavefunction antisymmetry,

  \items $\lrvert{\psi(\cdots, \rmxx_i, \cdots, \rmxx_j, \cdots)}^2 = \lrvert{\psi(\cdots, \rmxx_j, \cdots, \rmxx_i, \cdots)}^2$:
      the joint density is \emph{symmetric} (the electrons are indistinguishable);

    \itemb $\lrvert{\psi(\cdots, \rmxx, \cdots, \rmxx, \cdots)}^2 = 0$:
      events that ``electron $i$ is at $\rmxx$'' and that ``electron $j$ is at $\rmxx$'' for $i \ne j$ are \emph{mutually exclusive}, meaning the \textbf{Pauli exclusion principle}.

  \item \textbf{Density}:
    $\rho(\rmxx) := N \int \ud\bfxx_{2:N} \, \lrvert{\psi(\rmxx, \rmxx_2, \cdots, \rmxx_N)}^2$.
    May also denoted as $\rho_{[\psi]}(\rmxx)$, $\rho^{(1)}_{[\psi]}(\rmxx)$, or $\rho_1$.

    It is not the usual probability density function, but the \emph{electron number density} that takes all the $N$ electrons into account.
    Due to Pauli exclusion principle and the symmetry of $\lrvert{\psi(\bfxx)}^2$, we have:
    $\rho(\rmxx) = \sum_{i \in [N]} \int \ud\bfxx_{\neg i} \, \lrvert{\psi(\rmxx_1, \cdots, \rmxx_i = \rmxx, \cdots, \rmxx_N)}^2
    = \sum_{i \in [N]} \int \ud\bfxx_{\neg 1} \, \lrvert{\psi(\rmxx_1 = \rmxx, \rmxx_2, \cdots, \rmxx_N)}^2
    = N \int \bfxx_{\neg 1} \, \lrvert{\psi(\rmxx, \rmxx_2, \cdots, \rmxx_N)}^2$, as is defined here.

    \itemn It normalizes to $\int \ud\rmxx \, \rho(\rmxx) = N$, the total number of electrons (can also be seen from the definition).

      Define $N^s := \int \ud\rrv \, \rho(\rrv, s) = N p^s$, where $p^s := \int \ud\rrv \int \ud\bfxx_{2:N} \, \lrvert{\psi((\rrv, s), \rmxx_2, \cdots, \rmxx_N)}^2$.
      Due to joint density symmetry, this $p^s$ is the \emph{same for all electrons} (independent of $i$).
      Note $\sum_s p^s = 1$, and $\sum_s N^s = N$.

  \item \textbf{Pair density}:
    $\rho(\rmxx_1, \rmxx_2) := N (N-1) \int \ud\bfxx_{3:N} \, \lrvert{\psi(\rmxx_1, \rmxx_2, \rmxx_3, \cdots, \rmxx_N)}^2$.
    May also denoted as $\rho^{(2)}_{[\psi]}(\rmxx_1, \rmxx_2)$ or $\rho_{12}$.

    Similar to the density, this pair density is the \emph{electron-pair number density} of firstly finding an electron at $\rmxx_1$ and then finding another electron at $\rmxx_2$.
    Again due to Pauli exclusion principle and joint-density symmetry,
    $\rho(\rmxx_1^\star, \rmxx_2^\star) = \sum_{i \ne j} \int \bfxx_{\neg i, \neg j} \, \lrvert{\psi(\cdots, \rmxx_i = \rmxx_1^\star, \cdots, \rmxx_j = \rmxx_2^\star, \cdots)}^2
    = \sum_{i \ne j} \int \bfxx_{\neg 1, \neg 2} \, \lrvert{\psi(\rmxx_1 = \rmxx_1^\star, \rmxx_2 = \rmxx_2^\star, \cdots)}^2
    = N(N-1) \int \bfxx_{\neg 1, \neg 2} \, \lrvert{\psi(\rmxx_1 = \rmxx_1^\star, \rmxx_2 = \rmxx_2^\star, \cdots)}^2$, as is defined here.

    \itemn It normalizes to $N(N-1)$, the total number of ordered electron pairs (can also be seen from the definition).

    \itemm Marginalization yields $\int \ud\rmxx_2 \, \rho(\rmxx_1, \rmxx_2) = (N-1) \rho(\rmxx_1)$ from the definitions.
      This is to marginalize over other electrons, so the number of other electrons $N-1$ is derived.

    \itemi If two electrons are statistically independent, then $\rho(\rmxx_1, \rmxx_2) = \rho(\rmxx_1) \frac{N-1}{N} \rho(\rmxx_2)$,
      where the factor enters since given that one electron is found at $\rmxx_1$, the conditional electron density of another electron normalizes to $N-1$.

    \items Pair density is symmetric: $\rho(\rmxx_1, \rmxx_2) = \rho(\rmxx_2, \rmxx_1)$, due to the wavefunction antisymmetry,

    \itemb \textbf{Fermi hole} or \textbf{exchange hole} (or \textbf{Fermi/exchange correlation}): $\rho(\rmxx, \rmxx) = 0$. \\
      This indicates a correlation between two \emph{parallel-spin} electrons, \ie any two \emph{parallel-spin} electrons are \emph{not independent}.
      This correlation/hole comes from the antisymmetry of $\psi$, and only applies to parallel-spin electrons.
      It does not rely on any interaction (\eg, Coulomb potential), and is a global/distant correlation.
      It is the ``force'' that keeps a neutron star from collapse under gravity, where there is no Coulomb repulsion. Like the electron, a neutron has spin $\rfrac{1}{2}$.

      \textbf{Spacial versions}: $\rho(\rrv_1, \rrv_2) := \sum_{s_1, s_2} \rho((\rrv_1,s_1), (\rrv_2,s_2))$, $\rho^s(\rrv_1, \rrv_2) := \rho((\rrv_1,s), (\rrv_2,s))$.
      Then $\rho^s(\rrv,\rrv) = 0$, but $\rho(\rrv,\rrv)$ is unnecessarily zero.

  \item \textbf{One-particle reduced density matrix (1-RDM)}:
    $P(\rmxx; \rmxx') := N \int \ud\bfxx_{2:N} \, \psi(\rmxx, \rmxx_2, \cdots, \rmxx_N) \psi^*(\rmxx', \rmxx_2, \cdots, \rmxx_N)$.
    May also denoted as $P^{(1)}_{[\psi]}(\rmxx; \rmxx')$, $\obraket{\rmxx}{P^{(1)}}{\rmxx'}$, or $P^1_{1'}$.

    Note that $P(\rmxx; \rmxx) = \rho(\rmxx)$, and $P(\rmxx; \rmxx') = P^*(\rmxx'; \rmxx)$.

  \item \textbf{Two-particle reduced density matrix (2-RDM)}~\citep{tsuneda2014density,koch2001chemist}: \\
    $P(\rmxx_1, \rmxx_2; \rmxx'_1, \rmxx'_2) := N (N-1) \int \ud\bfxx_{3:N} \, \psi(\rmxx_1, \rmxx_2, \rmxx_3, \cdots, \rmxx_N) \psi^*(\rmxx'_1, \rmxx'_2, \rmxx_3, \cdots, \rmxx_N)$.
    May also denoted as $P^{(2)}_{[\psi]}(\rmxx_1, \rmxx_2; \rmxx'_1, \rmxx'_2)$, $\obraket{\rmxx_1, \rmxx_2}{P^{(2)}}{\rmxx'_1, \rmxx'_2}$, or $P^{12}_{1'2'}$.

    Note that $P(\rmxx_1, \rmxx_2; \rmxx_1, \rmxx_2) = \rho(\rmxx_1, \rmxx_2)$,
    $\int \ud\rmxx'_2 \, P(\rmxx_1, \rmxx'_2; \rmxx'_1, \rmxx'_2) = (N-1) P(\rmxx_1; \rmxx'_1)$, and
    $P(\rmxx_1, \rmxx_2; \rmxx'_1, \rmxx'_2) = P^*(\rmxx'_1, \rmxx'_2; \rmxx_1, \rmxx_2)$,
    $P(\rmxx_1, \rmxx_2; \rmxx'_1, \rmxx'_2) = -P(\rmxx_2, \rmxx_1; \rmxx'_1, \rmxx'_2) = -P(\rmxx_1, \rmxx_2; \rmxx'_2, \rmxx'_1) = P(\rmxx_2, \rmxx_1; \rmxx'_2, \rmxx'_1)$.
    Due to the Fermi/exchange hole, we have $P(\rmxx, \rmxx; \rmxx, \rmxx) = \rho(\rmxx, \rmxx) = 0$.
\end{itemize}

\subsubsection{Correlations and Hole Functions} \label{sec:corr-hole}

Define the \textbf{conditional probability} $\rho(\rmxx_2|\rmxx_1) := \rho(\rmxx_1,\rmxx_2) / \rho(\rmxx_1)$. \\
Again, it is actually the \emph{conditional electron number density} of other electrons given that there is already one electron found at $\rmxx_1$ out of $N$ electrons in total. \\
\itemn It normalizes to $\int \ud\rmxx_2 \, \rho(\rmxx_2|\rmxx_1) = N-1$, the number of other electrons (see also Property~(m) of pair density). \\
\itemi For independent electrons, $\rho(\rmxx_2|\rmxx_1) = \frac{N-1}{N} \rho(\rmxx_2)$.

Define the \textbf{correlation factor} $f(\rmxx_1,\rmxx_2) := \frac{\rho(\rmxx_1,\rmxx_2)}{\rho(\rmxx_1) \rho(\rmxx_2)} - 1$. \\
\items It is symmetric. \\
\itemi For independent electrons, $f(\rmxx_1,\rmxx_2) = -\frac{1}{N}$. \\
\itemone $\rho(\rmxx_1,\rmxx_2) = \rho(\rmxx_1) \rho(\rmxx_2) (1 + f(\rmxx_1,\rmxx_2))$.

Define the \textbf{exchange-correlation hole function} \\
$h_\XC(\rmxx_2|\rmxx_1) := \rho(\rmxx_2|\rmxx_1) - \rho(\rmxx_2)
= \frac{\rho(\rmxx_1,\rmxx_2)}{\rho(\rmxx_1)} - \rho(\rmxx_2)
= \frac{\rho(\rmxx_1,\rmxx_2) - \rho(\rmxx_1) \rho(\rmxx_2)}{\rho(\rmxx_1)}
= \rho(\rmxx_2) f(\rmxx_1,\rmxx_2)$. \\
This definition is for an $N$-free description of two-electron correlation. \\
\itemn By definition, it normalizes to $\int \ud\rmxx_2 \, h_\XC(\rmxx_2|\rmxx_1) = -1$, representing the removal of the given electron. \\
\itemb Since $\rho(\rmxx,\rmxx) = 0$, we have $h_\XC(\rmxx|\rmxx) = -\rho(\rmxx)$. \\
\itemi For independent electrons, $h_\XC(\rmxx_2|\rmxx_1) = -\frac{1}{N} \rho(\rmxx_2)$. \\
\itemone By definition, $\rho(\rmxx_1,\rmxx_2) = \rho(\rmxx_1) \rho(\rmxx_2) + \rho(\rmxx_1) h_\XC(\rmxx_2|\rmxx_1)$.

Define the \textbf{spin-independent exchange-correlation hole function} (or the \textbf{total hole function}) \\
$h_\XC(\rrv_2|\rrv_1) := \frac{\rho(\rrv_1,\rrv_2)}{\rho(\rrv_1)} - \rho(\rrv_2)
= \frac{\sum_{s_1,s_2} \rho((\rrv_1, s_1), (\rrv_2, s_2))}{\sum_{s'_1} \rho(\rrv_1, s'_1)} - \sum_{s'_2} \rho(\rrv_2, s'_2)
= \frac{\sum_{s_1,s_2} \rho((\rrv_1, s_1), (\rrv_2, s_2)) - \rho(\rrv_1, s_1) \rho(\rrv_2, s_2)}{\sum_{s'_1} \rho(\rrv_1, s'_1)}$. \\
Note that $h_\XC(\rrv_2|\rrv_1) \ne \sum_{s_1,s_2} h_\XC(\rrv_2, s_2 | \rrv_1, s_1) = \sum_{s_1,s_2} \frac{\rho((\rrv_1, s_1), (\rrv_2, s_2)) - \rho(\rrv_1, s_1) \rho(\rrv_2, s_2)}{\rho(\rrv_1, s_1)}$ in general. \\
\itemn It normalizes to $\int \ud\rrv_2 \, h_\XC(\rrv_2|\rrv_1) = \frac{(N-1) \rho(\rrv_1)}{\rho(\rrv_1)} - N = -1$. \\
\itemi For independent electrons, $h_\XC(\rrv_2|\rrv_1) = -\frac{1}{N} \rho(\rrv_2)$ (since $\rho(\rrv_1,\rrv_2) = \sum_{s_1,s_2} \rho(\rmxx_1,\rmxx_2) = \sum_{s_1,s_2} \rho(\rmxx_1) \frac{N-1}{N} \rho(\rmxx_2) = \rho(\rrv_1) \frac{N-1}{N} \rho(\rrv_2)$). \\
\itemone By definition, $\rho(\rrv_1,\rrv_2) = \rho(\rrv_1) \rho(\rrv_2) + \rho(\rrv_1) h_\XC(\rrv_2|\rrv_1)$.

\paragraph{The Fermi hole and the Coulomb hole}
The total hole $h_\XC(\rrv_2|\rrv_1)$ can be decomposed of two components:
\begin{align}
  h_\XC(\rrv_2|\rrv_1) = h_\tnX(\rrv_2|\rrv_1) + h_\tnC(\rrv_2|\rrv_1).
  \label{eqn:hole-decomp}
\end{align}
The \textbf{Fermi hole} $h_\tnX(\rrv_2|\rrv_1)$ is ``due to the Pauli exclusion principle, \ie, the antisymmetry of the wavefunction, and applies only to electrons with the same spin''~\citep{koch2001chemist}.
It describes the correlation between the given electron at $\rrv_1$ and other electrons with the same spin as the given electron. \\
\itemn It normalizes to $\int \ud\rrv_2 \, h_\tnX(\rrv_2|\rrv_1) = -1$, since excluding the given electron reduces the number of electrons with the same spin by 1.
  ``By this removal of one charge, the Fermi hole also takes care of the self-interaction problem''~\citep{koch2001chemist}. \\
\itemb Since the two electrons have the same spin, when they have the same spacial coordinate, the hole should reduce the density to zero. So $h_\tnX(\rrv_1|\rrv_1) = -\rho(\rrv_1)$. \\
\itemone ``$h_\tnX$ is negative everywhere, $h_\tnX(\rrv_2|\rrv_1) < 0$''~\citep{koch2001chemist}. \\
\itemtwo $h_\tnX(\rrv_2|\rrv_1)$ for the $\rmH_2$ molecule is independent of $\rrv_1$. Also, it is then delocalized. \\
\itemthr $h_\tnX(\rrv_2|\rrv_1)$ is not spherically symmetric in $\rrv_2$.
  It ``stays behind'' in the normal high electron density regions when $\rrv_1$ goes outside the regions~\citep{koch2001chemist}.

The \textbf{Coulomb hole} $h_\tnC(\rrv_2|\rrv_1)$ is due to dynamical interaction between electrons. \\
\itemn Due to the normalization of $h_\XC$ and $h_\tnX$, it normalizes to $\int \ud\rrv_2 \, h_\tnC(\rrv_2|\rrv_1) = 0$.
  It is natural since excluding the given electron does not change the number of electrons with a different spin. \\
\itemone It is thus positive in some regions and negative in others, and may also be delocalized. \\
\itemtwo Since this hole is due to physical interaction, $h_\tnC(\rrv_2|\rrv_1)$ must change with $\rrv_1$.

However, the exact definition of the two holes at this generality that satisfies the mentioned properties is not found.
\itemI The first possibility is $h_\tnX(\rrv_2|\rrv_1) = \big( \sum_s \rho((\rrv_1,s), (\rrv_2,s)) - \rho(\rrv_1,s) \rho(\rrv_2,s) \big) / \rho(\rrv_1)$,
  and $h_\tnC(\rrv_2|\rrv_1) = \big( \sum_{s \ne s'} \rho((\rrv_1,s), (\rrv_2,s')) - \rho(\rrv_1,s) \rho(\rrv_2,s') \big) / \rho(\rrv_1)$.
  This may best fit the conceptual definition, and complies with the HF case $h_\XC^\HF(\rrv_2|\rrv_1) = h_\tnX^\HF(\rrv_2|\rrv_1)$ (with A2.2; see \eqnref{hx-spacial-hf}).
  But it does not naturally satisfy Property~(n) $\int \ud\rrv_2 \, h_\tnX(\rrv_2|\rrv_1) = \big( \sum_s \int \ud\rrv_2 \, \rho((\rrv_1,s), (\rrv_2,s)) - N^s \rho(\rrv_1,s) \big) / \big( \sum_s \rho(\rrv_1,s) \big) \stackrel{?}{=} -1$
  nor Property~(b) $h_\tnX(\rrv_1|\rrv_1) = - \big( \sum_s \rho(\rrv_1,s)^2 \big) / \big( \sum_s \rho(\rrv_1,s) \big) \stackrel{?}{=} -\sum_s \rho(\rrv_1,s)$.
\itemII Another guess is $h_\tnX(\rrv_2|\rrv_1) = \big( \sum_s \rho((\rrv_1,s), (\rrv_2,s)) \big) / \rho(\rrv_1) - \rho(\rrv_2)$,
  and $h_\tnC(\rrv_2|\rrv_1) = \big( \sum_{s \ne s'} \rho((\rrv_1,s), (\rrv_2,s')) \big) / \rho(\rrv_1)$.
  This satisfies Property~(b) $h_\tnX(\rrv_1|\rrv_1) = -\rho(\rrv_1)$, but does not comply with the HF case $h_\XC^\HF(\rrv_2|\rrv_1) \ne h_\tnX^\HF(\rrv_2|\rrv_1)$ (even with A2.2)
  nor Property~(n) $\int \ud\rrv_2 \, h_\tnX(\rrv_2|\rrv_1) = \big( \sum_s \int \ud\rrv_2 \, \rho((\rrv_1,s), (\rrv_2,s)) \big) / \rho(\rrv_1) - N \stackrel{?}{=} -1$.
\itemIII Perhaps $h_\tnX(\rrv_2|\rrv_1)$ can be defined as $h_\tnX^\HF(\rrv_2|\rrv_1)$ (see \eqnref{hx-spacial-hf}), the total hole in the HF case under A2.2 (\ie, sum over parallel spins only), which only takes the exchange correlation into consideration.
  This seems adopted by [ref. \href{http://cmt.dur.ac.uk/sjc/thesis_mcg/node159.html}{webpage1}, \href{http://cmt.dur.ac.uk/sjc/thesis_mcg/node15.html}{webpage2}] and may also be the way to define $h_\tnX$ as $h_\XC^{\lambda=0}$ in the adiabatic connection.
  But the correspondence between the orbitals $\{\phi_i\}_{i \in [N]}$ and a general $N$-electron wavefunction is undetermined (effective potential is unknown), at least not explicit.
  Even in an HF system where the Slater determinant is exact, Properties~(n) and~(b) do not seem to hold (see \eqnref{pairdensity-hf}).
\itemIV Maybe the definition of the Fermi and Coulomb holes should not follow the spin-independent pair density.
  The usage of the holes comes from the exchange-correlation energy $E_\XC := \frac{1}{2} \int \ud\rrv_1 \ud\rrv_2 \, \frac{\sum_{s_1,s_2} \rho(\rmxx_1, \rmxx_2) - \rho(\rmxx_1) \rho(\rmxx_2)}{r_{12}}$ (\eqnref{exc}), where
  $\sum_{s_1,s_2} \rho(\rmxx_1, \rmxx_2) - \rho(\rmxx_1) \rho(\rmxx_2)
  = \big( \sum_s \rho((\rrv_1,s), (\rrv_2,s)) - \rho(\rrv_1,s) \rho(\rrv_2,s) \big) + \big( \sum_{s \ne s'} \rho((\rrv_1,s), (\rrv_2,s')) - \rho(\rrv_1,s) \rho(\rrv_2,s') \big)
  =: \sum_s \rho(\rrv_1,s) \big( h_\tnX^s(\rrv_2|\rrv_1) + h_\tnC^s(\rrv_2|\rrv_1) \big)$, where we have defined
  $h_\tnX^s(\rrv_2|\rrv_1) := h_\XC((\rrv_2,s)|(\rrv_1,s)) = \frac{\rho((\rrv_1,s), (\rrv_2,s))}{\rho(\rrv_1,s)} - \rho(\rrv_2,s)$, and
  $h_\tnC^s(\rrv_2|\rrv_1) := h_\XC((\rrv_2,\neg s)|(\rrv_1,s)) = \frac{\rho((\rrv_1,s), (\rrv_2,\neg s))}{\rho(\rrv_1,s)} - \rho(\rrv_2,\neg s)$.
  This $h_\tnX^s$ satisfies the counterpart of Property~(b) $h_\tnX^s(\rrv_1|\rrv_1) = -\rho(\rrv_1,s)$, and complies with the HF case $h_\tnC^s \equiv 0$ for either $s$ (with A2.2),
  But it is unknown for Property~(n) $\int \ud\rrv_2 \, h_\tnX^s(\rrv_2|\rrv_1) = \big( \int \ud\rrv_2 \, \rho((\rrv_1,s), (\rrv_2,s)) \big) / \rho(\rrv_1,s) - N^s \stackrel{?}{=} -1$.
  It also does not satisfy \eqnref{hole-decomp}, though the spin-dependent counterpart holds: $h_\XC(\rmxx_2|\rmxx_1) = \bbone_{s_2 = s_1} h_\tnX^{s_1}(\rrv_2|\rrv_1) + \bbone_{s_2 \ne s_1} h_\tnC^{s_1}(\rrv_2|\rrv_1)$.
  It seems to match \citet[Eq.~(2-20)]{koch2001chemist}, but cannot make the form $E_\XC = \frac{1}{2} \int \ud\rrv_1 \ud\rrv_2 \, \frac{\rho(\rrv_1) h_\XC(\rrv_2|\rrv_1)}{r_{12}}$.

Note that the above concepts and conclusions are general.
The Fermi (/Coulomb) hole describes any correlation between any two fermions of the same kind with the same spin (/with different spins) and the conclusions hold in general,
whatever the interaction is between the two fermions.
The Coulomb hole gets its name from the common case where the correlation between antiparallel-spin electrons mainly comes from their Coulomb (electrostatic) interaction.
The \textbf{total hole} is named exchange-correlation hole, since for interacting electrons, it ``describes the change in conditional probability caused by the correlation for self-interaction, exchange and Coulomb correlation, compared to the completely uncorrelated situation''~\citep{koch2001chemist}.
To determine the holes, inter-fermion interaction and global potential (\ie, the Hamiltonian) need to be specified, but in any setup, the conclusions here hold.

\subsubsection{Operators}

For a general one-particle operator $\Oh_\cdot$ and a general two-particle operator $\Wh_{\cdot\cdot}$, define their extensions $\Oh$, $\Wh$ to an $N$-particle wavefunction as:
\begin{align}
  \Oh \psi(\bfxx) := \sum_{k \in [N]} \Oh_k \psi(\rmxx_1, \cdots, \rmxx_N), \qquad
  \Wh \psi(\bfxx) := \frac{1}{2} \sum_{j \ne k} \Wh_{jk} \psi(\rmxx_1, \cdots, \rmxx_N).
  \label{eqn:opr-extension}
\end{align}
If $\psi(\rmxx_1, \cdots, \rmxx_N)$ is (anti)symmetric, then:
\begin{align}
  \obraket*{\psi}{\Oh}{\psi}
  ={} & \sum_{k \in [N]} \int \ud\rmxx_1 \cdots \ud\rmxx_N \, \psi^*(\rmxx_1, \cdots, \rmxx_N) \Oh_k \psi(\rmxx_1, \cdots, \rmxx_N) \\
  \stackrel{\text{(*)}}{=} {} & \sum_{k \in [N]} \int \ud\rmxx_k \ud\rmxx_1 \cdots \ud\rmxx_{\neg k} \cdots \ud\rmxx_N \, \psi^*(\rmxx_k, \rmxx_1, \cdots, \rmxx_{\neg k}, \cdots, \rmxx_N) \Oh_k \psi(\rmxx_k, \rmxx_1, \cdots, \rmxx_{\neg k}, \cdots, \rmxx_N) \\
  \stackrel{\text{(\#)}}{=} {} & \sum_{k \in [N]} \int \ud\rmxx_1 \cdots \ud\rmxx_N \, \psi^*(\rmxx_1, \cdots, \rmxx_N) \Oh_1 \psi(\rmxx_1, \cdots, \rmxx_N) \\
  ={} & N \obraket*{\psi}{\Oh_1}{\psi},
  \label{eqn:1-opr-avg-wavefn} \\
  \obraket*{\psi}{\Oh}{\psi}
  ={} & \int \ud\rmxx_1 \, \Oh_1 P^{(1)}_{[\psi]}(\rmxx_1; \rmxx'_1 = \rmxx_1)
  = \tr\big( \Oh_\cdot P^{(1)}_{[\psi]} \big),
  \label{eqn:1-opr-avg-rdm}
\end{align}
where (*) is due to that $\Oh$ is linear thus commutes with the sign, and (\#) is just a rename of dummy variables. Similarly,
\begin{align}
  \obraket*{\psi}{\Wh}{\psi} = \frac{1}{2} N (N-1) \obraket*{\psi}{\Wh_{12}}{\psi}
  = \frac{1}{2} \tr\big( \Wh_{\cdot\cdot} P^{(2)}_{[\psi]} \big).
  \label{eqn:2-opr-avg}
\end{align}

\subsection{Second Quantization} \label{sec:qmb-2ndq}

[\href{http://eduardo.physics.illinois.edu/phys561/non_relativistic_final.pdf}{ref. course note}].

\subsubsection{Ladder operators: Algebraic description}

\begin{figure}[h]
  \centering
  \vspace{-6pt}
  \includegraphics[width=5.0cm]{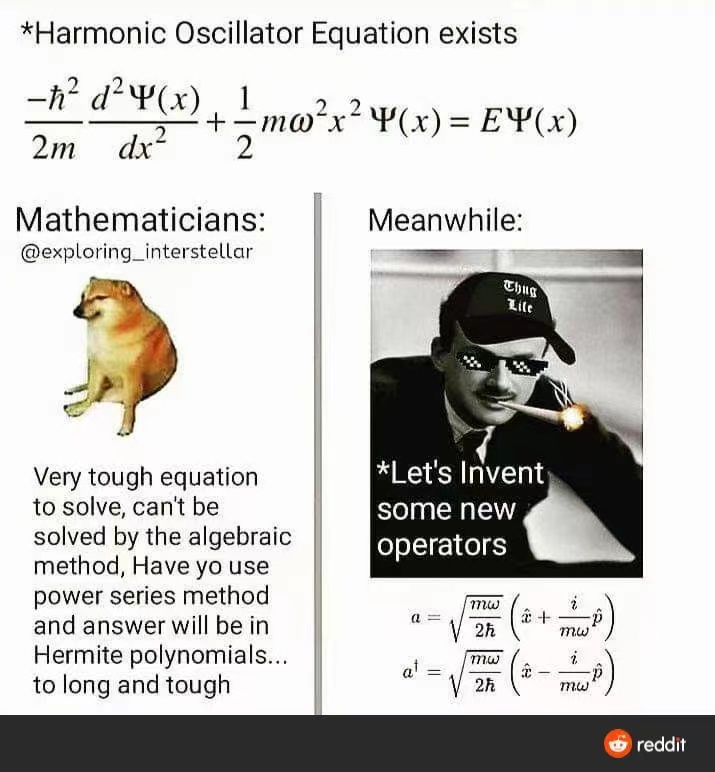}
  \vspace{-2pt}
  \caption{The ladder operators make things easier. [\href{https://www.reddit.com/r/physicsmemes/comments/rgd8do/ima_go_get_drunk_at_the_h_bar/}{source}]}
  \label{fig:sec-quant-meme}
  \vspace{-4pt}
\end{figure}

Let $\aah$ be an operator on an arbitrary Hilbert space of states, and $\aah^\dagger$ be its conjugate.
Then $\aah$ is called a \textbf{lowering/annihilation operator} and $\aah^\dagger$ a \textbf{raising/creation/destruction operator}),
if $[\aah, \aah^\dagger] = 1$ (identity operator).
Jointly, $\aah$ and $\aah^\dagger$ are called \textbf{ladder operators}.
The eigenstates of $\aah^\dagger \aah$ satisfy:
\begin{align}
  [\aah, \aah^\dagger] = 1
  ~~\Longrightarrow~~
  \left\{ \begin{array}{@{}r@{}l}
      \aah^\dagger \aah \, \ket{n} & {}= n \, \ket{n}, \\
      \aah^\dagger \, \ket{n} & {}= \sqrt{n+1} \, \ket{n+1}, \\
      \aah \, \ket{n} & {}= \sqrt{n} \, \ket{n-1},
  \end{array} \right.
  n \in \bbN.
\end{align}

For quantum harmonic oscillator, the Hamiltonian is $\Hh = \frac{\pph^2}{2m} + \frac{1}{2} m \omega^2 x^2$.
We can choose $\aah = \frac{1}{\sqrt{2}} \left( \sqrt{\frac{m \omega}{\hbar}} x + \ii \frac{\pph}{\sqrt{m \omega \hbar}} \right)$,\
then $[\aah, \aah^\dagger] = 1$, and $\Hh = \hbar \omega \left( \aah^\dagger \aah + \frac{1}{2} \right)$ (\figref{sec-quant-meme}).
The eigenstates are then given by those of the ladder operator.

\subsubsection{The Fock space}

\paragraph{Useful conclusions.}
Let $\rmPi_N$ be the set of permutations / symmetric group on $[N] := \{1, \dots, N\}$, and $\frpp \in \{-1, 1\}$ be the particle parity.
If $\frpp = -1$ (fermion), $\frpp^{\bmpi}$ is the sign of permutation $\bmpi$.
Let $\bigcirc$, $\bigsquare$ be two reducing operators (\eg, $\sum$, $\prod$).

For a function $f(\cdot,\cdot)$ that adopts two indices each within $[N]$, we have:
\begin{align}
  \bigsquare_{\bmpi, \bmpi' \in \rmPi_N} \frpp^{\bmpi} \frpp^{\bmpi'} \bigcirc_{i \in [N]} f(\bmpi_i, \bmpi'_i)
  = N! \bigsquare_{\bmpi \in \rmPi_N} \frpp^{\bmpi} \bigcirc_{i \in [N]} f(i, \bmpi_i).
  \label{eqn:perm-prod}
\end{align}
To see this, l.h.s is $\bigsquare_{\bmpi, \bmpi' \in \rmPi_N} \frpp^{\bmpi} \frpp^{\bmpi'} \bigcirc_{j \in [N]} f(j, \bmpi'_{\bmpi^{-1}_j})
= \bigsquare_{\bmpi \in \rmPi_N} \bigsquare_{\bmpi' \in \rmPi_N} \frpp^{\bmpi' \circ \bmpi^{-1}} \bigcirc_{j \in [N]} f(j, (\bmpi' \circ \bmpi^{-1})_j)$.
Since for a fixed $\bmpi$, $\bmpi' \circ \bmpi^{-1}$ traverses $\rmPi_N$ when $\bmpi'$ does, so this is
$\bigsquare_{\bmpi \in \rmPi_N} \bigsquare_{\bmpi'' \in \rmPi_N} \frpp^{\bmpi''} \bigcirc_{j \in [N]} f(j, \bmpi''_j)
= N! \bigsquare_{\bmpi'' \in \rmPi_N} \frpp^{\bmpi''} \bigcirc_{j \in [N]} f(j, \bmpi''_j)$.

For a function $f(\cdot, \cdots, \cdot)$ that adopts $N$ (usually distinct) indices each within $[N]$, and a fixed position $k \in [N]$, we have:
\begin{align}
  \bigcirc_{\bmpi \in \rmPi_N} \frpp^{\bmpi} f(\bmpi_1, \cdots, \bmpi_N)
  = \bigcirc_{j \in [N]} \bigcirc_{\bmpi' \in \rmPi_{[N] \setminus \{j\}}} \frpp^{j-k} \frpp^{\bmpi'} f(\bmpi'_1, \cdots, \underbrace{j}_{k\text{-th}}, \cdots, \bmpi'_{N-1}).
  \label{eqn:perm-decomp}
\end{align}
To see this, for $j > k$, note that permuting $\bmpi = (\bmpi'_1, \cdots, \bmpi'_{k-1}, j, \bmpi'_k, \cdots, \bmpi'_N)$ to $(1, \cdots, N)$ whose sign is $\frpp^{\bmpi}$, can be done by
first permuting $\bmpi'$ to $(1, \cdots, k-1, k, \cdots, j-1, j+1, \cdots, N)$ whose sign is $\frpp^{\bmpi'}$ (kind of by definition),
and then permuting $(1, \cdots, k-1, j, k, \cdots, j-1, j+1, \cdots, N)$ to $(1, \cdots, k-1, k, \cdots, j-1, j, j+1, \cdots, N)$ whose sign is $\frpp^{k-j} = \frpp^{j-k}$.
Things go similarly for $j \le k$.
So for such a $(\bmpi, \bmpi')$ pair, we have $\frpp^{\bmpi} = \frpp^{j-k} \frpp^{\bmpi'}$.
This leads to the determinant expansion along row/column $k$ when $\frpp = -1$.

\paragraph{Wavefunction/State description.}
Given a Hilbert space $\scH$ of one-particle wavefunctions/states, the Fock space is the state space of multiple non-interacting identical/indistinguishable $\frpp$-symmetric particles:
\begin{align}
  \scF_\frpp(\scH) := \overline{\bigoplus_{N=0}^\infty \scF_{\frpp,N}(\scH)}, \quad
  \scF_{\frpp,N}(\scH) := \clA_\frpp(\scH^N),
\end{align}
and $\scF_{\frpp,0} := \bbC$.
Here $\clA_\frpp$ is the $\frpp$-symmetrizer, which constructs a Fock state from $\{\phi_i\}_{i=1}^N \subset \scH$:
\begin{align}
  \ket{\phi_1 \cdots \phi_N}_\frpp
  &:= \clA_\frpp(\ket{\phi_1} \otimes \cdots \otimes \ket{\phi_N})
  := \frac{1}{\sqrt{N!}} \sum_{\bmpi \in \rmPi_N} \frpp^{\bmpi} \, \ket{\phi_{\bmpi_1}} \otimes \cdots \otimes \ket{\phi_{\bmpi_N}}
  \quad
  \in \scF_{\frpp,N}(\scH),
  \label{eqn:symmetrizer} \\
  \ket{\phi_1 \cdots \phi_N}_\frpp & (\rmxx_1, \cdots, \rmxx_N)
  = \braket*[\big]{\rmxx_1, \cdots, \rmxx_N}{\ket{\phi_1 \cdots \phi_N}_\frpp} \\
  &= \frac{1}{\sqrt{N!}} \sum_{\bmpi \in \rmPi_N} \frpp^{\bmpi} \phi_{\bmpi_1}(\rmxx_1) \cdots \phi_{\bmpi_N}(\rmxx_N)
  = \frac{1}{\sqrt{N!}} \sum_{\bmpi \in \rmPi_N} \frpp^{\bmpi} \phi_1(\rmxx_{\bmpi_1}) \cdots \phi_N(\rmxx_{\bmpi_N})
  = \frac{1}{\sqrt{N!}} \det\nolimits_\frpp \lrbrack{\phi_i(\rmxx_j)}_{ij}.
  \label{eqn:symmetrizer-det}
\end{align}
For bosons, $\frpp = 1$ and this is a \emph{permanent}.
For fermions, $\frpp = -1$ and this gives the \emph{Slater determinant} \eqnref{slater-determ}.
By definition, we also have:
\begin{align}
  \ket{\phi_1 \cdots \phi_k \cdots \phi_N}_\frpp = \frpp^{k-1} \ket{\phi_k \phi_1 \cdots \phi_{\neg k} \cdots \phi_N}_\frpp.
  \label{eqn:symmetrizer-interchange}
\end{align}
The inner product in $\scF_{\frpp,N}$ is:
\begin{align}
  \braket{\chi_1 \cdots \chi_N}{\phi_1 \cdots \phi_N}_\frpp
  = \frac{1}{N!} \sum_{\bmpi,\bmpi' \in \rmPi_N} \frpp^{\bmpi} \frpp^{\bmpi'} \prod_{i \in [N]} \braket{\chi_{\bmpi'_i}}{\phi_{\bmpi_i}}
  \stackrel{\text{\eqnref{perm-prod}}}{=} \sum_{\bmpi \in \rmPi_N} \frpp^{\bmpi} \prod_{i \in [N]} \braket{\chi_i}{\phi_{\bmpi_i}}
  = \det\nolimits_\frpp \lrbrack{\braket{\chi_i}{\phi_j}}_{ij}.
  \label{eqn:fock-inprod}
\end{align}

For the second-quantization representation of a Fock state, let $\{\ket{\alpha}\}_\alpha$ be a complete basis of $\scH$.
Then a state $\ket{\phi_1 \cdots \phi_N}_\frpp \in \scF_{\frpp,N}$ where $\{\phi_i\}_i \subseteq \{\ket{\alpha}\}_\alpha$ can also be expressed by counting the number of $\phi_i$ that occupies $\ket{1}, \ket{2}, \cdots$ in turn: $\ket{n_1, n_2, \cdots}$.
Due to the Pauli exclusion principle, for fermions, each $n_i$ can only be either $0$ or $1$.
So this expression is particularly interested for bosons, in which case each $\phi_i$ occupying $\ket{\alpha}$ (\ie, $\ket{\phi_i} = \ket{\alpha}$) is called a \textbf{phonon} of type $\alpha$,
and $\ket{n_1, n_2, \cdots} = \frac{1}{\sqrt{n_1! n_2! \cdots}} \ket*{\underbrace{1 \cdots 1}_{n_1} \underbrace{2 \cdots 2}_{n_2} \cdots}_{+1}$.

\paragraph{Reduced densities and RDMs of a Fock state.}
For the reduced density matrices (RDMs) of a Fock state $\psi_\frpp = \ket{\phi_1 \cdots \phi_N}_\frpp$, first note that:
\begin{align}
  & \psi_\frpp(\rmxx_1, \rmxx_2, \rmxx_3, \cdots \rmxx_N)
  \stackrel{\text{\eqnref{symmetrizer-det}}}{=} \frac{1}{\sqrt{N!}} \sum_{\bmpi \in \rmPi_N} \frpp^{\bmpi} \phi_{\bmpi_1}(\rmxx_1) \phi_{\bmpi_2}(\rmxx_2) \phi_{\bmpi_3}(\rmxx_3) \cdots \phi_{\bmpi_N}(\rmxx_N) \\
  \stackrel{\text{\eqnref{perm-decomp}}}{=} {} & \frac{1}{\sqrt{N!}} \sum_{j \in [N]} \frpp^{j-1} \sum_{\bmpi' \in \rmPi_{[N] \setminus \{j\}}} \frpp^{\bmpi'} \phi_j(\rmxx_1) \phi_{\bmpi'_1}(\rmxx_2) \phi_{\bmpi'_2}(\rmxx_3) \cdots \phi_{\bmpi'_{N-1}}(\rmxx_N) \\
  ={} & \frac{1}{\sqrt{N}} \sum_{j \in [N]} \frpp^{j-1} \phi_j(\rmxx_1) \ket{\cdots \phi_{\neg j} \cdots}_\frpp (\rmxx_2, \cdots, \rmxx_N)
  \label{eqn:fock-1-out-decomp} \\
  \stackrel[\text{(*)}]{\text{\eqnref{perm-decomp}}}{=} {} & \frac{1}{\sqrt{N!}} \sum_{j \in [N]} \frpp^{j-1} \sum_{k \in [N] \setminus \{j\}} \frpp^{k - \bbone_{k>j} - 1} \sum_{\bmpi'' \in \rmPi_{[N] \setminus \{j,k\}}} \frpp^{\bmpi''} \phi_j(\rmxx_1) \phi_k(\rmxx_2) \phi_{\bmpi''_1}(\rmxx_3) \cdots \phi_{\bmpi'_{N-2}}(\rmxx_N) \\
  ={} & \frac{1}{\sqrt{N(N-1)}} \sum_{j \ne k} \frpp^{j - k + \bbone_{k>j}} \phi_j(\rmxx_1) \phi_k(\rmxx_2) \ket{\cdots \phi_{\neg j} \cdots \phi_{\neg k} \cdots}_\frpp (\rmxx_3, \cdots, \rmxx_N)
  \label{eqn:fock-2-out-decomp-ne} \\
  ={} & \frac{1}{\sqrt{N(N-1)}} \sum_{j < k} \frpp^{j-k+1} \big( \phi_j(\rmxx_1) \phi_k(\rmxx_2) + \frpp \phi_k(\rmxx_1) \phi_j(\rmxx_2) \big) \ket{\cdots \phi_{\neg j} \cdots \phi_{\neg k} \cdots}_\frpp (\rmxx_3, \cdots, \rmxx_N),
  \label{eqn:fock-2-out-decomp}
\end{align}
where $\bbone_{k>j}$ arises in (*) since if $k>j$, then $k-1$ (instead of $k$) is the position (counted from 1) of the value $k$ in the ordered sequence $[1,\cdots,j-1,j+1,\cdots,N]$ of $\bmpi'$.
Since cavities are interchangeable, the symbol $\ket{\cdots \phi_{\neg j} \cdots \phi_{\neg k} \cdots}_\frpp$ means $\ket{\cdots \phi_{\neg \min\{j,k\}} \cdots \phi_{\neg \max\{j,k\}} \cdots}_\frpp$ and is the same as $\ket{\cdots \phi_{\neg k} \cdots \phi_{\neg j} \cdots}_\frpp$.
So the 1-RDM and 2-RDM are:
\begin{align}
  & \begin{aligned}
    P_\frpp(\rmxx; \rmxx') :={} & N \int \ud \rmxx_2 \cdots \ud \rmxx_N \, \psi_\frpp(\rmxx, \rmxx_2, \cdots \rmxx_N) \psi_\frpp^*(\rmxx', \rmxx_2, \cdots \rmxx_N) \\
    \stackrel{\text{\eqnref{fock-1-out-decomp}}}{=} {} & \sum_{j, j' \in [N]} \frpp^{j-j'} \phi_j(\rmxx) \phi_{j'}^*(\rmxx') \braket{\cdots \phi_{\neg j'} \cdots}{\cdots \phi_{\neg j} \cdots}_\frpp,
    \label{eqn:fock-1-rdm}
  \end{aligned} \\
  & P_\frpp(\rmxx_1, \rmxx_2; \rmxx'_1, \rmxx'_2) := N(N-1) \int \ud \rmxx_3 \cdots \ud \rmxx_N \, \psi_\frpp(\rmxx_1, \rmxx_2, \rmxx_3, \cdots \rmxx_N) \psi_\frpp^*(\rmxx'_1, \rmxx'_2, \rmxx_3, \cdots \rmxx_N) \\
  \stackrel{\text{\eqnref{fock-2-out-decomp-ne}}}{=} {} & \sum_{j \ne k} \sum_{j' \ne k'} \frpp^{j - k + \bbone_{k>j}} \frpp^{j' - k' + \bbone_{k'>j'}} \phi_j(\rmxx_1) \phi_k(\rmxx_2) \phi_{j'}^*(\rmxx'_1) \phi_{k'}^*(\rmxx'_2) \braket{\cdots \phi_{\neg j'} \cdots \phi_{\neg k'} \cdots}{\cdots \phi_{\neg j} \cdots \phi_{\neg k} \cdots}_\frpp
  \qquad \label{eqn:fock-2-rdm-ne} \\
  \stackrel{\text{\eqnref{fock-2-out-decomp}}}{=} {} & \sum_{j < k} \sum_{j' < k'} \frpp^{j-k} \frpp^{j'-k'}
  \big( \phi_j(\rmxx_1) \phi_k(\rmxx_2) + \frpp \phi_k(\rmxx_1) \phi_j(\rmxx_2) \big) \big( \phi_{j'}^*(\rmxx'_1) \phi_{k'}^*(\rmxx'_2) + \frpp \phi_{k'}^*(\rmxx'_1) \phi_{j'}^*(\rmxx'_2) \big) \cdots \\[-8pt]
  & \hspace{3.0cm} \braket{\cdots \phi_{\neg j'} \cdots \phi_{\neg k'} \cdots}{\cdots \phi_{\neg j} \cdots \phi_{\neg k} \cdots}_\frpp.
  \label{eqn:fock-2-rdm}
\end{align}
From \eqnref{fock-1-rdm}, since $\frpp^{j-j'} \phi_j(\rmxx) \phi_{j'}^*(\rmxx) \braket{\cdots \phi_{\neg j'} \cdots}{\cdots \phi_{\neg j} \cdots}_\frpp
= (\frpp^{j'-j} \phi_{j'}(\rmxx) \phi_j^*(\rmxx) \braket{\cdots \phi_{\neg j} \cdots}{\cdots \phi_{\neg j'} \cdots}_\frpp)^*$,
we have $\rho_\frpp(\rmxx) = \sum_{j \in [N]} \lrvert{\phi_j(\rmxx)}^2 \braket{\cdots \phi_{\neg j} \cdots}{\cdots \phi_{\neg j} \cdots}_\frpp
+ 2 \sum_{j < j'} \frpp^{j-j'} \Re\lrparen{ \phi_j(\rmxx) \phi_{j'}^*(\rmxx) \braket{\cdots \phi_{\neg j'} \cdots}{\cdots \phi_{\neg j} \cdots}_\frpp}$.

For \textbf{orthonormal} $\{\phi_i\}_{i=1}^N$, from \eqnref{fock-inprod}, we have $\braket{\cdots \phi_{\neg j'} \cdots}{\cdots \phi_{\neg j} \cdots}_\frpp = \delta_{j'j}$,
and $\braket{\cdots \phi_{\neg j'} \cdots \phi_{\neg k'} \cdots}{\cdots \phi_{\neg j} \cdots \phi_{\neg k} \cdots}_\frpp = \delta_{j'j} \delta_{k'k}$ for $j < k$ and $j' < k'$. So:
\begin{align}
  P_\frpp(\rmxx; \rmxx')
  ={} & \sum_{j \in [N]} \phi_j(\rmxx) \phi_j^*(\rmxx'),
  \label{eqn:fock-1-rdm-orthon} \\
  P_\frpp(\rmxx_1, \rmxx_2; \rmxx'_1, \rmxx'_2)
  ={} & \sum_{j < k} \big( \phi_j(\rmxx_1) \phi_k(\rmxx_2) + \frpp \phi_k(\rmxx_1) \phi_j(\rmxx_2) \big) \big( \phi_j^*(\rmxx'_1) \phi_k^*(\rmxx'_2) + \frpp \phi_k^*(\rmxx'_1) \phi_j^*(\rmxx'_2) \big) \\
  ={} & \sum_{j < k} \phi_j(\rmxx_1) \phi_k(\rmxx_2) \big( \phi_j^*(\rmxx'_1) \phi_k^*(\rmxx'_2) + \frpp \phi_k^*(\rmxx'_1) \phi_j^*(\rmxx'_2) \big)
  + \sum_{j < k} \frpp \phi_k(\rmxx_1) \phi_j(\rmxx_2) \big( \phi_j^*(\rmxx'_1) \phi_k^*(\rmxx'_2) + \frpp \phi_k^*(\rmxx'_1) \phi_j^*(\rmxx'_2) \big) \\
  \stackrel{\text{(*)}}{=} {} & \sum_{j < k} \phi_j(\rmxx_1) \phi_k(\rmxx_2) \big( \phi_j^*(\rmxx'_1) \phi_k^*(\rmxx'_2) + \frpp \phi_k^*(\rmxx'_1) \phi_j^*(\rmxx'_2) \big)
  + \sum_{k < j} \phi_j(\rmxx_1) \phi_k(\rmxx_2) \big( \frpp \phi_k^*(\rmxx'_1) \phi_j^*(\rmxx'_2) + \phi_j^*(\rmxx'_1) \phi_k^*(\rmxx'_2) \big) \\
  ={} & \sum_{j \ne k} \phi_j(\rmxx_1) \phi_k(\rmxx_2) \big( \phi_j^*(\rmxx'_1) \phi_k^*(\rmxx'_2) + \frpp \phi_k^*(\rmxx'_1) \phi_j^*(\rmxx'_2) \big) \\
  ={} & \sum_{j, k} \phi_j(\rmxx_1) \phi_k(\rmxx_2) \big( \phi_j^*(\rmxx'_1) \phi_k^*(\rmxx'_2) + \frpp \phi_k^*(\rmxx'_1) \phi_j^*(\rmxx'_2) \big)
  - \bbone_{\frpp = 1} 2 \sum_{j \in [N]} \phi_j(\rmxx_1) \phi_j(\rmxx_2) \phi_j^*(\rmxx'_1) \phi_j^*(\rmxx'_2),
  \label{eqn:fock-2-rdm-orthon}
\end{align}
where (*) just renames $(k,j) \asn (j,k)$ for the second term.
Note that for fermions $\frpp = -1$, the 2-RDM can be determined from the 1-RDM:
\begin{align}
  P_{\frpp=-1}(\rmxx_1, \rmxx_2; \rmxx'_1, \rmxx'_2)
  ={} & \sum_{j, k} \phi_j(\rmxx_1) \phi_j^*(\rmxx'_1) \phi_k(\rmxx_2) \phi_k^*(\rmxx'_2) - \sum_{j, k} \phi_j(\rmxx_1) \phi_j^*(\rmxx'_2) \phi_k(\rmxx_2) \phi_k^*(\rmxx'_1) \\
  ={} & P_{\frpp=-1}(\rmxx_1; \rmxx'_1) P_{\frpp=-1}(\rmxx_2; \rmxx'_2) - P_{\frpp=-1}(\rmxx_1; \rmxx'_2) P_{\frpp=-1}(\rmxx_2; \rmxx'_1).
  \label{eqn:fock-2-rdm-1-rdm}
\end{align}
Expressing 2-RDM using 1-RDM seems impossible for non-orthonormal $\{\phi_i\}_{i=1}^N$.
Note $\braket{\cdots \phi_{\neg j'} \cdots}{\cdots \phi_{\neg j} \cdots}_\frpp
= \sum_{k \in [N] \setminus \{j\}} \frpp^{(k - \bbone_{k>j}) - (k' - \bbone_{k'>j'})} \braket{\phi_{k'}}{\phi_k} \braket{\cdots \phi_{\neg j'} \cdots \phi_{\neg k'} \cdots}{\cdots \phi_{\neg j} \cdots \phi_{\neg k} \cdots}_\frpp$.

The density and pair-density functions are:
\begin{align}
  \rho_\frpp(\rmxx) ={} & \sum_{j \in [N]} \lrvert{\phi_j(\rmxx)}^2,
  \label{eqn:fock-1-den-orthon} \\
  \rho_\frpp(\rmxx_1, \rmxx_2)
  ={} & \sum_{j, k} \lrvert{\phi_j(\rmxx_1)}^2 \lrvert{\phi_k(\rmxx_2)}^2
  + \frpp \sum_{j, k} \phi_k^*(\rmxx_1) \phi_j(\rmxx_1) \phi_j^*(\rmxx_2) \phi_k(\rmxx_2)
  - \bbone_{\frpp = 1} 2 \sum_{j \in [N]} \lrvert{\phi_j(\rmxx_1)}^2 \lrvert{\phi_j(\rmxx_2)}^2 \\
  ={} & \rho_\frpp(\rmxx_1) \rho_\frpp(\rmxx_2) + \frpp \lrvert*[\Big]{\sum_{j \in [N]} \phi_j^*(\rmxx_1) \phi_j(\rmxx_2)}^2 - \bbone_{\frpp = 1} 2 \sum_{j \in [N]} \lrvert{\phi_j(\rmxx_1)}^2 \lrvert{\phi_j(\rmxx_2)}^2.
  \label{eqn:fock-2-den-orthon}
\end{align}

\subsubsection{Ladder operators in Fock space}
Define a handy notation, $[\phi_1, \cdots, \phi_N]_{\bmpi_1 \ldots \bmpi_N}$, as the tensor product where $\phi_i$ appears in the $\bmpi_i$-th position, \ie,
\begin{align}
  [\phi_1, \cdots, \phi_N]_{\bmpi_1 \ldots \bmpi_N} :={} & \ket*{\phi_{\bmpi^{-1}_1}} \otimes \cdots \otimes \ket*{\phi_{\bmpi^{-1}_N}}, \\
  \braket{\rmxx_1, \cdots, \rmxx_N}{[\phi_1, \cdots, \phi_N]_{\bmpi_1 \ldots \bmpi_N}} ={} & [\phi_1, \cdots, \phi_N]_{\bmpi_1 \ldots \bmpi_N}(\rmxx_1, \cdots, \rmxx_N)
  = \phi_1(\rmxx_{\bmpi_1}) \cdots \phi_N(\rmxx_{\bmpi_N}).
\end{align}
Since $\frpp^{\bmpi} = \frpp^{\bmpi^{-1}}$ and $\bmpi^{-1}$ traverses $\rmPi_N$ when $\bmpi$ does, we have:
\begin{align}
  \ket{\phi_1 \cdots \phi_N}_\frpp ={} & \frac{1}{\sqrt{N!}} \sum_{\bmpi \in \rmPi_N} \frpp^{\bmpi} \, [\phi_1, \cdots, \phi_N]_{\bmpi_1 \ldots \bmpi_N},
  \label{eqn:symmetrizer-alt} \\
  \ket*{\phi_1 \cdots \phi_k \cdots \phi_N}_\frpp
  \stackrel{\text{\eqnref{perm-decomp}}}{=} {} & \frac{1}{\sqrt{N!}} \sum_{j \in [N]} \frpp^{j-k} \sum_{\bmpi' \in \rmPi_{[N] \setminus \{j\}}} \frpp^{\bmpi'} \, [\phi_1, \cdots, \phi_k, \cdots, \phi_N]_{\bmpi'_1 \ldots j \ldots \bmpi'_{N-1}} \\
  ={} & \frac{1}{\sqrt{N!}} \sum_{j \in [N]} \frpp^{j-k} \sum_{\bmpi' \in \rmPi_{[N] \setminus \{j\}}} \frpp^{\bmpi'} \, [\phi_k, \phi_1, \cdots, \phi_{\neg k}, \cdots, \phi_N]_{j, \bmpi'_1 \ldots \bmpi'_{N-1}} \\
  ={} & \frac{1}{\sqrt{N}} \sum_{j \in [N]} \frpp^{j-k} \, [\phi_k, \ket{\phi_1 \cdots \phi_{\neg k} \cdots \phi_N}_\frpp]_{j, (1, \ldots, \neg j, \ldots, N)}.
  \label{eqn:symmetrizer-insertk}
\end{align}

Given $\eta \in \scH$, define the \textbf{creation operator} $\aah^\dagger(\eta): \scF_{\frpp,N-1} \to \scF_{\frpp,N}$ as:
\begin{align}
  \aah^\dagger(\eta) \ket{\phi_1 \cdots \phi_{N-1}}_\frpp
  := \ket{\eta \phi_1 \cdots \phi_{N-1}}_\frpp
  \stackrel{\text{\eqnref{symmetrizer-insertk}}}{=} \frac{1}{\sqrt{N}} \sum_{j \in [N]} \frpp^{j-1} \big[ \eta, \ket{\phi_1 \cdots \phi_{N-1}}_\frpp \big]_{j, (1, \ldots, \neg j, \ldots, N)}.
  \label{eqn:fock-creation}
\end{align}
Its adjoint in $\scF_\frpp$, the \textbf{annihilation operator} $\aah(\eta): \scF_{\frpp,N-1} \to \scF_{\frpp,N}$, is characterized by
\begin{align}
  \obraket{\phi_1 \cdots \phi_{N-1}}{\aah(\eta)}{\chi_1 \cdots \chi_N}_\frpp
  ={} & \obraket{\chi_1 \cdots \chi_N}{\aah^\dagger(\eta)}{\phi_1 \cdots \phi_{N-1}}_\frpp^* \\
  \stackrel{\text{\eqnref{fock-creation}}}{=} {} & \frac{1}{\sqrt{N}} \sum_{j \in [N]} \frpp^{j-1} \braket{\chi_1 \cdots \chi_N}{ \big[ \eta, \ket{\phi_1 \cdots \phi_{N-1}}_\frpp \big]_{j, (1, \ldots, \neg j, \ldots, N)} }_\frpp^* \\
  \stackrel{\text{\eqnref{symmetrizer-insertk}}}{=} {} & \frac{1}{N} \sum_{j \in [N]} \frpp^{j-1} \sum_{k \in [N]} \frpp^{k-j} \braket{\chi_k}{\eta}^* \braket{\chi_1 \cdots \chi_{\neg k} \cdots \chi_N}{\phi_1 \cdots \phi_{N-1}}_\frpp^* \\
  ={} & \sum_{k \in [N]} \frpp^{k-1} \braket{\eta}{\chi_k} \braket{\phi_1 \cdots \phi_{N-1}}{\chi_1 \cdots \chi_{\neg k} \cdots \chi_N}_\frpp,
\end{align}
which indicates:
\begin{align}
  \aah(\eta) \ket{\phi_1 \cdots \phi_N}_\frpp = \sum_{j \in [N]} \frpp^{j-1} \braket{\eta}{\phi_j} \ket{\phi_1 \cdots \phi_{\neg j} \cdots \phi_N}_\frpp.
  \label{eqn:fock-annih}
\end{align}

Alternatively, the two operators can also be defined recursively, in the form presented in \href{https://en.wikipedia.org/wiki/Second_quantization#Creation_and_annihilation_operators}{Wikipedia}.
Define the \textbf{insertion and deletion operators} recursively:
\begin{align}
  \ket{\phi_i} \otimes_\frpp \ket{1} := \ket{\phi_i}, \hspace{8pt}
  & \ket{\phi_i} \otimes_\frpp (\ket{\phi_j} \otimes \ket{\psi}) := \ket{\phi_i} \otimes \ket{\phi_j} \otimes \ket{\psi} + \frpp \ket{\phi_j} \otimes (\ket{\phi_i} \otimes_\frpp \ket{\psi}); \\
  \ket{\phi_i} \oslash_\frpp \ket{1} := 0, \hspace{8pt}
  & \ket{\phi_i} \oslash_\frpp (\ket{\phi_j} \otimes \ket{\psi}) := \delta_{ij} \ket{\psi} + \frpp \ket{\phi_j} \otimes (\ket{\phi_i} \oslash_\frpp \ket{\psi}).
\end{align}
Note $\otimes_{-1}$ is just the wedge product $\wedge$.
Then the two operators can be defined as:
\begin{align}
  \aah^\dagger(\eta) \ket{\psi} := \frac{1}{\sqrt{N+1}} \ket{\eta} \otimes_\frpp \ket{\psi}, \hspace{8pt}
  \aah(\eta) \ket{\psi} := \frac{1}{\sqrt{N}} \ket{\eta} \oslash_\frpp \ket{\psi}, \quad
  \forall \eta \in \scH, \psi \in \scF_{\frpp,N}.
\end{align}

We can then verify that for any $\ket{\eta_1}$ and $\ket{\eta_2}$,
\begin{align}
  \aah^\dagger(\eta_2) \aah(\eta_1) \ket{\phi_1 \cdots \phi_N}_\frpp
  \stackrel{\text{\eqnref{fock-annih}}}{=} {} & \sum_{j \in [N]} \frpp^{j-1} \braket{\eta_1}{\phi_j} \aah^\dagger(\eta_2) \ket{\phi_1 \cdots \phi_{\neg j} \cdots \phi_N}_\frpp
  \stackrel{\text{\eqnref{fock-creation}}}{=} \sum_{j \in [N]} \frpp^{j-1} \braket{\eta_1}{\phi_j} \ket*{\eta_2 \phi_1 \cdots \phi_{\neg j} \cdots \phi_N}_\frpp \\
  \stackrel{\text{\eqnref{symmetrizer-interchange}}}{=} {} & \sum_{j \in [N]} \braket{\eta_1}{\phi_j} \ket*{\phi_1 \cdots \underbrace{\eta_2}_\text{replace $\phi_j$} \cdots \phi_N}_\frpp, \qquad
  \label{eqn:count-op-fock} \\
  \aah(\eta_1) \aah^\dagger(\eta_2) \ket{\phi_1 \cdots \phi_N}_\frpp
  \stackrel{\text{\eqnref{fock-creation}}}{=} {} & \aah(\eta_1) \ket{\eta_2 \phi_1 \cdots \phi_N}_\frpp
  \stackrel{\text{\eqnref{fock-annih}}}{=} \braket{\eta_1}{\eta_2} \ket{\phi_1 \cdots \phi_N}_\frpp + \sum_{j \in [N]} \frpp^j \braket{\eta_1}{\phi_j} \ket*{\eta_2 \phi_1 \cdots \phi_{\neg j} \cdots \phi_N}_\frpp \\
  \stackrel{\text{\eqnref{symmetrizer-interchange}}}{=} {} & \braket{\eta_1}{\eta_2} \ket{\phi_1 \cdots \phi_N}_\frpp + \frpp \sum_{j \in [N]} \braket{\eta_1}{\phi_j} \ket*{\phi_1 \cdots \underbrace{\eta_2}_\text{replace $\phi_j$} \cdots \phi_N}_\frpp.
\end{align}
By defining $\lrbrack*{\Ah, \Bh}_{-\frpp} := \Ah \Bh - \frpp \Bh \Ah$, we have:
\begin{align}
  \lrbrack{\aah^\dagger(\eta_1), \aah^\dagger(\eta_2)}_{-\frpp} = \lrbrack{\aah(\eta_1), \aah(\eta_2)}_{-\frpp} = 0, \hspace{8pt}
  \lrbrack{\aah(\eta_1), \aah^\dagger(\eta_2)}_{-\frpp} = \braket{\eta_1}{\eta_2}.
\end{align}
So for bosons ($\frpp = 1$) and \textbf{orthonormal} $\{\eta_i\}_i$, they are the standard lowering and raising operators.
From \eqnref{count-op-fock}, we also have for \textbf{orthonormal} $\{\phi_i\}_i$,
\begin{align}
  \aah^\dagger(\phi_j) \aah(\phi_j) \ket{\phi_{i_1} \cdots \phi_{i_N}}_\frpp
  = \sum_{k \in [N]} \delta_{j i_k} \ket*{\phi_{i_1} \cdots \underbrace{\phi_j}_\text{replace $\phi_{i_k}$} \cdots \phi_{i_n}}_\frpp
  = \sum_{k \in [N]} \delta_{j i_k} \ket*{\phi_{i_1} \cdots \phi_{i_n}}_\frpp
  = n_j \ket{\phi_{i_1} \cdots \phi_{i_N}}_\frpp
\end{align}
counts the number of particles in $\ket{\phi_{i_1} \cdots \phi_{i_N}}_\frpp$ occupying state $\ket{\phi_j}$.

\subsubsection{Second quantization in Fock space}
For a general one-particle operator $\Oh_\cdot$, define its extension $\Oh$ to $N$ particles as:
  $\Oh \ket{\psi(\rmxx_1, \cdots, \rmxx_N)} := \sum_{k \in [N]} \Oh_k \ket{\psi(\rmxx_1, \cdots, \rmxx_N)},$
where $\Oh_k$ acts on particle $\rmxx_k$ (\eqnref{opr-extension}).
Its application on a Fock state is:
\begin{align}
  \Oh \ket{\phi_1 \cdots \phi_N}_\frpp
  ={} & \frac{1}{\sqrt{N!}} \sum_{k \in [N]} \sum_{\bmpi \in \rmPi_N} \frpp^{\bmpi} \ket{\phi_{\bmpi_1}} \otimes \cdots \otimes \Oh_\cdot \ket{\phi_{\bmpi_k}} \otimes \cdots \otimes \ket{\phi_{\bmpi_N}} \\
  \stackrel{\text{\eqnref{perm-decomp}}}{=} {} & \frac{1}{\sqrt{N!}} \sum_{k \in [N]} \sum_{j \in [N]} \frpp^{j-k} \sum_{\bmpi' \in \rmPi_{[N] \setminus \{j\}}} \frpp^{\bmpi'} \ket*{\phi_{\bmpi'_1}} \otimes \cdots \otimes \underbrace{\Oh_\cdot \ket{\phi_j}}_{k\text{-th}} \otimes \cdots \otimes \ket*{\phi_{\bmpi'_{N-1}}} \\
  \stackrel{\bmpi'' \asn {\bmpi'}^{-1}}{=} {} & \frac{1}{\sqrt{N!}} \sum_{j \in [N]} \sum_{k \in [N]} \frpp^{j-k} \sum_{\bmpi'' \in \rmPi_{[N] \setminus \{k\}}} \frpp^{\bmpi''} [\phi_1, \cdots, \Oh_\cdot \phi_j, \cdots, \phi_N]_{\bmpi''_1, \ldots, k, \ldots, \bmpi''_{N-1}} \\
  \stackrel{\text{\eqnref{perm-decomp}}}{=} {} & \frac{1}{\sqrt{N!}} \sum_{j \in [N]} \sum_{\bmpi \in \rmPi_N} \frpp^{\bmpi} [\phi_1, \cdots, \Oh_\cdot \phi_j, \cdots, \phi_N]_{\bmpi_1, \ldots, \bmpi_N} \\
  ={} & \sum_{j \in [N]} \ket*{\phi_1 \cdots (\Oh_\cdot \phi_j) \cdots \phi_N}_\frpp.
  \label{eqn:n-opr-fock}
\end{align}

For a basis $\{\ket{\alpha}\}_\alpha$ of $\scH$, we can construct a one-particle operator:
\begin{align}
  \Ah^{(\alpha\alpha')}_\cdot := \ket{\alpha}\bra{\alpha'}, \quad \scH \to \scH.
\end{align}
Its extension acting on a $\scF_{\frpp,N}$ state is given by:
\begin{align}
  \Ah^{(\alpha\alpha')} \ket{\phi_1 \cdots \phi_N}_\frpp
  \stackrel{\text{\eqnref{n-opr-fock}}}{=} \sum_{j \in [N]} \braket{\alpha'}{\phi_j} \ket*{\phi_1 \cdots \underbrace{\alpha}_\text{replace $\phi_j$} \cdots \phi_N}_\frpp.
\end{align}
Compared with \eqnref{count-op-fock}, we know:
\begin{align}
  \Ah^{(\alpha\alpha')} = \aah^\dagger(\alpha) \aah(\alpha').
\end{align}

If $\{\ket{\alpha}\}_\alpha$ is complete, then these operators $\{\Ah^{(\alpha\alpha')}_\cdot\}_{\alpha\alpha'}$ also form a complete basis of one-particle-state operators,
so for any one-particle operator $\Oh_\cdot$, we have $\Oh_\cdot = \sum_{\alpha, \alpha'} \ket{\alpha} \obraket*{\alpha}{\Oh_\cdot}{\alpha'} \bra{\alpha'}$.
So for its extension $\Oh$,
\begin{align}
  \Oh \ket{\phi_1 \cdots \phi_N}_\frpp
  \stackrel{\text{\eqnref{n-opr-fock}}}{=} {} & \sum_{j \in [N]} \ket*[\Big]{\phi_1 \cdots \lrparen*[\Big]{\sum_{\alpha, \alpha'} \ket{\alpha} \obraket*{\alpha}{\Oh_\cdot}{\alpha'} \braket{\alpha'}{\phi_j}} \cdots \phi_N}_\frpp
  = \sum_{\alpha, \alpha'} \obraket*{\alpha}{\Oh_\cdot}{\alpha'} \sum_{j \in [N]} \braket{\alpha'}{\phi_j} \ket*[\Big]{\phi_1 \cdots \underbrace{\alpha}_\text{replace $\phi_j$} \cdots \phi_N}_\frpp \\
  \stackrel{\text{\eqnref{count-op-fock}}}{=} {} & \sum_{\alpha, \alpha'} \obraket*{\alpha}{\Oh_\cdot}{\alpha'} \aah^\dagger(\alpha) \aah(\alpha') \ket{\phi_1 \cdots \phi_N}_\frpp,
\end{align}
which means:
\begin{align}
  \Oh = \sum_{\alpha, \alpha'} \obraket*{\alpha}{\Oh_\cdot}{\alpha'} \Ah^{(\alpha\alpha')}
  = \sum_{\alpha, \alpha'} \obraket*{\alpha}{\Oh_\cdot}{\alpha'} \aah^\dagger(\alpha) \aah(\alpha').
\end{align}

Similarly, for a two-particle operator $\Wh_{\cdot\cdot}$, we have
$\Wh_{\cdot\cdot} = \sum_{\alpha,\beta,\alpha',\beta'} \ket{\alpha \beta} \obraket*{\alpha \beta}{\Wh_{\cdot\cdot}}{\alpha' \beta'} \bra{\alpha' \beta'}$, and its extension
$\Wh \ket{\psi(\rmxx_1, \cdots, \rmxx_N)} := \frac{1}{2} \sum_{j \ne k} \Wh_{jk} \ket{\psi(\rmxx_1, \cdots, \rmxx_N)}$ (\eqnref{opr-extension}). So,
\begin{align}
  \Wh \ket{\phi_1 \cdots \phi_N}_\frpp
  ={} & \frac{1}{2} \sum_{j \ne k} \sum_{\alpha,\beta,\alpha',\beta'} \braket{\alpha'}{\phi_j} \braket{\beta'}{\phi_k} \obraket*{\alpha \beta}{\Wh_{\cdot\cdot}}{\alpha' \beta'} \ket*[\big]{\phi_1 \cdots \underbrace{\alpha}_\text{replace $\phi_j$} \cdots \underbrace{\beta}_\text{replace $\phi_k$} \cdots \phi_N}_\frpp,
\end{align}
which holds since $\braket{\alpha' \beta'}{\phi_j \phi_k} = \int \ud \rmxx_1 \ud \rmxx_2 \, {\alpha'}^*(\rmxx_1) {\beta'}^*(\rmxx_2) \phi_j(\rmxx_1) \phi_k(\rmxx_2)
= \braket{\alpha'}{\phi_j} \braket{\beta'}{\phi_k}$.
On the other hand,
\begin{align}
  & \aah^\dagger(\alpha) \aah^\dagger(\beta) \aah(\beta') \aah(\alpha') \ket{\phi_1 \cdots \phi_N}_\frpp
  \stackrel{\text{\eqnref{fock-annih}}}{=} \sum_{j \in [N]} \frpp^{j-1} \braket{\alpha'}{\phi_j} \aah^\dagger(\alpha) \aah^\dagger(\beta) \aah(\beta') \ket{\phi_1 \cdots \phi_{\neg j} \cdots \phi_N}_\frpp \\
  \stackrel{\text{\eqnref{count-op-fock}}}{=} {} & \sum_{j \in [N]} \frpp^{j-1} \braket{\alpha'}{\phi_j} \sum_{k \in [N] \setminus \{j\}} \braket{\beta'}{\phi_k} \aah^\dagger(\alpha) \ket*[\big]{\phi_1 \cdots \phi_{\neg j} \cdots \underbrace{\beta}_\text{replace $\phi_k$} \cdots \phi_N}_\frpp \\
  \stackrel{\text{\eqnref{fock-creation}}}{=} {} & \sum_{j \ne k} \frpp^{j-1} \braket{\alpha'}{\phi_j} \braket{\beta'}{\phi_k} \ket*[\big]{\alpha \phi_1 \cdots \phi_{\neg j} \cdots \underbrace{\beta}_\text{replace $\phi_k$} \cdots \phi_N}_\frpp \\
  \stackrel{\text{\eqnref{symmetrizer-interchange}}}{=} {} & \sum_{j \ne k} \braket{\alpha'}{\phi_j} \braket{\beta'}{\phi_k} \ket*[\big]{\phi_1 \cdots \underbrace{\alpha}_\text{replace $\phi_j$} \cdots \underbrace{\beta}_\text{replace $\phi_k$} \cdots \phi_N}_\frpp,
  \label{eqn:ladder-opr-double}
\end{align}
so we have:
\begin{align}
  \Wh = \frac{1}{2} \sum_{\alpha, \beta, \alpha', \beta'} \obraket*{\alpha \beta}{\Wh_{\cdot\cdot}}{\alpha' \beta'} \aah^\dagger(\alpha) \aah^\dagger(\beta) \aah(\beta') \aah(\alpha').
\end{align}

For an $N$-particle Hamiltonian
$\Hh = \Oh + \Wh
= \sum_{\alpha, \alpha'} \bfO^\alpha_{\alpha'} \aah^\dagger(\alpha) \aah(\alpha')
+ \frac{1}{2} \sum_{\alpha, \beta, \alpha', \beta'} \bfW^{\alpha\beta}_{\alpha'\beta'} \aah^\dagger(\alpha) \aah^\dagger(\beta) \aah(\beta') \aah(\alpha')$
where $\bfO^\alpha_{\alpha'} := \obraket*{\alpha}{\Oh_\cdot}{\alpha'}$ and $\bfW^{\alpha\beta}_{\alpha'\beta'} := \obraket*{\alpha \beta}{\Wh_{\cdot\cdot}}{\alpha' \beta'}$, we have:
\begin{align}
  \obraket*{\psi}{\Hh}{\psi} = \sum_{\alpha, \alpha'} \bfO^\alpha_{\alpha'} \bfP^{\alpha'}_\alpha + \frac{1}{2} \sum_{\alpha, \beta, \alpha', \beta'} \bfW^{\alpha\beta}_{\alpha'\beta'} \bfP^{\alpha'\beta'}_{\alpha\beta},
\end{align}
where $\bfP^{\alpha'}_\alpha := \obraket*{\psi}{\aah^\dagger(\alpha) \aah(\alpha')}{\psi}$
and $\bfP^{\alpha'\beta'}_{\alpha\beta} := \obraket*{\psi}{\aah^\dagger(\alpha) \aah^\dagger(\beta) \aah(\beta') \aah(\alpha')}{\psi}$
are the \textbf{one- and two-particle reduced density matrix (1- and 2-RDM)} of the $N$-particle wavefunction $\psi$.\footnote{
  They are the analogues of $\obraket*{\alpha'}{P^{(1)}}{\alpha}$ and $\obraket*{\alpha'\beta'}{P^{(2)}}{\alpha\beta}$.
}
This is the RDM-form of the Hamiltonian under the \emph{second quantization} formulation.
Particularly, for $\ket{\psi_\frpp} = \ket{\phi_1 \cdots \phi_N}_\frpp$,
\begin{align}
  & \obraket{\psi_\frpp}{\aah^\dagger(\rmxx'_1) \aah^\dagger(\rmxx'_2) \aah(\rmxx_2) \aah(\rmxx_1)}{\psi_\frpp}
  \stackrel{\text{\eqnref{ladder-opr-double}}}{=} \sum_{j \ne k} \braket{\rmxx_1}{\phi_j} \braket{\rmxx_2}{\phi_k} \braket*[\big]{\phi_1 \cdots \underbrace{\rmxx'_1}_\text{replace $\phi_j$} \cdots \underbrace{\rmxx'_2}_\text{replace $\phi_k$} \cdots \phi_N}{\phi_1 \cdots \phi_N}_\frpp^* \\
  \stackrel{\text{\eqnref{fock-inprod}}}{=} {} & \sum_{j \ne k} \phi_j(\rmxx_1) \phi_k(\rmxx_2) \sum_{\bmpi \in \rmPi_N} \frpp^{\bmpi} \prod_{i \notin \{j,k\}} \braket{\phi_i}{\phi_{\bmpi_i}}^* \phi_{\bmpi_j}^*(\rmxx'_1) \phi_{\bmpi_k}^*(\rmxx'_2) \\
  \stackrel{\text{\eqnref{perm-decomp}}}{=} {} & \sum_{j \ne k} \phi_j(\rmxx_1) \phi_k(\rmxx_2) \sum_{j' \in [N]} \frpp^{j'-j} \sum_{\bmpi' \in \rmPi_{[N] \setminus \{j'\}}} \frpp^{\bmpi'} \prod_{i \notin \{j,k\}} \braket*{\phi_i}{\phi_{\bmpi'_{i - \bbone_{i>j}}}}^* \phi_{j'}^*(\rmxx'_1) \phi_{\bmpi'_{k - \bbone_{k>j}}}^*(\rmxx'_2) \\
  \stackrel[\text{(*)}]{\text{\eqnref{perm-decomp}}}{=} {} & \sum_{j \ne k} \phi_j(\rmxx_1) \phi_k(\rmxx_2) \sum_{j' \in [N]} \frpp^{j'-j} \!\!\!\! \sum_{k' \in [N] \setminus \{j'\}} \!\! \frpp^{(k' - \bbone_{k'>j'}) - (k - \bbone_{k>j})} \!\!\!\!\!\! \sum_{\bmpi'' \in \rmPi_{[N] \setminus \{j',k'\}}} \!\!\!\! \frpp^{\bmpi''} \!\! \prod_{i \notin \{j,k\}} \braket*{\phi_i}{\phi_{\bmpi''_{i - \bbone_{i>j} - \bbone_{i>k}}}}^* \phi_{j'}^*(\rmxx'_1) \phi_{k'}^*(\rmxx'_2) \\
  \stackrel{\text{\eqnref{fock-inprod}}}{=} {} & \sum_{j \ne k} \sum_{j' \ne k'} \frpp^{j - k + \bbone_{k>j}} \frpp^{j' - k' + \bbone_{k'>j'}} \phi_j(\rmxx_1) \phi_k(\rmxx_2) \phi_{j'}^*(\rmxx'_1) \phi_{k'}^*(\rmxx'_2) \braket{\cdots \phi_{\neg j'} \cdots \phi_{\neg k'} \cdots}{\cdots \phi_{\neg j} \cdots \phi_{\neg k} \cdots}_\frpp \\
  \stackrel{\text{\eqnref{fock-2-rdm-ne}}}{=} {} & P_\frpp(\rmxx_1, \rmxx_2; \rmxx'_1, \rmxx'_2)
\end{align}
is the usual 2-RDM of $\psi_\frpp$ under the coordinate representation.
(Note in (*), $k' - \bbone_{k'>j'}$ represents the position of value $k'$ in the ordered sequence $[1,\cdots,j'-1,j'+1,\cdots,N]$ of $\bmpi'$, and $k - \bbone_{k>j}$ is the position at which $\bmpi'$ is to be replaced with value $k'$.)

\subsection{Coulomb Many-Body System}

\subsubsection{Hamiltonian}

\begin{figure}[t]
  \centering
  \includegraphics[width=.8\textwidth]{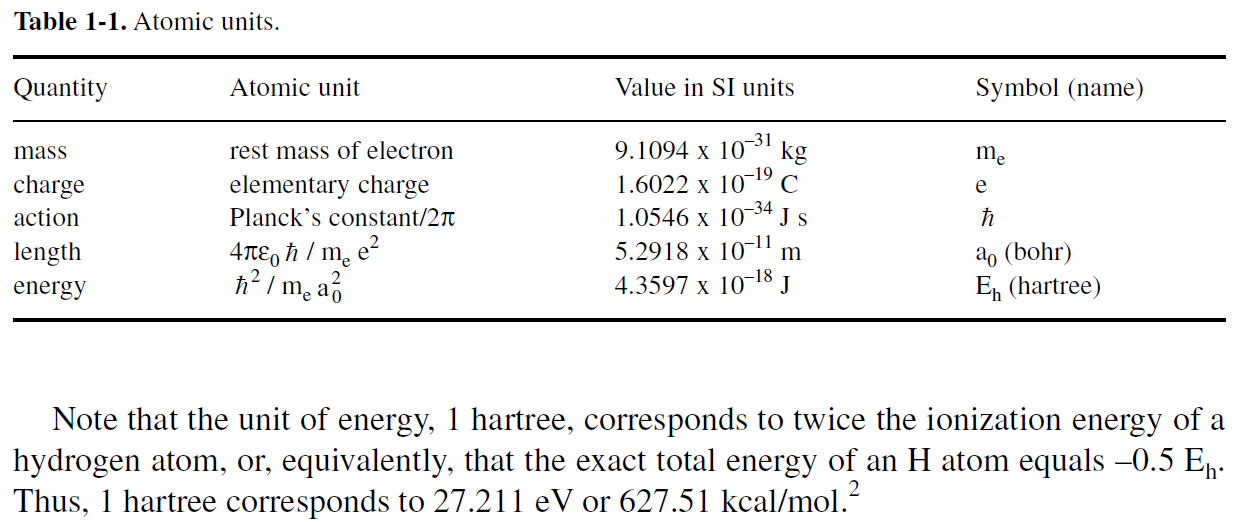}
  \caption{Atomic units (from~\citep{koch2001chemist}).}
  \label{fig:atomic-units}
\end{figure}

For a molecule system, under the Born-Oppenheimer approximation, the Hamiltonian for its $N$ electrons given the charges and positions of the $N_\tnA$ nuclei $\{(Z_A, \rrv_A)\}_{A \in [N_\tnA]}$ is (adopting the atomic units; see \figref{atomic-units}):
\begin{align}
  \Hh = \underbrace{\Th + \Vh_\Ne}_{=: \Hh^{(1)}} + \Wh_\ee
  = \sum_{i \in [N]} \lrparen*[\Big]{\Th_i + {\Vh_\Ne}_i} + \frac{1}{2} \sum_{i \ne j} {\Wh_\ee}_{ij},
  \label{eqn:hamiltonian}
\end{align}
where the terms correspond to the extensions of $\Th_i \psi(\bfxx) = -\frac{1}{2} \nabla^2_i \psi(\bfxx)$,
${\Vh_\Ne}_i \psi(\bfxx) = V_\Ne(\rrv_i) \psi(\bfxx) = -\sum_{A \in [N_\tnA]} \frac{Z_A}{r_{iA}} \psi(\bfxx)$ where $r_{iA} := \lrVert{\rrv_i - \rrv_A}$, and
${\Wh_\ee}_{ij} \psi(\bfxx) = \frac{1}{r_{ij}} \psi(\bfxx)$ where $r_{ij} := \lrVert{\rrv_i - \rrv_j}$.
From \eqnsref{1-opr-avg-wavefn,1-opr-avg-rdm,2-opr-avg}, the mean energy is:
\begin{align}
  E[\psi] = \obraket*{\psi}{\Hh}{\psi}
  ={} & -\frac{N}{2} \obraket*{\psi}{\nabla^2_1}{\psi} + N \obraket*{\psi}{{\Vh_\Ne}_1}{\psi} + \frac{1}{2} N (N-1) \obraket*{\psi}{1/r_{12}}{\psi} \\
  ={} & \underbrace{ -\frac{1}{2} \tr(\nabla^2 P^{(1)}_{[\psi]}) }_{T[P^{(1)}]}
  + \underbrace{ \vphantom{\frac{1}{2}} \bbE_{\rho_{[\psi]}} [V_\Ne] }_{E_\Ne[\rho]}
  + \underbrace{ \frac{1}{2} \bbE_{\rho^{(2)}_{[\psi]}(\rmxx_1, \rmxx_2)} [1/r_{12}] }_{E_\ee[\rho^{(2)}]}.
  \label{eqn:energy-decomp}
\end{align}
Symbolically,
\begin{align}
  E[P^{12}_{1'2'}] ={} & -\frac{1}{2} \sum\nolimits_{1'} \nabla^2_{1=1'} P^1_{1'} + \bbE_{\rho_1}[{V_\Ne}_1] + \frac{1}{2} \bbE_{\rho_{12}}[1/r_{12}], \text{where} \\
  P^1_{1'} ={} & \frac{1}{N-1} \sum\nolimits_{2'} P^{12'}_{1'2'}, \quad
  \rho_{12} = P^{12}_{12}, \quad
  \rho_1 = \frac{1}{N-1} \sum\nolimits_{2'} \rho_{12'}
\end{align}
can all be expressed using the 2-RDM.
Nevertheless, the proper set of 2-RDMs is hard to describe and only increasingly better necessary conditions and sufficient conditions are known.

Note that all contractions/reductions $\obraket{\cdot}{\cdot}{\cdot}$, $\tr(\cdot)$, $\bbE_\cdot[\cdot]$ also sum over spins.
As the operators do not contain/operate on spins, the same equations hold for the spacial counterparts of $\rho$, $\rho^{(2)}$, $P^{(1)}$, and $P^{(2)}$ that sum over spins beforehand.

By Property~(1) of the spin-independent exchange-correlation hole,
\begin{align}
  E_\ee[\rho^{(2)}] ={} & J[\rho] + E_\XC[\rho; h_\XC], \quad \text{where}
  \label{eqn:Eee-decomp} \\
  J[\rho] :={} & \frac{1}{2} \int \ud\rrv_1 \ud\rrv_2 \, \frac{\rho(\rrv_1) \rho(\rrv_2)}{r_{12}}
  = \frac{1}{2} \bbE_\rho[V_\tnJ]
  = \frac{1}{2} \bbE_{\rho_1 \rho_2} [1/r_{12}],
  \text{ where } V_\tnJ(\rrv) := \bbE_{\rho(\rrv')} [1/\lrVert{\rrv - \rrv'}],
  \label{eqn:ej} \\
  E_\XC[\rho; h_\XC] :={} & \frac{1}{2} \int \ud\rrv_1 \ud\rrv_2 \, \frac{\rho(\rrv_1,\rrv_2) - \rho(\rrv_1) \rho(\rrv_2)}{r_{12}}
  = \frac{1}{2} \int \ud\rrv_1 \ud\rrv_2 \, \frac{\rho(\rrv_1) h_\XC(\rrv_2|\rrv_1)}{r_{12}}
  = \frac{1}{2} \bbE_{\rho_1 {h_\XC}_{21}} [1/r_{12}].
  \label{eqn:exc}
\end{align}
Here, the Hartree energy $J[\rho]$ (may also be denoted as $E_\tnH$, $E_\textnormal{Hart}$) represents the \emph{classical} electrostatic/Coulomb energy of a charge distribution with itself.
It treats $\rho$ as a charge density but ignores its origin from correlated, distributed electrons.
It assumes the element forming the charge density $\rho$ is infinitely divisible so any infinitesimal bulk of $\rho$ can interact with other parts of $\rho$.
But this is not true since the charge of an electron cannot be divided and the Coulomb energy from the interaction of charges from the same electron should not be counted.
So there is an unphysical \emph{self-interaction} error in $J[\rho]$.
Particularly, if $\rho$ arises from just one electron distributed in space, $E_\ee$ should be zero while obviously $J[\rho]$ is not.
The exchange-correlation energy $E_\XC[\rho; h_\XC]$ handles the \emph{non-classical} portion, \ie, the effects of self-interaction, exchange (antisymmetry), and Coulomb correlations.

\subsubsection{Cusp condition}

\begin{figure}[h]
  \centering
  \includegraphics[width=.7\textwidth]{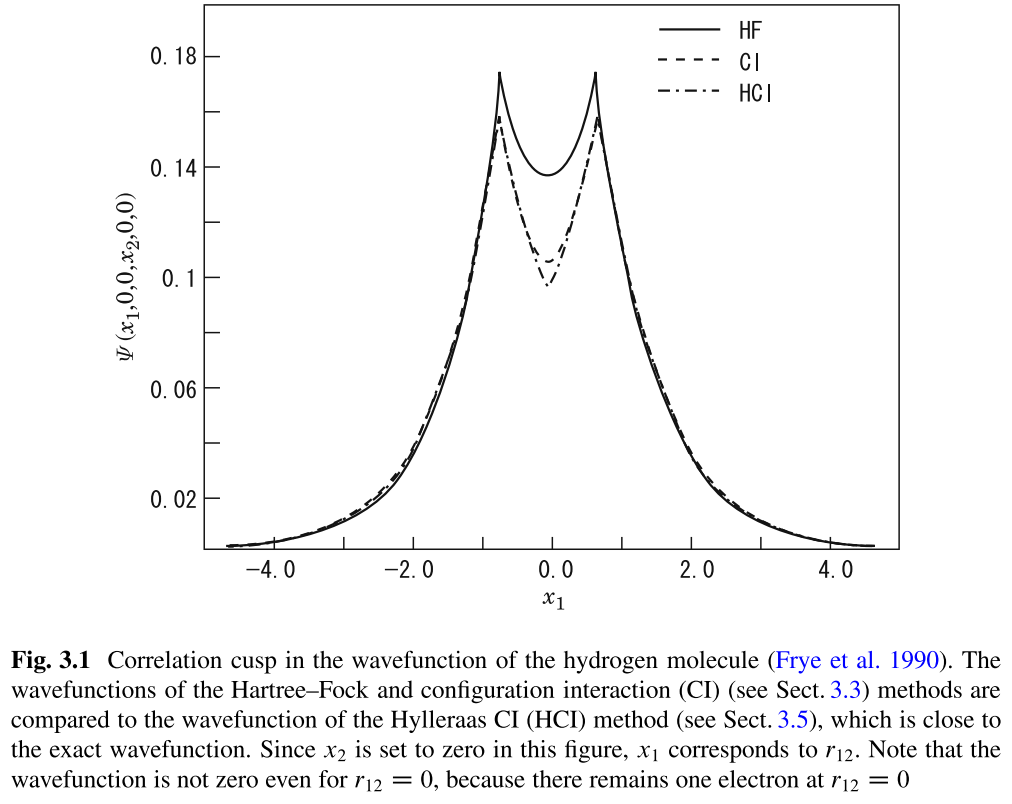}
  \caption{Ground state wavefunction of two antiparallel electrons in $\rmH_2$ (located at $-1$ and $+1$) when fixing $x_2 = 0$ (from~\citep{tsuneda2014density}).}
  \label{fig:antiparallel-cusp}
\end{figure}

Origin: Kato, 1957~\citep{kato1957eigenfunctions}.

\paragraph{Laplacian in spherical coordinates}
Let indices $i,j$ traverse over $\{x,y,z\}$, and $\alpha,\beta$ traverse over $\{r,\theta,\phi\}$.
\begin{align}
  &
  \begin{cases}
    x = r \sin \theta \cos \phi, \\
    y = r \sin \theta \sin \phi, \\
    z = r \cos \theta,
  \end{cases}
  \;
  \lrparen{\fracpartial{\rrv^i}{\rrv^\alpha}}_{i\alpha} =
  \begin{pmatrix}
    \sin \theta \cos \phi & r \cos \theta \cos \phi & -r \sin \theta \sin \phi \\
    \sin \theta \sin \phi & r \cos \theta \sin \phi & r \sin \theta \cos \phi \\
    \cos \theta & -r \sin \theta & 0
  \end{pmatrix},
  \\
  g_{\alpha\beta} &= g_{ij} \fracpartial{\rrv^i}{\rrv^\alpha} \fracpartial{\rrv^j}{\rrv^\beta} = \sum_i \fracpartial{\rrv^i}{\rrv^\alpha} \fracpartial{\rrv^i}{\rrv^\beta},
  \;
  (g_{\alpha\beta})_{\alpha\beta} =
  \begin{pmatrix}
    1 & 0 & 0 \\
    0 & r^2 & 0 \\
    0 & 0 & r^2 \sin^2 \theta
  \end{pmatrix},
  \;
  \sqrt{\lrvert{G}} (g^{\alpha\beta})_{\alpha\beta} =
  \begin{pmatrix}
    r^2 \sin \theta & 0 & 0 \\
    0 & \sin \theta & 0 \\
    0 & 0 & 1/\sin \theta
  \end{pmatrix}.
  \\
  \nabla^2 &= \partial_\alpha (\sqrt{\lrvert{G}} g^{\alpha\beta} \partial_\beta) / \sqrt{\lrvert{G}}
  = \partial_r (r^2 \sin \theta \partial_r) / (r^2 \sin \theta) + \partial_\theta (\sin \theta \partial_\theta) / (r^2 \sin \theta) + \partial_\phi ( (1/\sin \theta) \partial_\phi) / (r^2 \sin \theta) \\
  &= \frac{1}{r^2} \partial_r (r^2 \partial_r) + \frac{1}{r^2} \lrparen{ \frac{1}{\sin \theta} \partial_\theta (\sin \theta \partial_\theta) + \frac{1}{\sin^2 \theta} \partial_\phi^2}.
  \label{eqn:laplacian-sph-R3}
\end{align}
In $\bbR^d$, if denoting $\nabla^2_{\bbS^{d-1}}$ as the Laplacian on the $(d-1)$-dimensional unit sphere, we have:
\begin{align}
  \nabla^2 = \frac{1}{r^{d-1}} \partial_r (r^{d-1} \partial_r) + \frac{1}{r^2} \nabla^2_{\bbS^{d-1}}.
  \label{eqn:laplacian-sph-Rd}
\end{align}

\paragraph{Nuclear cusp condition}
Consider the $N$-electron Hamiltonian $\Hh = -\frac{1}{2} \sum_{i \in [N]} \nabla^2_i + \sum_{i \in [N]} {\Vh_\Ne}_i + \sum_{i < j} {\Wh_\ee}_{ij}$ (\eqnref{hamiltonian}).
For a solution $\psi$ to the Schr\"odinger equation $\Hh \psi = E \psi$, the local energy
\begin{align}
  \eps(\bmrrv) := \frac{(\Hh \psi)(\bmrrv)}{\psi(\bmrrv)}
  = -\frac{1}{2} \sum_{i \in [N]} \frac{\nabla^2_i \psi(\bmrrv)}{\psi(\bmrrv)} - \sum_{i \in [n]} \sum_{A' \in N_\tnA} \frac{Z_{A'}}{r_{iA'}} + \sum_{i < j} \frac{1}{r_{ij}}
  \equiv E
  \label{eqn:local-energy}
\end{align}
is a finite constant.
Note that the wavefunction $\psi$ is continuous everywhere (hence finite everywhere), and is continuously differentiable wherever the potential is finite and may have discontinuous derivatives elsewhere.
So for a configuration $\bmrrv = (\rrv_1, \rrv_2, \cdots, \rrv_N)$ where $\rrv_2, \cdots, \rrv_N$ makes all of $1/r_{iA'}$ and $1/r_{ij}$ finite while $\rrv_1 \to \rrv_A$,
all terms in $\eps(\bmrrv)$ are finite except $-\frac{1}{2} \frac{\nabla^2_1 \psi(\bmrrv)}{\psi(\bmrrv)}$ and $-\frac{Z_A}{r_{1A}}$.
To make $\eps(\bmrrv)$ still a finite constant $E$,
\begin{align}
  \frac{\nabla^2_1 \psi(\bmrrv)}{\psi(\bmrrv)} + \frac{2 Z_A}{r_{1A}}
  \stackrel{\text{\eqnref{laplacian-sph-Rd}}}{=}
  \frac{1}{r^{d-1}} \frac{\partial_r (r^{d-1} \partial_r \psi(r, \thetav, \rrv_{2:N}))}{\psi(r, \thetav, \rrv_{2:N})} + \frac{1}{r^2} \frac{\nabla^2_{\bbS^{d-1}} \psi(r, \thetav, \rrv_{2:N})}{\psi(r, \thetav, \rrv_{2:N})} + \frac{2 Z_A}{r}
  \label{eqn:cusp-wavefn-nuclear-deduction}
\end{align}
should remain finite, where we denoted $r_{1A}$ using $r$, and used the spherical coordinates $(r, \thetav)$ for $\rrv_1$ ($\thetav \in \bbS^{d-1}$).
Since $\psi$ is finite as $r \to 0$, it is equivalent to that $\frac{1}{r^{d-1}} \partial_r (r^{d-1} \partial_r \psi(r, \thetav, \rrv_{2:N})) + \frac{1}{r^2} \nabla^2_{\bbS^{d-1}} \psi(r, \thetav, \rrv_{2:N}) + \frac{2 Z_A}{r} \psi(r, \thetav, \rrv_{2:N})$ is finite.
Since $\psi(r, \thetav, \rrv_{2:N})$ is continuously differentiable in $\thetav$, we have:
\begin{align}
  \int_{\bbS^{d-1}} \ud\thetav \, \nabla^2_{\bbS^{d-1}} \psi(\thetav, \cdots)
  = \int_{\bbS^{d-1}} \ud\thetav \, \nabla_{\bbS^{d-1}} \cdot \nabla_{\bbS^{d-1}} \psi(\thetav, \cdots)
  = \int_{\partial \bbS^{d-1}} \ud\nnv \cdot \nabla_{\bbS^{d-1}} \psi(\thetav, \cdots)
\end{align}
due to Stokes' theorem, which is zero since $\partial \bbS^{d-1} = \emptyset$.
Hence, after taking the spherical average $\frac{1}{\int_{\bbS^{d-1}} \ud\thetav} \int_{\bbS^{d-1}} \ud\thetav \cdots$, it suffices that
\begin{align}
  \frac{1}{r^{d-1}} \partial_r (r^{d-1} \partial_r \psib(r, \rrv_{2:N})) + \frac{2 Z_A}{r} \psib(r, \rrv_{2:N})
  = \partial_r^2 \psib(r, \rrv_{2:N}) + \frac{d-1}{r} \partial_r \psib(r, \rrv_{2:N}) + \frac{2 Z_A}{r} \psib(r, \rrv_{2:N})
\end{align}
is finite as $r \to 0$.
If $\partial_r^2 \psib(r, \rrv_{2:N})$ is finite as $r \to 0$ ($d \ge 2$), we then have:
\begin{align}
  \partial_r \log \psib(r, \rrv_{2:N}) \big|_{r = 0^+} = -\frac{2}{d-1} Z_A.
  \label{eqn:cusp-wavefnmean-nuclear}
\end{align}
When $d = 3$, this is $\partial_r \log \psib(r, \rrv_{2:N}) \big|_{r = 0^+} = -Z_A$.

\paragraph{Schr\"odinger equation in the center-of-mass frame}
Let indices $i,j$ traverse over $\{x_1,y_1,z_1,x_2,y_2,z_2\}$, and $\alpha,\beta$ traverse over $\{x_\tndd,y_\tndd,z_\tndd,x_\tncc,y_\tncc,z_\tncc\}$.
\begin{align}
  &
  \begin{cases}
    \rrv_\tndd = \rrv_2 - \rrv_1, \\
    \rrv_\tncc = \frac{m_1}{m_1 + m_2} \rrv_1 + \frac{m_2}{m_1 + m_2} \rrv_2,
  \end{cases}
  \;
  \begin{cases}
    \rrv_1 = -\frac{m_2}{m_1 + m_2} \rrv_\tndd + \rrv_\tncc, \\
    \rrv_2 = \frac{m_1}{m_1 + m_2} \rrv_\tndd + \rrv_\tncc,
  \end{cases}
  \;
  \lrparen{\fracpartial{\rrv^i}{\rrv^\alpha}}_{i\alpha} =
  \begin{pmatrix}
    -\frac{m_2}{m_1 + m_2} \bfI & \bfI \\
    \frac{m_1}{m_1 + m_2} \bfI & \bfI
  \end{pmatrix},
  \;
  \lrvert{\fracpartial{\rrv^i}{\rrv^\alpha}}_{i\alpha} = 1.
  \\
  g_{\alpha\beta} &= g_{ij} \fracpartial{\rrv^i}{\rrv^\alpha} \fracpartial{\rrv^j}{\rrv^\beta} = 
  \begin{pmatrix}
    -\frac{m_2}{m_1 + m_2} \bfI & \frac{m_1}{m_1 + m_2} \bfI \\
    \bfI & \bfI
  \end{pmatrix}
  \begin{pmatrix}
    m_1 \bfI & 0 \\
    0 & m_2 \bfI
  \end{pmatrix}
  \begin{pmatrix}
    -\frac{m_2}{m_1 + m_2} \bfI & \bfI \\
    \frac{m_1}{m_1 + m_2} \bfI & \bfI
  \end{pmatrix}
  =
  \begin{pmatrix}
    \mu \bfI & 0 \\
    0 & (m_1 + m_2) \bfI
  \end{pmatrix},
  \;
  \mu := \frac{m_1 m_2}{m_1 + m_2}.
  \\
  \bmnabla^2 &= g^{ij} \partial_i \partial_j = \frac{1}{m_1} \nabla^2_1 + \frac{1}{m_2} \nabla^2_2
  = g^{\alpha\beta} \partial_\alpha \partial_\beta = \frac{1}{\mu} \nabla^2_\tndd + \frac{1}{m_1 + m_2} \nabla^2_\tncc.
  \\
  \Th &= -\nabla^2_\tndd - \frac{1}{4} \nabla^2_\tncc - \frac{1}{2} \sum_{i=3}^N \nabla^2_i \;\;\;\; \text{($i$ here indexes electrons again)}.
  \label{eqn:kinetic-op-center-of-mass}
\end{align}

\paragraph{Electronic cusp condition}
First consider two \emph{antiparallel-spin} electrons (no Pauli exclusion), numbered $1$ and $2$, so that $\psi$ remains finite and nonzero as $r_{12} \to 0$ while all other $1/r_{ij}$ and $1/r_{iA'}$ are finite.
The local energy (\eqnref{local-energy}) is finite iff $-\frac{1}{2} (\nabla^2_1 + \nabla^2_2) \psi(\bmrrv) / \psi(\bmrrv) + 1/r_{12}$ is.
Using \eqnref{kinetic-op-center-of-mass}, it requires that $(-\nabla^2_\tndd - \frac{1}{4} \nabla^2_\tncc) \psi(\rrv_\tndd, \rrv_\tncc, \rrv_{3:N}) / \psi(\rrv_\tndd, \rrv_\tncc, \rrv_{3:N}) + 1/r_\tndd$ is finite as $r_\tndd \to 0$,
where we denoted $r_{12}$ as $r_\tndd$ and used the center-of-mass coordinate system.
Since $\psi$ is finite (may be $0$) as $r_\tndd \to 0$, it suffices that $(-\nabla^2_\tndd - \frac{1}{4} \nabla^2_\tncc) \psi(\rrv_\tndd, \rrv_\tncc, \rrv_{3:N}) + \psi(\rrv_\tndd, \rrv_\tncc, \rrv_{3:N}) / r_\tndd$ is finite (may be $0$).
As the local kinetic energy of the center-of-mass / two-electron-system as a whole, $-\frac{1}{4} \nabla^2_\tncc \psi$ should remain finite as the relative distance $r_\tndd \to 0$.
So $-\nabla^2_\tndd \psi(\rrv_\tndd, \rrv_\tncc, \rrv_{3:N}) + \psi(\rrv_\tndd, \rrv_\tncc, \rrv_{3:N}) / r_\tndd$ should be finite.
Following a similar derivation for the nuclear cusp condition, we have:
\begin{align}
  \partial_{r_\tndd} \log \psib(r_\tndd, \rrv_\tncc, \rrv_{3:N}) \big|_{r_\tndd = 0^+} = \frac{1}{d-1},
  \label{eqn:cusp-wavefnmean-electron}
\end{align}
for $d \ge 2$, where $\psib(r_\tndd, \rrv_\tncc, \rrv_{3:N})$ is the spherical average of $\psi(\rrv_\tndd, \rrv_\tncc, \rrv_{3:N})$ over the direction of $\rrv_\tndd$.
When $d = 3$, this is $\partial_{r_\tndd} \log \psib(r_\tndd, \rrv_\tncc, \rrv_{3:N}) \big|_{r_\tndd = 0^+} = \frac{1}{2}$.
For a general expression, the cusp condition can be written as:
\begin{align}
  \partial_{r_{ij}} \log \psib(r_{ij}, {\rrv_\tncc}_{ij}, \rrv_{\neg \{i,j\}}) \big|_{r_{ij} = 0^+} = \frac{2 \mu_{ij} Z_i Z_j}{d-1}.
  \label{eqn:cusp-wavefnmean-general}
\end{align}

\figref{antiparallel-cusp} shows the cusp of the two-electron (with antiparallel spin) wavefunction in $\rmH_2$ molecule.
It shows two observations.
\itemi If Pauli exclusion principle is not explicitly enforced, the Coulomb repulsion between two electrons cannot enforce the wavefunction to go zero at coincidence, although the potential is infinitely large there.
\itemii Although electron coincidence is locally less probable, the region that the cusp condition plays a role may not be ``minor'' / ``less important'', since the neighbourhood of coincidence may take a high probability.

\emph{A concise reasoning} \hspace{4pt}
The cusp condition also applies to \emph{inter-electron Coulomb interaction} with more subtlety.
When $\rrv_i \to \rrv_j$ (for $i \ne j$), leaving only the dominating potential, from \eqnref{hamiltonian} the Schr\"{o}dinger equation becomes $-\frac{1}{2} \sum_{k \in [N]} \nabla^2_k \psi + \frac{1}{r_{ij}} \psi = E \psi$.
\itemI If $s_i = s_j$, then Pauli exclusion principle applies and $\psi(\dots, \rrv_i, \rrv_j, \dots) = -\psi(\dots, \rrv_j, \rrv_i, \dots)$.
  So $\lim_{\rrv_i \to \rrv_j} \nabla^2_i \psi(\dots, \rrv_i, \rrv_j, \dots) = \lim_{\rrv_j \to \rrv_i} \nabla^2_j \psi(\dots, \rrv_j, \rrv_i, \dots)
  = -\lim_{\rrv_j \to \rrv_i} \nabla^2_j \psi(\dots, \rrv_i, \rrv_j, \dots) = -\lim_{\rrv_i \to \rrv_j} \nabla^2_j \psi(\dots, \rrv_i, \rrv_j, \dots)$
  (need some continuity assumptions in the last equality?),
  and $\lim_{\rrv_i \to \rrv_j} \psi(\dots, \rrv_i, \rrv_j, \dots) = 0$ hence $\lim_{\rrv_i \to \rrv_j} \nabla^2_k \psi = 0$ for $k \ne i,j$.
  Taking the limit, the equation becomes $\lim_{\rrv_i \to \rrv_j} \psi / r_{ij} = 0$,
  which means that the Coulomb repulsion makes the antisymmetric wavefunction $\psi$ for \emph{parallel spin} decay faster than linear at the boundary.
\itemII If $s_i \ne s_j$, \ie for \emph{antiparallel spins}, there is no constraint on the spacial wavefunction {\it a priori} (the two electrons have different spins hence distinguishable).
  Apart from $\nabla^2_i \psi$, the term $\nabla^2_j \psi$ may also be infinite in the limit since $\rrv_i \to \rrv_j$ leads to the same singularity as $\rrv_j \to \rrv_i$.
  Other kinetic energy terms for $k \ne i,j$ as well as $E \psi$ remain finite in the limit.
  So leaving only dominating terms, the equation becomes $-\nabla^2_i \psi - \nabla^2_j \psi + \frac{2}{r_{ij}} \psi = 0$, or $(r_{ij} \nabla^2_i \psi - \psi) + (r_{ji} \nabla^2_j \psi - \psi) = 0$.
  Following the same deduction above for $r_{iA} \to 0$, we have $\fracpartial{\psib^{(j)}}{r_{ij}}(\dots, r_{ij}, \rrv_j, \dots) + \fracpartial{\psib^{(i)}}{r_{ji}}(\dots, \rrv_i, r_{ji}, \dots) - \psi(\dots, \rrv_i, \rrv_j, \dots) = 0$, where we made explicit the center for the spherical average.
  Since when $\rrv_i \to \rrv_j$, $\psib^{(j)}(\dots, r_{ij}=0, \rrv_j, \dots) = \psib^{(i)}(\dots, \rrv_i=\rrv_j, r_{ji}=0, \dots)$, the first two terms are equal (needs more investigation).
  So we have:
  \begin{align}
    \lim_{r_{ij} \to 0} \lrparen{\fracpartial{}{r_{ij}} - \frac{1}{2}} \psib(\dots, r_{ij}, \rrv_j, \dots)
    = \lim_{r_{ij} \to 0} \fracpartial{}{r_{ij}} \psib(\dots, r_{ij}, \rrv_j, \dots) - \frac{1}{2} \psi(\dots, \rrv_i = \rrv_j, \rrv_j, \dots) = 0,
  \end{align}
  which also holds when taking $i$ as the center of spherical average (\ie, swapping $i$ and $j$).

\emph{A more detailed and precise derivation} \hspace{4pt}
Ref.:~\citep{foulkes2001quantum}.
Denote $\rrv$ in place of the $\rrv_\tndd$ above as the relative displacement of two electrons in the center-of-mass system.
In the notation, omit other coordinates, $\rrv_\tncc, \rrv_3, \cdots, \rrv_N$, since they does not change or cause divergence as $r \to 0$.
If we let the wavefunction be in the form:
\begin{align}
  \psi(\rrv) = e^{-u(r)} f(\rrv),
  \label{eqn:slater-jastrow}
\end{align}
where $f(\rrv)$ is \emph{smooth} and \emph{antisymmetric} (\eg, a combination of some Slater determinants),
then by noting $\nabla r = \nabla \lrVert{\rrv} = \rrh$ and $\nabla \rrh\trs = (I - \rrh \rrh\trs) / r$ where $\rrh := \rrv / r$ is the unit vector in the direction of $\rrv$ which is a function only of $\thetav$, we have:
\begin{align}
  \nabla \psi(\rrv) ={} & -u'(r) e^{-u(r)} f(\rrv) \rrh + e^{-u(r)} \nabla f(\rrv),
  \label{eqn:slater-jastrow-grad} \\
  \nabla \nabla^\top \psi(\rrv) ={} & \big( -u''(r) + u'(r)^2 \big) e^{-u(r)} f(\rrv) \rrh \rrh\trs
  - u'(r) e^{-u(r)} \big( \rrh \nabla f(\rrv)^\top + \nabla f(\rrv) \rrh\trs \big) - \frac{u'(r)}{r} e^{-u(r)} f(\rrv) (I - \rrh \rrh\trs) + e^{-u(r)} \nabla \nabla\trs f(\rrv) \\
  ={} & \Big( -u''(r) + u'(r)^2 + \frac{u'(r)}{r} \Big) e^{-u(r)} f(\rrv) \rrh \rrh\trs - u'(r) e^{-u(r)} \Big( \rrh \nabla f(\rrv)^\top + \nabla f(\rrv) \rrh\trs + \frac{f(\rrv)}{r} I \Big) + e^{-u(r)} \nabla \nabla\trs f(\rrv),
  \label{eqn:slater-jastrow-hess} \\
  \nabla^2 \psi(\rrv) ={} & \Big( -u''(r) + u'(r)^2 + \frac{u'(r)}{r} \Big) e^{-u(r)} f(\rrv) - 2 u'(r) e^{-u(r)} \nabla f(\rrv) \cdot \rrh - 3 u'(r) e^{-u(r)} \frac{f(\rrv)}{r} + e^{-u(r)} \nabla^2 f(\rrv) \\
  ={} & \Big( -u''(r) + u'(r)^2 - 2 \frac{u'(r)}{r} \Big) e^{-u(r)} f(\rrv) - 2 u'(r) e^{-u(r)} \nabla f(\rrv) \cdot \rrh + e^{-u(r)} \nabla^2 f(\rrv).
  \label{eqn:slater-jastrow-laplacian}
\end{align}
Electronic cusp condition is derived from the finiteness of:
\begin{align}
  -\frac{\nabla^2 \psi(\rrv)}{\psi(\rrv)} + \frac{1}{r}
  = u''(r) - u'(r)^2 + 2 \frac{u'(r)}{r} + 2 u'(r) \frac{\nabla f(\rrv) \cdot \rrh}{f(\rrv)} - \frac{\nabla^2 f(\rrv)}{f(\rrv)} + \frac{1}{r}
\end{align}
(reduced mass $\mu = \frac{1}{2}$ is used) when $r \to 0$.
For antiparallel-spin electrons, $f(\zrov) \ne 0$ in general, so we require
$2 \frac{u'(r)}{r} + \frac{1}{r} = 0$ to avoid $\frac{1}{r}$ divergence.
This leads to $u'(0) = -\frac{1}{2}$.
For parallel-spin electrons,
\begin{align}
  f(\rrv) = \nabla f(\zrov) \cdot \rrv + O(r^3) = (\nabla f(\zrov) \cdot \rrh) r + O(r^3).
\end{align}
This indicates $\nabla^2 f(\rrv) = O(r)$, so $\frac{\nabla^2 f(\rrv)}{f(\rrv)} = O(1)$ hence finite. So we require
$2 \frac{u'(r)}{r} + 2 \frac{u'(r)}{r} + \frac{1}{r} = 0$ to avoid $\frac{1}{r}$ divergence.
This leads to $u'(0) = -\frac{1}{4}$.

Using this result, we can expand $\psi$ near $\zrov$.
For antiparallel-spin electrons, $\nabla \psi(\zrov) = \big( \frac{1}{2} \rrh + \frac{\nabla f(\zrov)}{f(\zrov)} \big) e^{-u(0)} f(\zrov)$, so:
\begin{align}
  \psi(\rrv) = \psi(\zrov) \Big( 1 + \frac{1}{2} r + \frac{\nabla f(\zrov) \cdot \rrh}{f(\zrov)} r \Big) + O(r^2).
  \label{eqn:cusp-wavefn-electron-antipara}
\end{align}
Taking spherical average gives $\psib(r) = \psi(\zrov) \big( 1 + \frac{1}{2} r \big) + O(r^2)$, which agrees with \eqnref{cusp-wavefnmean-electron}.
For parallel-spin electrons,
\begin{align}
  \psi(\rrv) = e^{-u(0)} \nabla f(\zrov) \cdot \rrv + \frac{1}{4} e^{-u(0)} (\nabla f(\zrov) \cdot \rrv) r + O(r^3)
  = e^{-u(0)} (\nabla f(\zrov) \cdot \rrh) \Big( r + \frac{1}{4} r^2 \Big) + O(r^3).
  \label{eqn:cusp-wavefn-electron-para}
\end{align}
Taking spherical average gives $\psib(r) = O(r^3)$.
This also agrees with \eqnref{cusp-wavefnmean-electron}, which indicates $\psib'(0) = 0$.
Note it also holds that $\psib''(0) = 0$.

\eqnsref{cusp-wavefn-electron-antipara,cusp-wavefn-electron-para} can also be seen from another way.
Let $\psi(\rrv)$ keep $-\frac{\nabla^2 \psi(\rrv)}{\psi(\rrv)} + \frac{1}{r}$ finite.
Separate the variable $\rrv = (r, \thetav)$ in $\psi$ and expand the $\thetav$ factor with spherical harmonics:
\begin{align}
  \psi(\rrv) = \psi(r, \thetav) = \sum_{l=l_0}^\infty \psi_l(r, \thetav), \qquad
  \psi_l(r, \thetav) := R_l(r) \sum_{m=-l}^l b_{lm} Y_{lm}(\thetav).
  \label{eqn:r-theta-separate-sph-harmonic}
\end{align}
If we impose the $r \to 0$ finiteness of local energy for each $\psi_l$, %
by noting \eqnref{laplacian-sph-Rd} and that $\nabla^2_{\bbS^2} = \Lh^2$ is the squared angular momentum operator, this means the $r \to 0$ finiteness of:
\begin{align}
  -\frac{\nabla^2 \psi_l(r, \thetav)}{\psi_l(r, \thetav)} + \frac{1}{r}
  = -\frac{R_l''}{R_l} - \frac{2 R_l'}{r R_l} - \frac{l (l+1)}{r^2} + \frac{1}{r}.
\end{align}
This is similar to the radial equation for hydrogen atoms.
To keep it finite as $r \to 0$, we have:
\begin{align}
  R_l(r) = C_l r^l \Big( 1 + \frac{Z_i Z_j \mu_{ij}}{l + 1} r + O(r^2) \Big) = C_l r^l \Big( 1 + \frac{1}{2 (l+1)} r + O(r^2) \Big).
\end{align}
Since $\Lh^2 = \Sh^2 + \Lh^2_\textnormal{orbit}$, the total angular momentum is contributed from the spins and the spacial (orbital) motion,
we know that $l \ge s$, so $l_0 = s$, so for antiparallel-spin electrons, $l_0 = 0$ (singlet state), and for parallel-spin electrons, $l_0 = 1$ (triplet state).
So for antiparallel-spin electrons, noting $Y_{00}(\thetav) = \frac{1}{2 \sqrt{\pi}}$ is constant, we have:
\begin{align}
  \psi(\rrv) = C_0 \Big( 1 + \frac{1}{2} r \Big) + C_1 r \sum\nolimits_{m=-1}^1 b_{1m} Y_{1m}(\thetav) + O(r^2).
  \label{eqn:cusp-wavefn-electron-sph-harmonic-antipara}
\end{align}
This agrees with \eqnref{cusp-wavefn-electron-antipara}.
For the latter, not only does the summation index $l$ need to be larger than $l_0 = 1$.
Since two parallel-spin electrons are indistinguishable, the wavefunction needs to be antisymmetric, so only odd $l$ are allowed, whose corresponding $Y_{lm}(\thetav)$ has odd parity.
So the wavefunction near coincidence is:
\begin{align}
  \psi(\rrv) = C_1 \Big( r + \frac{1}{4} r^2 \Big) \sum\nolimits_{m=-1}^1 b_{1m} Y_{1m}(\thetav) + O(r^3).
  \label{eqn:cusp-wavefn-electron-sph-harmonic-para}
\end{align}
This agrees with \eqnref{cusp-wavefn-electron-para}.
Note that for antiparallel-spin electrons, the wavefunction is not symmetric in general, except \eg, there are only these two electrons.
For Slater-Jastrow ans\"atz, the two coordinates are input into two different Slater determinants, so switching their values (even if they are close) does not yield the same (close) determinant values. So $\nabla f(\zrov) \ne \zrov$.
This is why \eqnsref{cusp-wavefn-electron-antipara,cusp-wavefn-electron-sph-harmonic-antipara} has odd-parity terms.
In all, the electronic cusp condition is:
\begin{align}
  & \psib'(r=0) \stackrel{\text{\eqnref{cusp-wavefnmean-electron}}}{=} \frac{1}{2} \psib(r=0), \\
  & \psi(\rrv) \propto
  \begin{cases}
    1 + \frac{1}{2} r + (\aav \cdot \rrh) r + O(r^2), & \text{antiparallel-spin, \eqnsref{cusp-wavefn-electron-antipara,cusp-wavefn-electron-sph-harmonic-antipara}}, \\
    (\aav \cdot \rrh) \Big( r + \frac{1}{4} r^2 \Big) + O(r^3), & \text{parallel-spin, \eqnsref{cusp-wavefn-electron-para,cusp-wavefn-electron-sph-harmonic-para}}.
  \end{cases}
  \label{eqn:cusp-wavefn-electron}
\end{align}

\paragraph{Cusp condition for density}
``At any position of an atom $\rrv_A$, $\rho(\rrv)$ exhibits a maximum with a finite value, while its gradient there has a discontinuity and a cusp result''~\citep{koch2001chemist}:
\begin{align}
  \partial_r \log \rhob(r) \big|_{r = 0^+} = -2 Z_A,
\end{align}
where $r$ is the radial coordinate of the spherical coordinates centered at $\rrv_A$, and $\rhob(r)$ is the spherical average of $\rho(\rrv) = \rho(r, \thetav)$, \ie $\rhob(r) = \frac{1}{\Omega(\bbS^{d-1})} \int_{\bbS^{d-1}} \ud\thetav \, \rho(r, \thetav)$.
This can be seen from \eqnref{cusp-wavefnmean-nuclear}:
$\partial_r \rhob(r) \big|_{r=0^+}
= \frac{N}{\Omega(\bbS^2)} \partial_r \int_{\bbS^2} \ud\thetav \int \ud\rrv_{2:N} \lrvert{\psi(r, \thetav, \rrv_{2:N})}^2 \big|_{r=0^+}
\stackrel{\text{\eqnref{general-complex-deriv}}}{=} \frac{2 N}{\Omega(\bbS^2)} \int_{\bbS^2} \ud\thetav \int \ud\rrv_{2:N} \psi \partial_r \psi^* \big|_{r=0^+}
= $\footnote{\label{ftn:cusp-directional}
  This deduction requires the cusp condition to hold in any direction: $\partial_r \log \psi(r, \thetav, \rrv_{2:N}) \big|_{r=0^+} = -Z_A$ for any $\thetav \in \bbS^2$.
  This is somehow not obvious from the above condition \eqnref{cusp-wavefn-nuclear-deduction}:
  $\partial_r^2 \psi(r, \thetav, \rrv_{2:N}) + \frac{2}{r} \partial_r \psi(r, \thetav, \rrv_{2:N}) + \frac{1}{r^2} \nabla^2_{\bbS^2} \psi(r, \thetav, \rrv_{2:N}) + \frac{2 Z_A}{r} \psi(r, \thetav, \rrv_{2:N})$ is finite,
  since if $\nabla^2_{\bbS^2} \psi(r, \thetav, \rrv_{2:N}) = O(r)$, then it is $2 \partial_r \psi + 2 Z_A \psi + \frac{1}{r} \nabla^2_{\bbS^2} \psi = 0$ that should vanish.
  Nevertheless, it can be argued that as $r \to 0$, the dominating potential $\frac{1}{r}$ is spherically symmetric, so is $\psi(r, \thetav, \rrv_{2:N})$ as $r \to 0$, hence $\frac{1}{r} \nabla^2_{\bbS^2} \psi \to 0$ as $r \to 0$.
}
$-2 Z_A \frac{N}{\Omega(\bbS^2)} \int_{\bbS^2} \ud\thetav \int \ud\rrv_{2:N} \lrvert{\psi(r, \thetav, \rrv_{2:N})}^2 = -2 Z_A \rhob(0)$.
Since the Hamiltonian does not explicitly depend on spin, the deduction applies to $\psi^{\bmss}(\cdots)$ and $\rho^s(\rrv)$ as well.
(Even when the nucleus $A$ also has a half-integer spin, it is anyway a different, distinguishable particle from electron $i$, so Pauli exclusion principle does not apply.)

For one-electron density $\rho(\rrv)$, only the nuclear cusp is relevant.
For pair density and holes, ``since no two electrons of parallel spin can occupy the same point in space, this cusp condition occurs only for electrons of antiparallel spins''~\citep{koch2001chemist},
so it leads to a cusp condition for the Coulomb hole $h_\tnC(\rrv_2|\rrv_1)$.
For this, we have
$\partial_{r_{12}} \rho_{s, \neg s}(\rrv_1=\rrv_2, \rrv_2)
:= \partial_{r_{12}} \rhob_{s, \neg s}({\rrv_\tncc}_{12}, r_{12}=0)
:= \partial_{r_{12}} \frac{1}{\Omega(\bbS^2)} \int_{\bbS^2} \ud\thetav_{12} \, \rho_{s, \neg s}({\rrv_\tncc}_{12}, r_{12}=0, \thetav_{12})
= \frac{N(N-1)}{\Omega(\bbS^2)} \partial_{r_{12}} \int_{\bbS^2} \ud\thetav_{12} \int \ud\rrv_{3:N} \lrvert{\psi({\rrv_\tncc}_{12}, r_{12}=0, \thetav_{12}, \rrv_{3:N})}^2
= \frac{N(N-1)}{\Omega(\bbS^2)} \int_{\bbS^2} \ud\thetav_{12} \int \ud\rrv_{3:N} \big( \psi^* \partial_{r_{12}} \psi + \psi \partial_{r_{12}} \psi^* \big) = $\footref{ftn:cusp-directional}
$ \frac{N(N-1)}{\Omega(\bbS^2)} \int_{\bbS^2} \ud\thetav_{12} \int \ud\rrv_{3:N} \lrvert{\psi({\rrv_\tncc}_{12}, r_{12}=0, \thetav_{12}, \rrv_{3:N})}^2
= \frac{1}{\Omega(\bbS^2)} \int_{\bbS^2} \ud\thetav_{12} \rho_{s, \neg s}({\rrv_\tncc}_{12}, r_{12}=0, \thetav_{12})
= \rhob_{s, \neg s}({\rrv_\tncc}_{12}, r_{12}=0)
= \rho_{s, \neg s}(\rrv_1=\rrv_2, \rrv_2)$.

\subsubsection{Large-distance asymptotic behavior}
When walking far away from all nuclei $r \to \infty$, the inter-electron potential can be neglected%
, so the Schr\"{o}dinger equation from \eqnref{hamiltonian} becomes $\sum_{i \in [N]} -\frac{1}{2} \nabla^2_i \psi - \frac{Z}{r_i} \psi = E \psi$, where $Z := \sum_{A \in [N_\tnA]} Z_A$ is the total charge.
The Hamiltonian in this case is separable, so we consider the one-electron equation, $-\frac{1}{2} \nabla^2 \psi - \frac{Z}{r} \psi = -E_\tnI \psi$,
where $E_\tnI$ is the exact first ionization energy of the system.
Again, leveraging the Laplacian spherical decomposition and taking the spherical average, we have
$-\frac{1}{2} \fracdiff[2]{\psib}{r} - \frac{1}{r} \fracdiff{\psib}{r} - \frac{Z}{r} \psib = -E_\tnI \psib$,
which gives $\frac{1}{2} \fracdiff[2]{\psib}{r} = E_\tnI \psib$ in the limit $r \to \infty$, whose solution is (omitting the oscillating solution and only take the decaying solution):
\begin{align}
  \psib(r) \propto \exp\{- \sqrt{2 E_\tnI} \, r\}, \hspace{8pt}
  \rho(\rrv) \propto \exp\{-2 \sqrt{2 E_\tnI} \, r\}.
\end{align}

%% file: hf.tex
\section{Hartree-Fock Methods} \label{sec:hf}

\subsection{Wavefunction Assumptions}

\paragraph{A1: Linearly-dependent spacial wavefunction}

\begin{align}
  \psi^{\downarrow, s_2 \ldots s_N} (\rrv_1, \cdots, \rrv_N) = S \psi^{\uparrow, s_2 \ldots s_N} (\rrv_1, \cdots, \rrv_N),
\end{align}
where $S$ is a constant (spin-flipping coefficient).
Roughly means that position $\rrv_1$ change does not change the probability of taking a spin value; which roughly means there is no spin-orbit coupling.

This assumption means $\psi^{s_1 \ldots \downarrow \ldots} (\rrv_1, \cdots, \rrv_i, \cdots) = -\psi^{\downarrow \ldots s_1 \ldots} (\rrv_i, \cdots, \rrv_1, \cdots)
= -S \psi^{\uparrow \ldots s_1 \ldots} (\rrv_i, \cdots, \rrv_1, \cdots) = S \psi^{s_1 \ldots \uparrow \ldots} (\rrv_1, \cdots, \rrv_i, \cdots)$,
so the \emph{spin-flipping coefficient $S$ is the same for any electron}.

Moreover, $\psi^{s_1 s_2} = \psi^{s_2 s_1}$ if $s_1 = s_2$.
If not, suppose $s_1 = \downarrow$, $s_2 = \uparrow$, then $\psi^{s_1 s_2} = S \psi^{\uparrow \uparrow} = \psi^{\uparrow \downarrow} = \psi^{s_2 s_1}$.
So the \emph{wavefunction is invariant against spin permutation}.

\paragraph{Explanation:}
From \href{https://en.wikipedia.org/wiki/Wave_function#One-particle_states_in_3d_position_space}{Wikipedia},
``$\psi(\rrv, s, t) = \psi(\rrv,t) \xi(s,t)$. The tensor product factorization is only possible if the orbital and spin angular momenta of the particle are separable in the Hamiltonian operator underlying the system's dynamics (in other words, the Hamiltonian can be split into the sum of orbital and spin terms[28]). The time dependence can be placed in either factor, and time evolution of each can be studied separately. The factorization is not possible for those interactions where an external field or any space-dependent quantity couples to the spin; examples include a particle in a magnetic field, and spin–orbit coupling.''
If only eigenstates are considered, in that case, $\psi(\rrv,s) = \psi(\rrv) \xi(s)$, so $\psi(\rrv,\downarrow) = \psi(\rrv) \xi(\uparrow) \frac{\xi(\downarrow)}{\xi(\uparrow)} = S \psi(\rrv,\uparrow)$.

\subsubsection{A2: Slater determinant (Hartree-Fock wavefunction) (UHF)} \label{sec:hf-wavefn-uhf}

It is the Fock state \eqnref{symmetrizer} for fermions.
Results in \secref{qmb-2ndq} hold for subscript ``\tnS'' (for ``Slater'') as for $\frpp = -1$.
\begin{align}
  \psi_\tnS(\bfxx) := \ket{\phi_1 \cdots \phi_N}_{\frpp = -1}(\bfxx)
  = \frac{1}{\sqrt{N!}} \det \lrbrack{\phi_i(\rmxx_j)}_{ij}
  = \frac{1}{\sqrt{N!}} \sum_{\bmpi \in \rmPi_N} (-1)^{\bmpi} \phi_{\bmpi_1}(\rmxx_1) \cdots \phi_{\bmpi_N}(\rmxx_N).
  \label{eqn:slater-determ}
\end{align}
Further assume the \textbf{orthonormality} of the spin-orbitals $\{\phi_i\}_{i=1}^N$ (or just called orbitals; \ie, one-electron wavefunctions):
$\braket{\phi_i}{\phi_j} = \int \ud\rmxx \, \phi_i(\rmxx)^* \phi_j(\rmxx) = \delta_{ij}$.

The determinant makes $\psi_\tnS(\bfxx)$ exactly antisymmetric,
so it naturally \emph{handles the Fermi/exchange correlation, \ie $\rho_\tnS(\rmxx, \rmxx) = 0$, for any pair of electrons with \emph{parallel} spins} (see \eqnref{pairdensity-hf} below).
The \textbf{orthonormality} of $\{\phi_i\}_{i=1}^N$ makes $\psi_\tnS(\bfxx)$ normalized:
  $\braket{\psi_\tnS}{\psi_\tnS}
  \stackrel{\text{\eqnref{fock-inprod}}}{=} \det \lrbrack{\braket{\phi_i}{\phi_j}}_{ij} = 1.$

Note that for each spin-orbital $\phi_k$, $\phi_k^\uparrow(\rrv)$ and $\phi_k^\downarrow(\rrv)$ are two different spacial orbitals.
So for open-shell systems this assumption does not guarantee being an eigenfunction of the total spin operator $\Svh^2$ (it is still eigenfunction of $\Sh_z$) [\href{https://chemistry.stackexchange.com/questions/70643/whats-the-difference-between-rhf-and-uhf}{source}].
This corresponds to the \textbf{Unrestricted Hartree-Fock method (UHF)}~\citep{fukutome1981unrestricted}.

\paragraph{Densities, correlations and holes.}
The 1- and 2-RDMs and the density
and the pair density\footnote{
  Direct derivation using \textbf{orthonormality}:
  $\rho_\tnS(\rmxx_1, \rmxx_2) := N(N-1) \int \ud\rmxx_3 \cdots \ud\rmxx_N \, \lrvert{\psi_\tnS(\bfxx)}^2$
  \begin{align}
    ={} & \frac{N(N-1)}{N!} \int \ud\rmxx_3 \cdots \ud\rmxx_N \, \sum_{\bmpi', \bmpi \in \rmPi_N} (-1)^{\bmpi'} (-1)^{\bmpi}
      \phi_{\bmpi'_1}(\rmxx_1)^* \phi_{\bmpi_1}(\rmxx_1) \cdots \phi_{\bmpi'_N}(\rmxx_N)^* \phi_{\bmpi_N}(\rmxx_N) \\
    ={} & \frac{N(N-1)}{N!} \sum_{\bmpi', \bmpi \in \rmPi_N} (-1)^{\bmpi'} (-1)^{\bmpi} \phi_{\bmpi'_1}(\rmxx_1)^* \phi_{\bmpi_1}(\rmxx_1) \phi_{\bmpi'_2}(\rmxx_2)^* \phi_{\bmpi_2}(\rmxx_2) \delta_{\bmpi'_3 \bmpi_3} \cdots \delta_{\bmpi'_N \bmpi_N} \\
    ={} & \frac{N(N-1)}{N!} \sum_{\bmpi \in \rmPi_N} \lrparen{
      \phi_{\bmpi_1}(\rmxx_1)^* \phi_{\bmpi_2}(\rmxx_2)^* \phi_{\bmpi_1}(\rmxx_1) \phi_{\bmpi_2}(\rmxx_2) -
      \phi_{\bmpi_2}(\rmxx_1)^* \phi_{\bmpi_1}(\rmxx_2)^* \phi_{\bmpi_1}(\rmxx_1) \phi_{\bmpi_2}(\rmxx_2)
    } \\
    ={} & \sum_{k \ne l} \lrparen{
      \phi_k(\rmxx_1)^* \phi_l(\rmxx_2)^* \phi_k(\rmxx_1) \phi_l(\rmxx_2) -
      \phi_l(\rmxx_1)^* \phi_k(\rmxx_2)^* \phi_k(\rmxx_1) \phi_l(\rmxx_2)
    }
    =     \sum_{k,l} \lrparen{
      \phi_k(\rmxx_1)^* \phi_l(\rmxx_2)^* \phi_k(\rmxx_1) \phi_l(\rmxx_2) -
      \phi_l(\rmxx_1)^* \phi_k(\rmxx_2)^* \phi_k(\rmxx_1) \phi_l(\rmxx_2)
    } \\
    ={} & \sum_{k \in [N]} \phi_k(\rmxx_1)^* \phi_k(\rmxx_1) \sum_{l \in [N]} \phi_l(\rmxx_2)^* \phi_l(\rmxx_2)
        - \sum_{l \in [N]} \phi_l(\rmxx_1)^* \phi_l(\rmxx_2) \sum_{k \in [N]} \big( \phi_k(\rmxx_1)^* \phi_k(\rmxx_2) \big)^*
    =     \rho_\tnS(\rmxx_1) \rho_\tnS(\rmxx_2) - \lrvert*[\Big]{\sum_{k \in [N]} \phi_k(\rmxx_1)^* \phi_k(\rmxx_2)}^2.
  \end{align}
} under the Slater determinant with \textbf{orthonormal} $\{\phi_i\}_{i=1}^N$ are:
\begin{align}
  P_\tnS(\rmxx; \rmxx')
  \stackrel{\text{\eqnref{fock-1-rdm-orthon}}}{=} {} & \sum_{k \in [N]} \phi_k(\rmxx) \phi_k^*(\rmxx'),
  \label{eqn:1rdm-hf} \\
  P_\tnS(\rmxx_1, \rmxx_2; \rmxx'_1, \rmxx'_2)
  \stackrel{\text{\eqnref{fock-2-rdm-orthon}}}{=} {} & \sum_{j, k} \phi_j(\rmxx_1) \phi_k(\rmxx_2) \big( \phi_j^*(\rmxx'_1) \phi_k^*(\rmxx'_2) - \phi_k^*(\rmxx'_1) \phi_j^*(\rmxx'_2) \big)
  \stackrel{\text{\eqnref{fock-2-rdm-1-rdm}}}{=} P_\tnS(\rmxx_1; \rmxx'_1) P_\tnS(\rmxx_2; \rmxx'_2) - P_\tnS(\rmxx_1; \rmxx'_2) P_\tnS(\rmxx_2; \rmxx'_1),
  \label{eqn:2rdm-hf} \\
  \rho_\tnS(\rmxx) \stackrel{\text{\eqnref{fock-1-den-orthon}}}{=} {} & \sum_{k \in [N]} \lrvert{\phi_k(\rmxx)}^2,
  \label{eqn:density-hf} \\
  \rho_\tnS(\rmxx_1, \rmxx_2)
  \stackrel{\text{\eqnref{fock-2-den-orthon}}}{=} {} & \rho_\tnS(\rmxx_1) \rho_\tnS(\rmxx_2) - \lrvert*[\Big]{\sum_{k \in [N]} \phi_k(\rmxx_1)^* \phi_k(\rmxx_2)}^2.
  \label{eqn:pairdensity-hf}
\end{align}
We see indeed $\rho_\tnS^{s,s}(\rrv,\rrv) = 0$ for any parallel-spin electron pair.

For the pair density of an antiparallel-spin electron pair, \textbf{assume A2.2:} \emph{for any $k \in [N]$, either $\phi_k^\uparrow \equiv 0$ or $\phi_k^\downarrow \equiv 0$;
or equivalently, for any $k \in [N]$, $\phi_k^s(\rrv_1)^* \phi_k^{\neg s}(\rrv_2) \equiv 0$}
(\eg, under A2.1 below; but A2.1 can be made more general: only impose the orthonormality of spin basis but not the specific spin basis).
Then we have:
\begin{align}
  \rho_\tnS^{s, \neg s}(\rrv_1,\rrv_2) \stackrel{\text{A2.2}}{=} \rho_\tnS^s(\rrv_1) \rho_\tnS^{\neg s}(\rrv_2),
  \label{eqn:pairdensity-hf-antiparallel}
\end{align}
which indicates two antiparallel-spin electrons distribute \emph{independently} (unnecessarily implies they do not interact).

By definition, the exchange-correlation hole function is:
\begin{align}
  h_\XC^\HF(\rmxx_2|\rmxx_1) := \frac{\rho_\tnS(\rmxx_1,\rmxx_2)}{\rho_\tnS(\rmxx_1)} - \rho_\tnS(\rmxx_2)
  = -\frac{1}{\rho_\tnS(\rmxx_1)} \lrvert*[\Big]{\sum_{k \in [N]} \phi_k(\rmxx_1)^* \phi_k(\rmxx_2)}^2,
\end{align}
and the correlation factor is $
f(\rmxx_1,\rmxx_2) = - \frac{\lrvert{\sum_{k \in [N]} \phi_k(\rmxx_1)^* \phi_k(\rmxx_2)}^2}{\rho_\tnS(\rmxx_1) \rho_\tnS(\rmxx_2)}
$. The spin-independent hole (total hole) is:
\begin{align}
  & h_\XC^\HF(\rrv_2|\rrv_1) := \frac{\rho_\tnS(\rrv_1, \rrv_2)}{\rho_\tnS(\rrv_1)} - \rho_\tnS(\rrv_2)
  = -\frac{1}{\rho_\tnS(\rrv_1)} \sum_{s_1,s_2} \lrvert*[\Big]{\sum_{k \in [N]} \phi_k^{s_1}(\rrv_1)^* \phi_k^{s_2}(\rrv_2)}^2
  \label{eqn:hxc-spacial-hf} \\
  \stackrel{\text{A2.2}}{=} {} &
  h_\tnX^\HF(\rrv_2|\rrv_1) := -\frac{1}{\rho_\tnS(\rrv_1)} \sum_s \lrvert*[\Big]{\sum_{k \in [N]} \phi_k^s(\rrv_1)^* \phi_k^s(\rrv_2)}^2.
  \label{eqn:hx-spacial-hf}
\end{align}
Under A2.2, the summation is over parallel spins only, so the HF total hole is seen as composed of the exchange hole only, \ie $h_\XC^\HF = h_\tnX^\HF$.
It well handles self-interaction and exchange correlation (it cancels the self-interaction term and is the full consequence of antisymmetry (determinant)), so it is the exact exchange hole in HF.
This may also be a definition of the exchange hole (see \secref{corr-hole}, trial~(III), though don't know the correspondence of a general wavefunction to the orbitals).
But the Coulomb hole is zero, which should be there based on a qualitative analysis.
So HF with A2.2 is seen to have missed the Coulomb correlation (and the kinetic energy error).

\paragraph{Explanation:}
The Slater determinant $\psi_\tnS$ is the form of the exact ground-state solution (except for some degenerate cases~\citep{lieb1983density}) of a system of $N$ \emph{non-interacting} fermions (no direct/physical inter-particle interaction like the Coulomb repulsion) moving in some maybe-effective, one-particle potential $V_\tnS(\rrv)$.
The Hamiltonian is:
\begin{align}
  \Hh = \Th + \Vh_\tnS
  = \sum_{i \in [N]} -\frac{1}{2} \nabla^2_i + {\Vh_\tnS}_i,
  \quad \text{where }
  {\Vh_\tnS}_i \psi(\bfxx) := V_\tnS(\rrv_i) \psi(\bfxx).
  \label{eqn:hamiltonian-nonintera}
\end{align}

\subsubsection{A2.1: Open- and Closed-Shell Systems (R(O)HF)} \label{sec:hf-wavefn-rhf}

In many cases, two antiparallel-spin electrons can be seen to occupy the same spacial orbital.
So we can pair two spin-orbitals and make them share the same spacial part:
\begin{align}
  \{\phi_i(\rmxx)\}_{i=1}^N = \{\varphi_a(\rrv) \delta^\uparrow_s\}_{a=1}^{N^\uparrow} \cup \{\varphi_b(\rrv) \delta^\downarrow_s\}_{b=1}^{N^\downarrow},
  \label{eqn:hf-wavefn-rhf}
\end{align}
where $N^\uparrow + N^\downarrow = N$, $\{\varphi_a(\rrv)\}_{a=1}^{\Nsp}$ with $\Nsp := \max\{N^\uparrow, N^\downarrow\}$ is a set of spacial orbitals,
and the spin basis $\{\delta^\uparrow_s, \delta^\downarrow_s\}$ are \textbf{orthonormal}:
$\braket*{\delta^\alpha_\cdot}{\delta^\beta_\cdot} = \sum_s \delta^\alpha_s \delta^\beta_s = \delta_{\alpha \beta}$.
The spin-orbital \textbf{orthonormality} then indicates the \textbf{orthonormality} of spacial orbitals: $\int \ud\rrv \, \varphi^*_a(\rrv) \varphi_{a'}(\rrv) = \delta_{a a'}, \forall a,a' \in [\Nsp]$.
Note that for any spin-orbital $\phi_i = \varphi_a \delta^\uparrow_\cdot$ or $= \varphi_b \delta^\downarrow_\cdot$, either $\phi_i^\uparrow \equiv 0$ or $\phi_i^\downarrow \equiv 0$, so A2.2 holds.

\textbf{Closed-shell} system: $N^\uparrow = N^\downarrow = \Nsp$.
This corresponds to the \textbf{Restricted Hartree-Fock method (RHF)}.
Otherwise, it is called an \textbf{Open-shell} system: there is unpaired electron (orbital), and the multiplicity / spin polarization number (the number of unpaired electrons, the number of singly-occupied orbitals) $\Nsp - \Npr$ is not zero, where $\Npr := \min\{N^\uparrow,N^\downarrow\}$ is the pair number of paired electrons (number of doubly-occupied orbitals).
This corresponds to the \textbf{Restricted Open-shell Hartree-Fock method (ROHF)}~\citep{roothaan1960self}.
``In contrast to UHF, the R(O)HF wavefunction \eqnref{hf-wavefn-rhf} is a satisfactory eigenfunction of the total spin operator $\Svh^2$ (no spin contamination).'' [\href{https://en.wikipedia.org/wiki/Restricted_open-shell_Hartree\%E2\%80\%93Fock}{Wikipedia}]
\footnote{Words based on \citep{koch2001chemist}:
  Closed-shell systems correspond to restricted Hartree-Fock (RHF).
  The open-shell Hartree-Fock method is called unrestricted Hartree-Fock (UHF).
  The major disadvantage of UHF is that it is not an eigenfunction of the total spin operator $\Svh^2$.
  A variant is the restricted open-shell HF scheme (ROHF), but it is much more complicated than UHF (ROHF wavefunction is a limited linear combination of a few determinants), though it does not meet the problem of not being an eigenfunction of $\Svh^2$.
}

The wavefunction then becomes:
$\psi_\tnS(\bfxx)$
\begin{align}
  ={} & \frac{1}{\sqrt{N!}} \sum_{\bmpi \in \rmPi_N} (-1)^{\bmpi} \phi_1^{s_{\bmpi_1}}(\rrv_{\bmpi_1}) \cdots \phi_N^{s_{\bmpi_N}}(\rrv_{\bmpi_N})
  =     \frac{1}{\sqrt{N!}} \sum_{\bmpi \in \rmPi_N} (-1)^{\bmpi}
  \lrparen*[\bigg]{ \prod_{a \in [N^\uparrow]} \varphi_a(\rrv_{\bmpi_a}) \delta^\uparrow_{s_{\bmpi_a}} }
  \lrparen*[\bigg]{ \prod_{b \in [N^\downarrow]} \varphi_b(\rrv_{\bmpi_{b+N^\uparrow}}) \delta^\downarrow_{s_{\bmpi_{b+N^\uparrow}}} } \\
  ={} & \frac{(-1)^{\bmpi^\star}}{\sqrt{N!}} \sum_{\bmpi \in \rmPi_{N^\uparrow}} \sum_{\bmpi' \in \rmPi_{N^\downarrow}} (-1)^{\bmpi} (-1)^{\bmpi'}
  \lrparen*[\bigg]{ \prod_{a \in [N^\uparrow]} \varphi_a(\rrv_{\bmpi^\star_{\bmpi_a}}) } \lrparen*[\bigg]{ \prod_{b \in [N^\downarrow]} \varphi_b(\rrv_{\bmpi^\star_{\bmpi'_b + N^\uparrow}}) } \\
  ={} & \frac{(-1)^{\bmpi^\star}}{\sqrt{N!}} \det[\varphi_a(\rrv_{\bmpi^\star_{a'}})]_{a,a' \in [N^\uparrow]} \det[\varphi_b(\rrv_{\bmpi^\star_{b' + N^\uparrow}})]_{b,b' \in [N^\downarrow]}
  \label{eqn:slater-determ-rhf}
\end{align}
if $\#\{i \in [N] \mid s_i = \uparrow\} = N^\uparrow$ in which case $\bmpi^\star$ is a permutation that makes $s_{\bmpi^\star_a} = \uparrow$ for all $a \in [N^\uparrow]$,
otherwise $\psi_\tnS(\bfxx) = 0$.
This indicates that either $\psi_\tnS^{\uparrow, s_2 \ldots s_N}$ or $\psi_\tnS^{\downarrow, s_2 \ldots s_N}$ must be zero.
If it is the former, then A1 holds with $S = 0$; otherwise A1 does not hold (or holds in another direction with $S = 0$).
In either case, swapping different spins changes the sign.

\paragraph{Explanation:}
This is adopted when the Hamiltonian does not involve spins.
Note however that it is \emph{not} the case when spin-orbit coupling, or spin-spin interaction, are considered, \eg in Hund's rule.
Due to the explanation for A1, in this case, the wavefunction can be separated into a spacial and a spin part: $\phi_i(\rmxx) = \phi_i(\rrv) \xi_i(s)$.
Moreover, due to the spacial-spin disentanglement, the wavefunction basis can be taken as the Cartesian product of spacial orbitals and spin basis,
$\{\varphi_a(\rrv)\}_{a=1}^{\Nsp} \times \{\delta^\uparrow_s, \delta^\downarrow_s\}$.
This is possible since a spacial orbital does not have a preference on the spin due to the disentanglement (no spin-orbit or cross-orbit spin-spin interaction),
and any orbit-dependent, non-eigenstate spin basis $\{\varphi_a(\rrv) \xi_{a,1}(s), \varphi_a(\rrv) \xi_{a,2}(s)\}$ can be expressed under this orbit-free, eigenstate basis.

\subsection{Energy Expressions}

\subsubsection{Energy under Slater determinant A2 (UHF)} \label{sec:energy-hf}

The electronic Hamiltonian of a molecular system is given by \eqnref{hamiltonian}.
From \eqnsref{energy-decomp,Eee-decomp}, we have the energy under the Slater determinant with \textbf{orthonormal} $\{\phi_i\}_i$: \footnote{
  Direct derivation using \textbf{orthonormality}:
    $\obraket*{\psi_\tnS}{\Hh^{(1)}_i}{\psi_\tnS}
    =    \int \ud\bfxx_1 \cdots \ud\bfxx_N \, \psi_\tnS(\bfxx)^* \Hh^{(1)}_i \psi_\tnS(\bfxx)$
  \begin{align}
    \stackrel{\text{A2}}{=} {} & \frac{1}{N!} \int \ud\rmxx_1 \cdots \ud\rmxx_N \sum_{\bmpi', \bmpi \in \rmPi_N} (-1)^{\bmpi'} (-1)^{\bmpi}
    \phi_{\bmpi'_1}(\rmxx_1)^* \phi_{\bmpi_1}(\rmxx_1) \cdots \phi_{\bmpi'_i}(\rmxx_i)^* \Hh^{(1)}_i \phi_{\bmpi_i}(\rmxx_i) \cdots \phi_{\bmpi'_N}(\rmxx_N)^* \phi_{\bmpi_N}(\rmxx_N) \\
    ={} & \frac{1}{N!} \sum_{\bmpi', \bmpi \in \rmPi_N} (-1)^{\bmpi'} (-1)^{\bmpi} \delta_{\bmpi'_1 \bmpi_1} \cdots \delta_{\bmpi'_{\neg i} \bmpi_{\neg i}} \cdots \delta_{\bmpi'_N \bmpi_N}
    \int \ud \rmxx_i \, \phi_{\bmpi'_i}(\rmxx_i)^* \Hh^{(1)}_i \phi_{\bmpi_i}(\rmxx_i) \\
    ={} & \frac{1}{N!} \sum_{\bmpi \in \rmPi_N} (-1)^{\bmpi^2}
    \int \ud \rmxx_i \, \phi_{\bmpi_i}(\rmxx_i)^* \Hh^{(1)}_i \phi_{\bmpi_i}(\rmxx_i)
    =     \frac{1}{N} \sum_{k \in [N]} \int \ud\rmxx_i \, \phi_k(\rmxx_i)^* \Hh^{(1)}_i \phi_k(\rmxx_i)
    =      \frac{1}{N} \sum_{k \in [N]} \obraket*{\phi_k}{\Hh^{(1)}}{\phi_k},
  \end{align}
  \begin{align}
    & \obraket*{\psi_\tnS}{(1/r_{ij})}{\psi_\tnS}
    =     \int \ud\rmxx_1 \cdots \ud\rmxx_N \, \psi_\tnS(\bfxx)^* (1/r_{ij}) \psi_\tnS(\bfxx) \\
    \stackrel{\text{A2}}{=} {} & \frac{1}{N!} \int \ud\rmxx_1 \cdots \ud\rmxx_N \sum_{\bmpi', \bmpi \in \rmPi_N} (-1)^{\bmpi'} (-1)^{\bmpi}
    \frac{1}{r_{ij}} \phi_{\bmpi'_i}(\rmxx_i)^* \phi_{\bmpi_i}(\rmxx_i) \phi_{\bmpi'_j}(\rmxx_j)^* \phi_{\bmpi_j}(\rmxx_j) \cdots \phi_{\bmpi'_N}(\rmxx_N)^* \phi_{\bmpi_N}(\rmxx_N) \\
    ={} & \frac{1}{N!} \int \ud\rmxx_i \ud\rmxx_j \sum_{\bmpi', \bmpi \in \rmPi_N} (-1)^{\bmpi'} (-1)^{\bmpi}
    \frac{1}{r_{ij}} \phi_{\bmpi'_i}(\rmxx_i)^* \phi_{\bmpi_i}(\rmxx_i) \phi_{\bmpi'_j}(\rmxx_j)^* \phi_{\bmpi_j}(\rmxx_j)
    \delta_{\bmpi'_1 \bmpi_1} \cdots \delta_{\bmpi'_{\neg i} \bmpi_{\neg i}} \delta_{\bmpi'_{\neg j} \bmpi_{\neg j}} \cdots \delta_{\bmpi'_N \bmpi_N} \\
    ={} & \frac{1}{N!} \int \ud\rmxx_i \ud\rmxx_j \sum_{\bmpi \in \rmPi_N}
    \frac{1}{r_{ij}} \lrparen{
      \phi_{\bmpi_i}(\rmxx_i)^* \phi_{\bmpi_i}(\rmxx_i) \phi_{\bmpi_j}(\rmxx_j)^* \phi_{\bmpi_j}(\rmxx_j) -
      \phi_{\bmpi_j}(\rmxx_i)^* \phi_{\bmpi_i}(\rmxx_i) \phi_{\bmpi_i}(\rmxx_j)^* \phi_{\bmpi_j}(\rmxx_j)
    } \\
    ={} & \frac{1}{N(N-1)} \int \ud\rmxx_i \ud\rmxx_j \sum_{k \ne l}
    \frac{1}{r_{ij}} \lrparen{
      \phi_k(\rmxx_i)^* \phi_k(\rmxx_i) \phi_l(\rmxx_j)^* \phi_l(\rmxx_j) -
      \phi_l(\rmxx_i)^* \phi_k(\rmxx_i) \phi_k(\rmxx_j)^* \phi_l(\rmxx_j)
    } \\
    ={} & \frac{1}{N(N-1)} \int \ud\rmxx_i \ud\rmxx_j \sum_{k,l \in [N]}
    \frac{1}{r_{ij}} \lrparen{
      \phi_k(\rmxx_i)^* \phi_k(\rmxx_i) \phi_l(\rmxx_j)^* \phi_l(\rmxx_j) -
      \phi_l(\rmxx_i)^* \phi_k(\rmxx_i) \phi_k(\rmxx_j)^* \phi_l(\rmxx_j)
    } \\
    ={} & \frac{1}{N(N-1)} \sum_{k,l \in [N]} \sum_{s, s'} (I^{s s'}_{kl | kl} - I^{s s'}_{lk | kl})
    =     \frac{1}{N(N-1)} \sum_{k,l \in [N]} (I_{kl | kl} - I_{lk | kl})
    =     \frac{1}{N(N-1)} \sum_{k,l \in [N]} (J_{kl} - K_{kl}).
  \end{align}
}
\begin{align}
  & \hspace{1.0cm} E[\psi_\tnS] = T[P^{(1)}_\tnS] + E_\Ne[\rho_\tnS] + J[\rho_\tnS] + E_\XC^\HF[\psi_\tnS], \quad \text{where} \\
  T[\psi_\tnS] & \stackrel{\text{\eqnref{1rdm-hf}}}{=}
  -\frac{1}{2} \sum_{k \in [N]} \obraket{\phi_k}{\nabla^2}{\phi_k}
  = -\frac{1}{2} \sum_{k \in [N]} \sum_s \obraket{\phi_k^s}{\nabla^2}{\phi_k^s},
  \label{eqn:kinetic-hf} \\
  E_\Ne[\rho_\tnS] &=
  \bbE_{\rho_\tnS}[V_\Ne]
  \stackrel{\text{\eqnref{density-hf}}}{=} \sum_{k \in [N]} \bbE_{\lrvert{\phi_k(\rmxx)}^2} [V_\Ne(\rrv)]
  = \sum_{k \in [N]} \sum_s \bbE_{\lrvert{\phi_k^s(\rrv)}^2} [V_\Ne(\rrv)],
  \label{eqn:ene-hf} \\
  J[\rho_\tnS] & \stackrel{\text{\eqnref{ej}}}{=}
  \frac{1}{2} \bbE_{\rho_\tnS(\rmxx_1) \rho_\tnS(\rmxx_2)} \Big[ \frac{1}{r_{12}} \Big]
  \stackrel{\text{\eqnref{density-hf}}}{=} \frac{1}{2} \sum_{k,l \in [N]} \int \ud\rmxx_1 \ud\rmxx_2 \, \frac{\lrvert{\phi_k(\rmxx_1)}^2 \lrvert{\phi_l(\rmxx_2)}^2}{r_{12}}
  = \frac{1}{2} \sum_{k,l} J_{kl},
  \label{eqn:ej-hf} \\
  E_\XC^\HF[\psi_\tnS] & \stackrel{\text{\eqnsref{exc,hxc-spacial-hf}}}{=}
  -\frac{1}{2} \int \ud\rmxx_1 \ud\rmxx_2 \, \frac{\lrvert*{\sum_k \phi_k(\rmxx_1)^* \phi_k(\rmxx_2)}^2}{r_{12}}
  = -\frac{1}{2} \sum_{k,l} \int \ud\rmxx_1 \ud\rmxx_2 \, \frac{\phi_k(\rmxx_1)^* \phi_l(\rmxx_2)^* \phi_l(\rmxx_1) \phi_k(\rmxx_2)}{r_{12}}
  = -\frac{1}{2} \sum_{k,l} K_{kl}.
  \label{eqn:exc-hf}
\end{align}
In sum,
\begin{align}
  E[\psi_\tnS] = -\frac{1}{2} \sum_{k \in [N]} \obraket*{\phi_k}{\nabla^2}{\phi_k} + \sum_{k \in [N]} \bbE_{\lrvert{\phi_k}^2} [V_\Ne]
  + \frac{1}{2} \sum_{k,l \in [N]} J_{kl} - \frac{1}{2} \sum_{k,l \in [N]} K_{kl}.
  \hspace{1.3cm} \text{(UHF)} \hspace{-1.3cm}
  \label{eqn:energy-hf}
\end{align}
In the above equations, by defining \textbf{4-center-2-electron integral}:
\begin{align}
  I_{ij | kl} := \int \ud\rmxx_1 \ud\rmxx_2 \, \phi_i(\rmxx_1)^* \phi_j(\rmxx_2)^* \frac{1}{r_{12}} \phi_k(\rmxx_1) \phi_l(\rmxx_2)
  = \obraket{\phi_i \otimes \phi_j}{\Wh_\ee}{\phi_k \otimes \phi_l},
\end{align}
which by construction satisfies:
\begin{align}
  I_{ij | kl} = I^*_{kl | ij} = I_{ji | lk},
  \label{eqn:4c2e-symmetry}
\end{align}
we have used the \textbf{Coulomb integral}:
\begin{align}
  J_{kl} := I_{kl | kl}
  = \int \ud\rmxx_1 \ud\rmxx_2 \, \frac{\phi_k(\rmxx_1)^* \phi_l(\rmxx_2)^* \phi_k(\rmxx_1) \phi_l(\rmxx_2)}{r_{12}}
  = \int \ud\rmxx_1 \ud\rmxx_2 \, \frac{\lrvert{\phi_k(\rmxx_1)}^2 \lrvert*{\phi_l(\rmxx_2)}^2}{r_{12}}
  \label{eqn:J-int}
\end{align}
that describes the Coulomb interaction between a pair of electrons occupying orbitals $k$ and $l$ with any possible spin configurations, and the \textbf{exchange integral}:
\begin{align}
  K_{kl} := I_{kl | lk}
  = \int \ud\rmxx_1 \ud\rmxx_2 \, \frac{\phi_k(\rmxx_1)^* \phi_l(\rmxx_2)^* \phi_l(\rmxx_1) \phi_k(\rmxx_2)}{r_{12}}
  \label{eqn:K-int}
\end{align}
that arises from the permutation in the Slater determinant, and comes in the form of exchanging the two electrons.
By definition, $J_{kl} = J_{lk}$, $K_{kl} = K_{lk}$.
Note in \eqnsref{ej-hf,exc-hf}, $k = l$ is allowed, since the unphysical \textbf{self-interaction} $J_{kk} \ne 0$ is cancelled by $K_{kk} = J_{kk}$.

\subsubsection{Energy under A2.1 (R(O)HF)} \label{sec:energy-rhf}

Let $\Iupr := \{\Npr+1, \cdots, \Nsp\}$ denote the index set of unpaired electrons / singly-occupied spacial orbitals.
Under A2.1 \eqnref{hf-wavefn-rhf}, for each $\phi_k$, only one value of $s$ is activated, so from \eqnsref{kinetic-hf,ene-hf},
\begin{align}
  T[\psi_\tnS] &= \sum_{a \in [N^\uparrow]} \obraket*{\varphi_a}{-\frac{1}{2} \nabla^2}{\varphi_a} + \sum_{b \in [N^\downarrow]} \obraket*{\varphi_b}{-\frac{1}{2} \nabla^2}{\varphi_b}
  = 2 \sum_{p \in [\Npr]} \obraket*{\varphi_p}{-\frac{1}{2} \nabla^2}{\varphi_p} + \sum_{u \in \Iupr} \obraket*{\varphi_u}{-\frac{1}{2} \nabla^2}{\varphi_u},
  \label{eqn:kinetic-rhf} \\
  E_\Ne[\psi_\tnS] &= \sum_{a \in [N^\uparrow]} \bbE_{\lrvert{\varphi_a}^2}[V_\Ne] + \sum_{b \in [N^\downarrow]} \bbE_{\lrvert{\varphi_b}^2}[V_\Ne]
  = 2 \sum_{p \in [\Npr]} \bbE_{\lrvert{\varphi_p}^2}[V_\Ne] + \sum_{u \in \Iupr} \bbE_{\lrvert{\varphi_u}^2}[V_\Ne].
  \label{eqn:ene-rhf}
\end{align}

For the classical Coulomb energy (Hartree energy), from \eqnref{ej-hf},
\begin{align}
  & 2 J[\rho_\tnS] = \int \frac{\ud\rrv_1 \ud\rrv_2}{r_{12}} \sum_k \sum_{s_1} \lrvert{\phi_k^{s_1}(\rrv_1)}^2 \sum_l \sum_{s_2} \lrvert{\phi_l^{s_2}(\rrv_2)}^2 \\
  ={} & \int \frac{\ud\rrv_1 \ud\rrv_2}{r_{12}}
    \lrparen*[\bigg]{ \sum_{a \in [N^\uparrow]} \sum_{s_1} \lrvert{\varphi_a(\rrv_1) \delta^\uparrow_{s_1}}^2 + \sum_{b \in [N^\downarrow]} \sum_{s_1} \lrvert{\varphi_b(\rrv_1) \delta^\downarrow_{s_1}}^2 }
    \lrparen*[\bigg]{ \sum_{a' \in [N^\uparrow]} \sum_{s_2} \lrvert{\varphi_{a'}(\rrv_2) \delta^\uparrow_{s_2}}^2 + \sum_{b' \in [N^\downarrow]} \sum_{s_2} \lrvert{\varphi_{b'}(\rrv_2) \delta^\downarrow_{s_2}}^2 } \\
  ={} & \int \frac{\ud\rrv_1 \ud\rrv_2}{r_{12}}
    \lrparen*[\bigg]{ \sum_{a \in [N^\uparrow]} \lrvert{\varphi_a(\rrv_1)}^2 + \sum_{b \in [N^\downarrow]} \lrvert{\varphi_b(\rrv_1)}^2 }
    \lrparen*[\bigg]{ \sum_{a' \in [N^\uparrow]} \lrvert{\varphi_{a'}(\rrv_2)}^2 + \sum_{b' \in [N^\downarrow]} \lrvert{\varphi_{b'}(\rrv_2)}^2 } \\
  ={} & \sum_{a,a' \in [N^\uparrow]} \Jt_{a a'} + 2 \sum_{a \in [N^\uparrow]} \sum_{b \in [N^\downarrow]} \Jt_{ab} + \sum_{b,b' \in [N^\downarrow]} \Jt_{b b'} \\
  ={} & 4 \sum_{p,p' \in [\Npr]} \Jt_{p p'} + 4 \sum_{p \in [\Npr]} \sum_{u \in \Iupr} \Jt_{pu} + \sum_{u,u' \in \Iupr} \Jt_{u u'},
  \label{eqn:ej-rhf}
\end{align}
where $\Jt_{ab} := \int \ud\rrv_1 \ud\rrv_2 \, \varphi^*_a(\rrv_1) \varphi^*_b(\rrv_2) \frac{1}{r_{12}} \varphi_a(\rrv_1) \varphi_b(\rrv_2) = \int \ud\rrv_1 \ud\rrv_2 \, \frac{\lrvert{\varphi_a(\rrv_1)}^2 \lrvert{\varphi_b(\rrv_2)}^2}{r_{12}}$ describes the Coulomb interaction between two electrons occupying spacial orbitals $a$ and $b$.
By definition, $\Jt_{ab} = \Jt_{ba}$.

For the exchange-correlation energy, from \eqnref{exc-hf},
\begin{align}
  & -2 E_\XC^\HF[\psi_\tnS] = \int \frac{\ud\rrv_1 \ud\rrv_2}{r_{12}} \sum_{s_1,s_2} \sum_k \phi_k^{s_1}(\rrv_1)^* \phi_k^{s_2}(\rrv_2) \sum_l \phi_l^{s_1}(\rrv_1) \phi_l^{s_2}(\rrv_2)^* \\
  ={} & \int \frac{\ud\rrv_1 \ud\rrv_2}{r_{12}} \sum_{s_1,s_2}
    \lrparen*[\bigg]{ \sum_{a \in [N^\uparrow]} \varphi_a(\rrv_1)^* \delta^{\uparrow*}_{s_1} \varphi_a(\rrv_2) \delta^\uparrow_{s_2} + \sum_{b \in [N^\downarrow]} \varphi_b(\rrv_1)^* \delta^{\downarrow*}_{s_1} \varphi_b(\rrv_2) \delta^\downarrow_{s_2} } \cdots \\*
    & \hspace{2.3cm} \lrparen*[\bigg]{ \sum_{a' \in [N^\uparrow]} \varphi_{a'}(\rrv_1) \delta^\uparrow_{s_1} \varphi_{a'}(\rrv_2)^* \delta^{\uparrow*}_{s_2} + \sum_{b' \in [N^\downarrow]} \varphi_{b'}(\rrv_1) \delta^\downarrow_{s_1} \varphi_{b'}(\rrv_2)^* \delta^{\downarrow*}_{s_2} } \\
  ={} & \sum_{a,a' \in [N^\uparrow]} \It_{aa' | a'a} \braket*{\delta^\uparrow}{\delta^\uparrow} \braket*{\delta^\uparrow}{\delta^\uparrow}
    + \sum_{a \in [N^\uparrow]} \sum_{b' \in [N^\downarrow]} \It_{ab' | b'a} \braket*{\delta^\uparrow}{\delta^\downarrow} \braket*{\delta^\downarrow}{\delta^\uparrow} \\*
    & {} + \sum_{a' \in [N^\uparrow]} \sum_{b' \in [N^\downarrow]} \It_{ba' | a'b} \braket*{\delta^\downarrow}{\delta^\uparrow} \braket*{\delta^\uparrow}{\delta^\downarrow}
    + \sum_{b,b' \in [N^\downarrow]} \It_{bb' | b'b} \braket*{\delta^\downarrow}{\delta^\downarrow} \braket*{\delta^\downarrow}{\delta^\downarrow} \\
  \stackrel{\text{(*)}}{=} {} & \sum_{a,a' \in [N^\uparrow]} \It_{aa' | a'a} + \sum_{b,b' \in [N^\downarrow]} \It_{bb' | b'b}
  =     \sum_{a,a' \in [N^\uparrow]} \Kt_{aa'} + \sum_{b,b' \in [N^\downarrow]} \Kt_{bb'} \\
  ={} & 2 \sum_{p,p' \in [\Npr]} \Kt_{pp'} + \sum_{u,u' \in \Iupr} \Kt_{uu'},
  \label{eqn:exc-rhf}
\end{align}
where (*) is due to the orthonormality of the spin basis under A2.1, and
$\Kt_{ab} := \It_{ab|ba} = \int \ud\rrv_1 \ud\rrv_2 \, \varphi^*_a(\rrv_1) \varphi^*_b(\rrv_2) \frac{1}{r_{12}} \varphi_b(\rrv_1) \varphi_a(\rrv_2)$.
By definition, $\Kt_{ab} = \Kt_{ba}$.
Note that different from the Coulomb term, orbitals only with parallel spins contribute to the exchange term.
Finally, the energy for ROHF is:
\begin{align}
  E[\psi_\tnS]
  ={} & 2 \sum_{p \in [\Npr]} \lrparen*[\Big]{ \obraket*{\varphi_p}{-\frac{1}{2} \nabla^2}{\varphi_p} + \bbE_{\lrvert{\varphi_p}^2}[V_\Ne] }
    + \sum_{p,p' \in [\Npr]} \lrparen*[\Big]{ 2 \Jt_{p p'} - \Kt_{pp'} } \cdots
    & \text{(RHF)} \\
  & {} + \sum_{u \in \Iupr} \lrparen*[\Big]{ \obraket*{\varphi_u}{-\frac{1}{2} \nabla^2}{\varphi_u} + \bbE_{\lrvert{\varphi_u}^2}[V_\Ne] }
    + 2 \sum_{p \in [\Npr]} \sum_{u \in \Iupr} \Jt_{pu} + \frac{1}{2} \sum_{u,u' \in \Iupr} \lrparen*[\Big]{ \Jt_{u u'} - \Kt_{uu'} }.
    & \text{(ROHF)}
  \label{eqn:energy-rhf}
\end{align}
For RHF, $N^\uparrow = N^\downarrow = \Nsp = \Npr$ and $\Iupr = \emptyset$, so only the first line of the above equation remains.

If defining running $i \in [N]$ as running $a \in [N^\uparrow]$ and $b \in [N^\downarrow]$ (or running $p \in [\Npr]$ \textbf{twice} and $u \in \Iupr$ once), then:
\begin{align}
  E[\psi_\tnS]
  ={} & \sum_{i \in [N]} \lrparen*[\Big]{ \obraket*{\varphi_i}{-\frac{1}{2} \nabla^2}{\varphi_i} + \bbE_{\lrvert{\varphi_i}^2}[V_\Ne] }
    + \frac{1}{2} \sum_{i,j \in [N]} \Jt_{ij}
    - \begin{dcases}
      \frac{1}{4} \sum_{i,j \in [N]} \Kt_{ij},
      & \text{\phantom{O}(RHF)} \\
      \frac{1}{2} \lrparen*[\Big]{ 2 \sum_{p,p' \in [\Npr]} \Kt_{pp'} + \sum_{u,u' \in \Iupr} \Kt_{uu'} },
      & \text{(ROHF)}
    \end{dcases}
  \label{eqn:energy-rhf-N}
\end{align}
which defers from \eqnref{energy-hf} only in the $\Kh$ term where the summation is only over parallel-spin spin-orbital pairs.

\subsection{Hartree-Fock Equation} \label{sec:hf-hf}

To minimize the energy \eqnref{energy-hf} as a functional of the orbitals $\{\phi_i\}_{i=1}^N$ subject to the \textbf{orthonormal} constraint,
taking the variation w.r.t $\phi_i^*$ (complex derivative in the sense of \eqnref{general-complex-deriv}, similar to the derivation from \eqnref{schrodinger-variation-hje}) with Lagrange multiplier (\textbf{orbital energy}) $\veps_i$ for the constraint $\int \ud\rmxx \, \phi_i(\rmxx)^* \phi_i(\rmxx) = 1$ (the orthogonality is naturally satisfied if there is no degeneracy) yields the \textbf{Hartree-Fock equation}:
\begin{align}
  \ffh^\HF \phi_i :={} & \big( \Th + \Vh_\tnS^\HF \big) \phi_i = \veps_i \phi_i, \quad \text{where} \;
  \Vh_\tnS^\HF := \Vh_\Ne + \sum\nolimits_{j \in [N]} \Jh^{(j)} - \sum\nolimits_{j \in [N]} \Kh^{(j)},
  \label{eqn:hf-eigen} \\
  & \text{and} \;
  \Jh^{(j)} \phi_i (\rmxx_1) := \int \ud\rmxx_2 \, \frac{\lrvert{\phi_j(\rmxx_2)}^2}{r_{12}} \phi_i(\rmxx_1),
  \qquad \Kh^{(j)} \phi_i (\rmxx_1) := \int \ud\rmxx_2 \, \frac{\phi_j(\rmxx_2)^* \phi_i(\rmxx_2)}{r_{12}} \phi_j(\rmxx_1).
  \label{eqn:JK-opr}
\end{align}
The \textbf{Fock operator} $\ffh^\HF$ involves the \textbf{Hartree-Fock potential operator} $\Vh_\tnS^\HF$, which reforms the inter-electron interactions into an equivalent, \emph{effective} potential that acts on each single electron (represented as its action onto each one-electron orbital).
Note that both $\Jh^{(j)}$ and $\Kh^{(j)}$ hence $\Vh_\tnS^\HF$ and $\ffh^\HF$ are operators determined by the orbitals $\{\phi_i\}_{i=1}^N$.
Also note that due to $\Kh^{(j)}$, the effective potential operator $\Vh_\tnS^\HF$ in HF is \emph{non-local}.
Compared with \eqnref{energy-hf}, the factor $\frac{1}{2}$ is dropped since the index $i$ appears twice in the scans of $k$ and of $l$.

For the \textbf{Coulomb operator} $\Jh^{(j)}$ and the \textbf{exchange operator} $\Kh^{(j)}$, we have $\obraket*{\phi_i}{\Jh^{(j)}}{\phi_i} = J_{ij}$, $\obraket*{\phi_i}{\Kh^{(j)}}{\phi_i} = K_{ij}$.
The term $\Jh^{(j)} \phi_i$ represents the effect of the Coulomb potential from all other electrons (with all possible spins), and also from itself ($j = i$ is included in the summation).
The unphysical self-interaction $\Jh^{(i)} \phi_i$ is offset by $\Kh^{(i)} \phi_i$, so \emph{the self-interaction is well-handled} in Hartree-Fock methods.
The term $\Kh^{(j)} \phi_i$ for $j \ne i$ accounts for the antisymmetry of wavefunction, \ie correcting for the proper exchange property.
So \emph{the exchange correlation (Fermi hole) is also well-handled}.

The operator $\sum_j \Jh^{(j)}$ can also be written as the classical potential $V_\tnJ$ from the electron charge density:
\begin{gather}
  \sum\nolimits_{j \in [N]} \Jh^{(j)} \phi_i(\rmxx_1)
  =: \Vh_\tnJ \phi_i(\rmxx_1) = \int \ud\rmxx_1 \, V_\tnJ(\rrv_1) \phi_i(\rmxx_1), \quad \text{where} \\
  V_\tnJ(\rrv_1) := \int \ud\rmxx_2 \, \frac{\sum_{j \in [N]} \lrvert{\phi_j(\rmxx_2)}^2}{r_{12}}
  \stackrel{\text{\eqnref{density-hf}}}{=} \int \ud\rmxx_2 \, \frac{\rho_\tnS(\rmxx_2)}{r_{12}} = \int \ud\rrv_2 \, \frac{\rho_\tnS(\rrv_2)}{r_{12}}.
  \label{eqn:VJ}
\end{gather}

Using the new operators, for \textbf{orthonormal} orbitals $\{\phi_i\}_{i=1}^N$, the energy \eqnref{energy-hf} can also be expressed as:
\begin{align}
  E[\psi_\tnS] = \sum_{k \in [N]} \obraket{\phi_k}{ \Th + \Vh_\Ne + \frac{1}{2} \sum\nolimits_{l \in [N]} \Jh^{(l)} - \frac{1}{2} \sum\nolimits_{l \in [N]} \Kh^{(l)} }{\phi_k}.
  \label{eqn:energy-hf-JKopr}
\end{align}
Using the \textbf{orthonormal} solution to the eigenvalue problem \eqnref{hf-eigen}, the HF ground-state energy is:
\begin{align}
  E^\HF = \frac{1}{2} \sum_{i \in [N]} \veps_i + \frac{1}{2} \sum_{i \in [N]} \obraket*{\phi_i}{\Th + \Vh_\Ne}{\phi_i}
  = \sum_{i \in [N]} \veps_i - \sum_{i \in [N]} \obraket{\phi_i}{ \frac{1}{2} \sum\nolimits_{j \in [N]} \Jh^{(j)} - \frac{1}{2} \sum\nolimits_{j \in [N]} \Kh^{(j)} }{\phi_i},
  \label{eqn:energy-hf-eigen}
\end{align}
but is not $\sum_{i \in [N]} \obraket*{\phi_i}{\ffh^\HF}{\phi_i} = \sum_{i \in [N]} \veps_i$.

\eqnref{hf-eigen} is a \emph{pseudo-eigenvalue problem}, as the Fock operator $\ffh^\HF$ itself depends on the solution $\{\phi_i\}_{i=1}^N$.
So an iterated procedure is used until convergence (\textbf{self-consistent field, SCF}).

\subsubsection{Hartree-Fock equation for R(O)HF (under A2.1)}

Taking the variation of \eqnref{energy-rhf} w.r.t $\varphi_p^*$ and $\varphi_u^*$ (see \eqnsref{general-complex-deriv,schrodinger-variation-hje}) for $p \in [\Npr]$ and $u \in \Iupr$ yields:
\begin{align}
  \ffh^\HF \varphi_p := \big( \Th + \Vh_\tnS^\HF \big) \varphi_p = \veps_p \varphi_p, & \; \text{where} \;
  \Vh_\tnS^\HF = \Vh_\Ne + \lrparen*[\Big]{ \sum\nolimits_{p' \in [\Npr]} 2 \Jth^{(p')} + \sum\nolimits_{u' \in \Iupr} \Jth^{(u')} } - \sum\nolimits_{p' \in [\Npr]} \Kth^{(p')},
  \label{eqn:rhf-eigen-p} \\
  \ffh^\HF \varphi_u := \big( \Th + \Vh_\tnS^\HF \big) \varphi_u = \veps_u \varphi_u, & \; \text{where} \;
  \Vh_\tnS^\HF = \Vh_\Ne + \lrparen*[\Big]{ \sum\nolimits_{p' \in [\Npr]} 2 \Jth^{(p')} + \sum\nolimits_{u' \in \Iupr} \Jth^{(u')} } - \sum\nolimits_{u' \in \Iupr} \Kth^{(u')}.
  \label{eqn:rhf-eigen-u}
\end{align}
For the sets of equations in the first line, the Lagrange multiplier for $\fracdelta{}{\varphi_p^*}$ is $2 \veps_p$, and the common factor of 2 is canceled from both sides.
For a more concise expression in the spirit of \eqnref{energy-rhf-N}, for $a \in [\Nsp]$,
\begin{align}
  & \ffh^\HF \varphi_a := \big( \Th + \Vh_\tnS^\HF \big) \varphi_a = \veps_a \varphi_a, \quad \text{where}
  \label{eqn:rhf-eigen-a} \\
  \Vh_\tnS^\HF ={} & \Vh_\Ne + \sum\nolimits_{j \in [N]} \Jth^{(j)} - \bbI_{\cdot \in [\Npr]} \sum\nolimits_{p' \in [\Npr]} \Kth^{(p')} - \bbI_{\cdot \in \Iupr} \sum\nolimits_{u' \in \Iupr} \Kth^{(u')} \\
  ={} & \Vh_\Ne + \sum\nolimits_{j \in [N]} \Jth^{(j)} - \sum\nolimits_{j \in [N]} \Kbh^{(j)}, \quad \text{where} \;
  \Kbh^{(b)} \varphi_a := \frac{1}{2} \bbI_{a,b \in [\Npr]} \Kth^{(b)} \varphi_a + \bbI_{a,b \in \Iupr} \Kth^{(b)} \varphi_a.
\end{align}
Note that this is not the variational equation directly from \eqnref{energy-rhf-N}, as the equations for $\varphi_p, p \in [\Npr]$ are divided by 2.
Using the \textbf{orthonormal} solution to \eqnref{rhf-eigen-a}, the energy can be expressed as follows by comparing $\ffh^\HF$ in \eqnref{rhf-eigen-a} with \eqnref{energy-rhf-N}:
\begin{align}
  E^\HF
  ={} & \sum_{i \in [N]} \obraket{\varphi_i}{\ffh^\HF}{\varphi_i}
  - \frac{1}{2} \sum_{i \in [N]} \underbrace{ \sum\nolimits_{j \in [N]} \Jt_{ij} }_{ \obraket{\varphi_i}{ \sum\nolimits_{j \in [N]} \Jth^{(j)} }{\varphi_i} }
  + \sum_{p \in [\Npr]} \underbrace{ \sum\nolimits_{p' \in [\Npr]} \Kt_{pp'} }_{ \obraket{\varphi_p}{ \sum\nolimits_{p' \in [\Npr]} \Kth^{(p')} }{\varphi_p} }
  + \frac{1}{2} \sum_{u \in \Iupr} \underbrace{ \sum\nolimits_{u' \in \Iupr} \Kt_{uu'} }_{ \obraket{\varphi_u}{ \sum\nolimits_{u' \in \Iupr} \Kth^{(u')} }{\varphi_u} } \\
  ={} & \sum_{i \in [N]} \obraket{\varphi_i}{\ffh^\HF}{\varphi_i}
  - \frac{1}{2} \sum_{i \in [N]} \obraket{\varphi_i}{ \sum\nolimits_{j \in [N]} \Jth^{(j)} }{\varphi_i}
  + \frac{1}{2} \sum_{i \in [N]} \obraket{\varphi_i}{ \bbI_{\cdot \in [\Npr]} \sum\nolimits_{p' \in [\Npr]} \Kth^{(p')} + \bbI_{\cdot \in \Iupr} \sum\nolimits_{u' \in \Iupr} \Kth^{(u')} }{\varphi_i} \\
  ={} & \sum_{i \in [N]} \obraket{\varphi_i}{ \ffh^\HF - \frac{1}{2} \sum\nolimits_{j \in [N]} \Jth^{(j)} + \frac{1}{2} \sum\nolimits_{j \in [N]} \Kbh^{(j)} }{\varphi_i} \label{eqn:energy-rhf-eigen-deduction} \\
  ={} & \sum_{i \in [N]} \veps_i - \sum_{i \in [N]} \obraket{\varphi_i}{ \frac{1}{2} \sum\nolimits_{j \in [N]} \Jth^{(j)} - \frac{1}{2} \sum\nolimits_{j \in [N]} \Kbh^{(j)} }{\varphi_i} \\
  ={} & 2 \sum_{p \in [\Npr]} \veps_p + \sum_{u \in \Iupr} \veps_u
  - \sum_{p,p' \in [\Npr]} \lrparen*[\Big]{ 2 \Jt_{p p'} - \Kt_{pp'} }
  - 2 \sum_{p \in [\Npr]} \sum_{u \in \Iupr} \Jt_{pu} - \frac{1}{2} \sum_{u,u' \in \Iupr} \lrparen*[\Big]{ \Jt_{u u'} - \Kt_{uu'} }.
\end{align}
Alternatively, by comparing $\ffh^\HF$ in \eqnref{rhf-eigen-a} with \eqnref{energy-rhf-eigen-deduction}, we have:
\begin{align}
  E^\HF
  ={} & \sum_{i \in [N]} \obraket{\varphi_i}{ \frac{1}{2} \ffh^\HF + \frac{1}{2} \Th + \frac{1}{2} \Vh_\Ne }{\varphi_i}
  = \frac{1}{2} \sum_{i \in [N]} \veps_i + \frac{1}{2} \sum_{i \in [N]} \obraket{\varphi_i}{ \Th + \Vh_\Ne }{\varphi_i} \label{eqn:energy-rhf-eigen} \\
  ={} & \sum_{p \in [\Npr]} \veps_p + \frac{1}{2} \sum_{u \in \Iupr} \veps_u
    + \sum_{p \in [\Npr]} \obraket{\varphi_p}{ \Th + \Vh_\Ne }{\varphi_p} + \frac{1}{2} \sum_{u \in \Iupr} \obraket{\varphi_u}{ \Th + \Vh_\Ne }{\varphi_u}.
\end{align}
Again, this is not just $\sum_{i \in [N]} \obraket*{\varphi_i}{\ffh^\HF}{\varphi_i} = \sum_{i \in [N]} \veps_i = 2 \sum_{p \in [\Npr]} \veps_p + \sum_{u \in \Iupr} \veps_u$.

For RHF, the equation reduces to:
\begin{gather}
  \ffh^\HF \varphi_p := \big( \Th + \Vh_\tnS^\HF \big) \varphi_p = \veps_p \varphi_p, \quad \text{where}
  \label{eqn:rrhf-eigen} \\
  \Vh_\tnS^\HF = \Vh_\Ne + \sum\nolimits_{p' \in [\Npr]} \lrparen*[\big]{ 2 \Jth^{(p')} - \Kth^{(p')} }
  = \Vh_\Ne + \sum\nolimits_{j \in [N]} \Jth^{(j)} - \frac{1}{2} \sum\nolimits_{j \in [N]} \Kth^{(j)}.
\end{gather}
For the energy, \eqnref{energy-rhf-eigen} simplifies to:
\begin{align}
  E^\HF
  ={} & \frac{1}{2} \sum_{i \in [N]} \veps_i + \frac{1}{2} \sum_{i \in [N]} \obraket{\varphi_i}{ \Th + \Vh_\Ne }{\varphi_i}
  = \sum_{p \in [\Npr]} \veps_p + \sum_{p \in [\Npr]} \obraket*{\varphi_p}{\Th + \Vh_\Ne}{\varphi_p} \label{eqn:energy-rrhf-eigen} \\
  ={} & 2 \sum_{p \in [\Npr]} \veps_p - \frac{1}{2} \sum_{i \in [N]} \obraket{\varphi_i}{ \sum\nolimits_{j \in [N]} \Jh^{(j)} - \frac{1}{2} \sum\nolimits_{j \in [N]} \Kh^{(j)} }{\varphi_i}
  = 2 \sum_{p \in [\Npr]} \veps_p - \sum_{p,p' \in [\Npr]} \lrparen*[\big]{ 2 \Jt_{pp'} - \Kt_{pp'} },
\end{align}
but is not $\sum_{i \in [N]} \obraket*{\varphi_i}{\ffh^\HF}{\varphi_i} = 2 \sum_{p \in [\Npr]} \veps_p$.

\subsection{Roothaan Equation} \label{sec:hf-roothaan}

Let $\{\ket{\alpha}\}_{\alpha \in [N_\tnB]}$ be a basis set for one-electron wavefunctions.
The orbitals can then be written as: $\phi_i = \sum_\alpha \bfC^\alpha_i \ket{\alpha}$ (the matrix form takes the superscript as the row index).
The Hartree-Fock equation (any case) then becomes $\sum_\alpha \bfC^\alpha_i \ffh^\HF \ket{\alpha} = \veps_i \sum_\beta \bfC^\beta_i \ket{\beta}$.
Taking inner product with $\ket{\alpha'}$, we have:
\begin{align}
  \bfff^\HF \bfC = \bfS \bfC \bfveps: \quad
  \sum_\alpha (\bfff^\HF)^{\alpha'}_\alpha \bfC^\alpha_i = \veps_i \sum_\beta \bfS^{\alpha'}_\beta \bfC^\beta_i, \; \forall \alpha' \in [N_\tnB], i \in [N],
  \label{eqn:roothaan}
\end{align}
where $(\bfff^\HF)^{\alpha'}_\alpha := \obraket*{\alpha'}{\ffh^\HF}{\alpha}$ is the Fock matrix (the Fock operator $\ffh^\HF$ is defined in \eqnref{hf-eigen}),
$\bfS^{\alpha'}_\alpha := \braket*{\alpha'}{\alpha}$ is the overlap matrix,
and $\bfveps := \Diag[\veps_1, \cdots, \veps_N]$.
The orthonormality constraint on the orbitals translates into $\delta_{ij} = \braket{\phi_i}{\phi_j} = \sum_{\alpha,\beta} {\bfC^\alpha_i}^* \braket{\alpha}{\beta} \bfC^\beta_j$,
or $\bfC^\dagger \bfS \bfC = \bfI$.

\eqnref{roothaan} is a generalized eigenvalue problem: it is equivalent to $\bfS^{-1} \bfff^\HF \bfC = \bfC \bfveps$.
Using the orthonormality, we also have:
\begin{align}
  \bfC^\dagger \bfff^\HF \bfC = \bfveps,
  \label{eqn:roothaan-diag}
\end{align}
so solving the problem is equivalent to diagonalizing the Fock matrix $\bfff^\HF$.
This gives $N_\tnB$ eigenstates while only $N \le N_\tnB$ is needed as (occupied) orbitals, but all eigenstates need to be solved so as to collect $N$ eigenstates with largest eigenvalues as the orbitals.
Other eigenstates may also be used as virtual/excited orbitals in post-HF methods.

\paragraph{Calculation of the Fock matrix.}
First note from \eqnref{1rdm-hf}, the density matrix is
$P_\tnS(\rmxx; \rmxx') = \sum_{i \in [N]} \sum_\alpha \bfC^\alpha_i \ket{\alpha}(\rmxx) \sum_{\alpha'} {\bfC^{\alpha'}_i}^* \ket{\alpha'}(\rmxx')^*
= \sum_{\alpha,\alpha'} \sum_{i \in [N]} \bfC^\alpha_i {\bfC^{\alpha'}_i}^* \ket{\alpha}\bra{\alpha'} (\rmxx, \rmxx')$,
so the density matrix under this basis is $\bfP^\alpha_{\alpha'} = \sum_{i \in [N]} \bfC^\alpha_i {\bfC^{\alpha'}_i}^*$,
or $\bfP = \bfC \bfC^\dagger$, so that
$P_\tnS(\rmxx; \rmxx') = \sum_{\alpha,\alpha'} \bfP^\alpha_{\alpha'} \ket{\alpha}\bra{\alpha'} (\rmxx, \rmxx') = \sum_{\alpha,\alpha'} \bfP^\alpha_{\alpha'} \braket*{\rmxx}{\alpha} \braket*{\alpha'}{\rmxx'}$.
Note that this $\bfP$ corresponds to the $\bfGamma$ (not $\bfGammat$) in \eqnref{dm-matrix}.
From \eqnref{density-hf}, the density is
$\rho_\tnS(\rmxx) = \sum_{\alpha,\alpha'} \big( \sum_{i \in [N]} \bfC^\alpha_i {\bfC^{\alpha'}_i}^* \big) \braket*{\rmxx}{\alpha} \braket*{\alpha'}{\rmxx}
= \sum_{\alpha,\alpha'} \bfP^\alpha_{\alpha'} \braket*{\rmxx}{\alpha} \braket*{\alpha'}{\rmxx} = \tr\Big( \bfP \big(\ket{\alpha}\bra{\alpha'}\big)^{\alpha'}_\alpha \Big)(\rmxx)$.
The normalization constraint is $\int \ud\rmxx \, \rho_\tnS(\rmxx) = \tr(\bfP \bfS) = N$.
This can also be seen from \eqnref{mean-dm-matrix}, where $\bfGamma$ is taken as $\bfP$, and $\Ah$ is taken as the identity so $\bfAt^{\alpha'}_\alpha := \braket{\alpha'}{\alpha} = \bfS^{\alpha'}_\alpha$ is the overlap matrix according to \eqnref{opr-matrix}.
Mean value of other operators can also be computed using \eqnref{mean-dm-matrix}.

The last two terms in $\bfff^\HF$ in \eqnref{hf-eigen} lead to the Hartree matrix $\bfJ$ and the exchange matrix $\bfK$:
\begin{align}
  \bfJ^{\alpha'}_\alpha :={} & \obraket*{\alpha'}{\sum_j \Jh^{(j)}}{\alpha}
  \stackrel{\text{\eqnref{JK-opr}}}{=}
  \int \ud\rmxx_1 \ud\rmxx_2 \, \frac{\sum_{\beta,\beta'} \bfP^\beta_{\beta'} \braket*{\rmxx_2}{\beta} \braket*{\beta'}{\rmxx_2}}{r_{12}} \braket{\rmxx_1}{\alpha} \braket*{\alpha'}{\rmxx_1}
  = \sum_{\beta,\beta'} \bfP^\beta_{\beta'} I_{\alpha'\beta' | \alpha\beta}
  = \tr(\bfP \sfbJ^{\alpha'}_\alpha), \\
  \bfK^{\alpha'}_\alpha :={} & \obraket*{\alpha'}{\sum_j \Kh^{(j)}}{\alpha}
  \stackrel{\text{\eqnref{JK-opr}}}{=}
  \int \ud\rmxx_1 \ud\rmxx_2 \, \sum_{j \in [N]} \frac{\sum_{\beta'} {\bfC^{\beta'}_j}^* \braket*{\beta'}{\rmxx_2} \braket*{\rmxx_2}{\alpha}}{r_{12}} \sum_\beta \bfC^\beta_j \braket{\rmxx_1}{\beta} \braket*{\alpha'}{\rmxx_1} \\
  ={} & \sum_{\beta,\beta'} \bfP^\beta_{\beta'} \int \ud\rmxx_1 \ud\rmxx_2 \, \frac{\braket*{\beta'}{\rmxx_2} \braket*{\rmxx_2}{\alpha}}{r_{12}} \braket{\rmxx_1}{\beta} \braket*{\alpha'}{\rmxx_1}
  = \sum_{\beta,\beta'} \bfP^\beta_{\beta'} I_{\alpha'\beta' | \beta\alpha}
  = \tr(\bfP \sfbK^{\alpha'}_\alpha),
  \label{eqn:K-matrix}
\end{align}
where $(\sfbJ^{\alpha'}_\alpha)^{\beta'}_\beta := I_{\alpha'\beta' | \alpha\beta}$, and $(\sfbK^{\alpha'}_\alpha)^{\beta'}_\beta := I_{\alpha'\beta' | \beta\alpha}$.
Note that $\bfJ$ and $\bfK$ depend on $\bfP$, which is taken as the previous-step $\bfP$ while solving \eqnref{roothaan} for the current step.
Constructing $\bfJ$ and $\bfK$ requires 4-center-2-electron integrals and incurs an $O(N^4)$ complexity, but density fitting techniques could make it $O(N^3)$.

The Fock matrix can then be expressed as:
\begin{align}
  (\bfff^\HF)^{\alpha'}_\alpha = \bfT^{\alpha'}_\alpha + (\bfV_\Ne)^{\alpha'}_\alpha + \tr(\bfP \sfbJ^{\alpha'}_\alpha) - \tr(\bfP \sfbK^{\alpha'}_\alpha)
  = \bfT^{\alpha'}_\alpha + (\bfV_\Ne)^{\alpha'}_\alpha + \bfJ^{\alpha'}_\alpha - \bfK^{\alpha'}_\alpha,
  \label{eqn:fock-matrix-decomp}
\end{align}
where $\bfT^{\alpha'}_\alpha := \tr(\Th \ket{\alpha} \bra{\alpha'}) = -\frac{1}{2} \obraket*{\alpha'}{\nabla^2}{\alpha}$, and
$(\bfV_\Ne)^{\alpha'}_\alpha := \tr(\Vh_\Ne \ket{\alpha} \bra{\alpha'}) = \obraket*{\alpha'}{\Vh_\Ne}{\alpha} = \int \ud\rmxx \, \alpha'(\rmxx)^* \alpha(\rmxx) V_\Ne(\rrv)$.
Using the density matrix, \eqnref{roothaan-diag} indicates $\tr(\bfP \bfff^\HF) = \tr(\bfveps)$,
so the energy of the system can be estimated by
$E[\psi_\tnS] = \frac{1}{2} \tr(\bfP \bfff^\HF) + \frac{1}{2} \tr\big( \bfP (\bfT + \bfV_\Ne) \big)$.
This is the common expression for UHF and R(O)HF, \eqnsref{energy-hf-eigen,energy-rhf-eigen,energy-rrhf-eigen}.
For R(O)HF, just use the appropriate index set for summing over $\Kh^{(j)}$.

\paragraph{Orbital orthonormality.}
Obviously the overlap matrix $\bfS$ and the density matrix $\bfP = \bfC \bfC^\dagger$ are Hermitian.
Note that the Fock operator hence the Fock matrix is also Hermitian, $(\bfff^\HF)^\dagger = \bfff^\HF$:
from \eqnref{hf-eigen}, we know that $\Th$ is Hermitian, also are $\Vh_\Ne$ and $\Jh^{(j)}$ since they are real-valued, multiplicative operators (sometimes also called local operators);
for $\sum_j \Kh^{(j)}$, from \eqnref{K-matrix} and noting \eqnref{4c2e-symmetry}, $\bfK^{\alpha'}_\alpha = \obraket*{\alpha'}{\sum_j \Kh^{(j)}}{\alpha}
= \sum_{\beta,\beta'} \bfP^\beta_{\beta'} I_{\alpha'\beta' | \beta\alpha}
= \sum_{\beta,\beta'} (\bfP^{\beta'}_\beta)^* I_{\beta\alpha | \alpha'\beta'}^*
= \sum_{\beta,\beta'} (\bfP^{\beta'}_\beta)^* I_{\alpha\beta | \beta'\alpha'}^*
= \obraket*{\alpha}{\sum_j \Kh^{(j)}}{\alpha'}^* = (\bfK^\alpha_{\alpha'})^*$.

We now show that $\bfC_i^\dagger \bfS \bfC_j = 0$ for $i \ne j$ if given non-degeneracy $\veps_i \ne \veps_j$, so to realize the orbital orthonormality requirement $\bfC^\dagger \bfS \bfC = \bfI$, we only need to normalize each orbital coefficient: $\bfC_i \asn \bfC_i / \sqrt{\bfC_i^\dagger \bfS \bfC_i}$.

From the Roothaan equation \eqnref{roothaan} $\bfff^\HF \bfC_i = \veps_i \bfS \bfC_i$, we have
$\bfC_i^\dagger \bfS \bfC_j = \bfC_i^\dagger \frac{1}{\veps_j} \bfff^\HF \bfC_j = \frac{1}{\veps_j} (\bfff^\HF \bfC_i)^\dagger \bfC_j = \frac{\veps_i}{\veps_j} \bfC_i^\dagger \bfS \bfC_j$ (since $\bfff^\HF$ is Hermitian),
so $\lrparen*[\big]{1 - \frac{\veps_i}{\veps_j}} \bfC_i^\dagger \bfS \bfC_j = 0$, which implies $\bfC_i^\dagger \bfS \bfC_j = 0$.

Alternatively, as $\bfS$ is positive definite and Hermitian, let $\bfM$ be a non-singular matrix s.t. $\bfS = \bfM \bfM^\dagger$.
Then the Roothaan equation \eqnref{roothaan} $\bfff^\HF \bfC_i = \veps_i \bfS \bfC_i$ implies $\bfff^\HF \bfC_i = \veps_i \bfM \bfM^\dagger \bfC_i$,
which is $\bfM^{-1} \bfff^\HF \bfM^{-\dagger} \bfM^\dagger \bfC_i = \veps_i \bfM^\dagger \bfC_i$, \ie, $\{(\veps_i, \bfM^\dagger \bfC_i)\}_i$ are eigenpairs of $\bfM^{-1} \bfff^\HF \bfM^{-\dagger}$.
Since $(\bfM^{-1} \bfff^\HF \bfM^{-\dagger})^\dagger = \bfM^{-1} \bfff^\HF \bfM^{-\dagger}$ is Hermitian, once $\veps_i \ne \veps_j$, the corresponding eigenvectors are orthogonal:
$0 = (\bfM^\dagger \bfC_i)^\dagger (\bfM^\dagger \bfC_j) = \bfC_i^\dagger \bfM \bfM^\dagger \bfC_j = \bfC_i^\dagger \bfS \bfC_j$.
Note that although the equation can be reformulated as $\bfS^{-1} \bfff^\HF \bfC_i = \veps_i \bfC_i$, this unnecessarily implies $\bfC_i^\dagger \bfC_j = 0$, since $\bfS^{-1} \bfff^\HF$ is not Hermitian.

\paragraph{Density matrix optimization.}
For large systems, it is more efficient to only finding the lowest $N$ eigenvectors instead of all of them.
Density matrix (referring to the 1-RDM) optimization is one approach.
Noting \eqnsref{energy-decomp,energy-hf-JKopr} and \eqnref{mean-dm-matrix} and the above argument, the variational energy in terms of density matrix $\bfP$ is:
\begin{align}
  E(\bfP) ={} & \tr( \bfP \bfT ) + \tr( \bfP \bfV_\Ne ) + \frac{1}{2} \tr\Big( \bfP \bfJ(\bfP) \Big) - \frac{1}{2} \tr\Big( \bfP \bfK(\bfP) \Big) \\
  ={} & \tr\Big( \bfP (\bfT + \bfV_\Ne) \Big) + \frac{1}{2} \tr\Big( \bfP \tr\big( \bfP (\sfbJ^{\alpha'}_\alpha - \sfbK^{\alpha'}_\alpha) \big)^{\alpha'}_\alpha \Big).
  \label{eqn:energy-dm-matrix}
\end{align}
Here we explicitly write the dependency of $\bfJ$, $\bfK$ on $\bfP$.
It can also be derived by plugging the basis expansion of orbitals $\phi_i = \sum_\alpha \bfC^\alpha_i \ket{\alpha}$ into \eqnref{energy-decomp} (also noting \eqnref{pairdensity-hf}).

Note that $\tr(\bfA^\dagger \bfB) = \sum_{\alpha,\alpha'} (\bfA^\dagger)^\alpha_{\alpha'} \bfB^{\alpha'}_\alpha
= \sum_{\alpha,\alpha'} (\bfA^{\alpha'}_\alpha)^* \bfB^{\alpha'}_\alpha
= \sum_{\alpha'\alpha} (\bfAb^{(\alpha'\alpha)})^* \bfBb^{(\alpha'\alpha)}
= \bfAb^\dagger \bfBb$,
where $\bfAb^{(\alpha'\alpha)} := \bfA^{\alpha'}_\alpha$ (sim. $\bfBb$) is the flattened vector.
So the first term $\tr\big( \bfP (\bfT + \bfV_\Ne) \big) = \tr\big( (\bfP^\dagger)^\dagger (\bfT + \bfV_\Ne) \big) = \bfPb^\dagger (\bfTb + \bfVb_\Ne)$
if we let $\bfTb^{(\alpha'\alpha)} := \bfT^{\alpha'}_\alpha$ (sim. $\bfVb_\Ne$),
and $\bfPb^{(\alpha'\alpha)} := (\bfP^\dagger)^{\alpha'}_\alpha$, which is also $\bfP^{\alpha'}_\alpha = (\bfP^\alpha_{\alpha'})^*$ since $\bfP = \bfP^\dagger$.

Note also that $(\bfAb^\dagger)_{(\alpha'\alpha)} = (\bfAb^{(\alpha'\alpha)})^* = (\bfA^{\alpha'}_\alpha)^* = (\bfA^\dagger)^\alpha_{\alpha'}$,
which indicates if $\tr(\bfL \bfB) = \bfLb \bfBb$, then $\bfLb_{(\alpha'\alpha)} := \bfL^\alpha_{\alpha'}$.
So the second term $\tr\Big( \bfP \tr\big( \bfP (\sfbJ^{\alpha'}_\alpha - \sfbK^{\alpha'}_\alpha) \big)^{\alpha'}_\alpha \Big)
= \tr\Big( (\bfP^\dagger)^\dagger \tr\big( (\sfbJ^{\alpha'}_\alpha - \sfbK^{\alpha'}_\alpha) \bfP \big)^{\alpha'}_\alpha \Big)
= \bfPb^\dagger (\sfbJb - \sfbKb) \bfPb$,
if we let $\sfbJb^{(\alpha'\alpha)}_{(\beta\beta')} := (\sfbJ^{\alpha'}_\alpha)^{\beta'}_\beta = I_{\alpha'\beta' | \alpha\beta}$,
and $\sfbKb^{(\alpha'\alpha)}_{(\beta\beta')} := (\sfbK^{\alpha'}_\alpha)^{\beta'}_\beta = I_{\alpha'\beta' | \beta\alpha}$.
Note that $\sfbJb^{(\beta\beta')}_{(\alpha'\alpha)} = I_{\beta\alpha | \beta'\alpha'} = I_{\beta'\alpha' | \beta\alpha}^* = I_{\alpha'\beta' | \alpha\beta}^* = (\sfbJb^{(\alpha'\alpha)}_{(\beta\beta')})^*$,
and $\sfbKb^{(\beta\beta')}_{(\alpha'\alpha)} = I_{\beta\alpha | \alpha'\beta'} = I_{\alpha'\beta' | \beta\alpha}^* = (\sfbKb^{(\alpha'\alpha)}_{(\beta\beta')})^*$,
so both $\sfbJb$ and $\sfbKb$ are Hermitian matrices.
\eqnref{energy-dm-matrix} then becomes:
\begin{align}
  E(\bfPb) = \bfPb^\dagger (\bfTb + \bfVb_\Ne) + \frac{1}{2} \bfPb^\dagger (\sfbJb - \sfbKb) \bfPb,
  \label{eqn:energy-dm-matrix-flatten}
\end{align}
which is a quadratic form of $\bfPb$.
This directly gives the gradient (see \secref{pre-complex-grad}) w.r.t the density matrix:
\begin{align}
  \nabla_{\bfPb} E(\bfPb)
  \stackrel{\text{\eqnsref{complex-grad-inprod,complex-grad-quadratic}}}{=}
  {} & \bfTb + \bfVb_\Ne + (\sfbJb - \sfbKb) \bfPb, \quad \text{or} \\
  \nabla_\bfP E(\bfP) ={} & \bfT + \bfV_\Ne + \Big( \tr\Big( (\sfbJ^{\alpha'}_\alpha - \sfbK^{\alpha'}_\alpha) \bfP \Big) \Big)^{\alpha'}_\alpha
  = \bfT + \bfV_\Ne + \bfJ(\bfP) - \bfK(\bfP)
  = \bfff^\HF(\bfP),
  \label{eqn:grad-energy-dm-matrix}
\end{align}
where the last equality is due to \eqnref{fock-matrix-decomp}

In the optimization, orbital orthonormality $\bfC^\dagger \bfS \bfC = \bfI$ is still required, which is necessary for the derivation of the energy expression \eqnref{energy-dm-matrix}.
The constraint indicates $\bfC \bfC^\dagger \bfS \bfC \bfC^\dagger = \bfC \bfC^\dagger$, which is $\bfP \bfS \bfP = \bfP$, meaning that the density matrix $\bfP$ is \emph{idempotent}.
(Note that $\bfP$ is likely degenerate, so this does not mean $\bfP \bfS = \bfI$ or $\bfP = \bfS^{-1}$.)
Introduce Lagrange multipliers $\bfLambda$ for $\bfP \bfS \bfP - \bfP$, which is Hermitian since $\bfP$ and $\bfS$ are.
The gradient of the constraint term is:
$\nabla_\bfP \tr(\bfLambda^\dagger (\bfP \bfS \bfP - \bfP)) = \bfLambda \bfP^\dagger \bfS^\dagger + \bfS^\dagger \bfP^\dagger \bfLambda - \bfLambda = \bfLambda \bfP \bfS + \bfS \bfP \bfLambda - \bfLambda$.
The optimized $\bfP^\star$ then satisfies:
\begin{align}
  \bfff^\HF(\bfP^\star) ={} & \bfT + \bfV_\Ne + \bfJ(\bfP^\star) - \bfK(\bfP^\star)
  = \bfT + \bfV_\Ne + \tr\lrparen*[\big]{ \bfP^\star (\sfbJ^{\alpha'}_\alpha - \sfbK^{\alpha'}_\alpha) }^{\alpha'}_\alpha \\
  ={} & \bfLambda^\star \bfP^\star \bfS + \bfS \bfP^\star \bfLambda^\star - \bfLambda^\star, \quad \text{and} \quad
  \bfP^\star \bfS \bfP^\star = \bfP^\star.
  \label{eqn:dm-optim-eq}
\end{align}
Multiplying both sides with $\bfP^\star$ yields:
\begin{align}
  \bfff^\HF(\bfP^\star) \bfP^\star = \bfS \bfP^\star \bfLambda^\star \bfP^\star, \quad \text{and} \quad
  \bfP^\star \bfS \bfP^\star = \bfP^\star.
  \label{eqn:dm-eigen}
\end{align}
Taking the trace, the l.h.s becomes $\tr({\bfP^\star}^\dagger \bfff^\HF(\bfP^\star))$, the r.h.s becomes $\tr(\bfS \bfP^\star \bfLambda^\star \bfP^\star) = \tr(\bfP^\star \bfS \bfP^\star \bfLambda^\star) = \tr(\bfP^\star \bfLambda^\star) = \tr({\bfP^\star}^\dagger \bfLambda^\star)$. So:
\begin{align}
  {\bfPb^\star}^\dagger \bfffb^\HF(\bfPb^\star)
  = {\bfPb^\star}^\dagger \bfTb + {\bfPb^\star}^\dagger \bfVb_\Ne + {\bfPb^\star}^\dagger (\sfbJb - \sfbKb) \bfPb^\star
  = {\bfPb^\star}^\dagger \bfLambdab^\star, \quad \text{and} \quad
  \bfP^\star \bfS \bfP^\star = \bfP^\star.
  \label{eqn:dm-vector-eigen}
\end{align}
The normalization constraint $\tr(\bfP \bfS) = N$ can be formulated as ${\bfPb^\star}^\dagger \bfSb = N$.
If ignoring the constraint $\bfP^\star \bfS \bfP^\star = \bfP^\star$ and only imposing the constraint ${\bfPb^\star}^\dagger \bfSb = N$ with Lagrange multiplier $\lambda$, taking the gradient w.r.t $\bfPb^\star$ yields:
\begin{align}
  2 (\sfbJb - \sfbKb) \bfPb^\star = \bfLambdab^\star + \lambda \bfSb - \bfTb - \bfVb_\Ne, \quad \text{and} \quad
  \bfPb^\star = \frac12 (\sfbJb - \sfbKb)^{-1} (\bfLambdab^\star + \lambda \bfSb - \bfTb - \bfVb_\Ne),
  \label{eqn:dm-vector-solution}
\end{align}
if assuming $\sfbJb - \sfbKb$ to be non-singular.

\subsection{Energy Gradient and Relation Among the Methods} \label{sec:hf-grad}
Given a conformation $\bfR := \{(Z_A, \rrv_A)\}_A$ and an atomic basis expansion of orbitals $\phi_i = \phi_{\bfR,\bfC_i} := \sum_\alpha \bfC^\alpha_i \ket{\alpha}_\bfR$, there are two equivalent ways from $E_\bfR[\{\phi_i\}_i]$ \eqnref{energy-hf} to Roothaan equation \eqnref{roothaan}.
First ignore the dependency on $\bfR$.

\itemone From $E[\{\phi_i\}_i]$ \eqnref{energy-hf}, by taking the variation, we have:
  \begin{align}
    \fracdelta{}{(\phi_j)^*} E[\{\phi_i\}_i] = 2 \ffh^\HF \phi_j,
    \label{eqn:variation-energy-orbital}
  \end{align}
  as in \eqnref{hf-eigen}.
  Leveraging the relation between derivative and variation, we have $\partial_{(\bfC^\beta_k)^*} E(\bfC)
  := \partial_{(\bfC^\beta_k)^*} E[\{\phi_{\bfC_i}\}_i]
  = \int \ud\rmxx \, \sum_j \fracdelta{}{(\phi_j)^*} E[\{\phi_i\}_i](\rmxx) \partial_{(\bfC^\beta_k)^*} \phi_{\bfC_j}^*(\rmxx)
  = 2 \int \ud\rmxx \, \sum_j \ffh^\HF \phi_{\bfC_j}(\rmxx) \partial_{(\bfC^\beta_k)^*} \sum_{\alpha'} (\bfC^{\alpha'}_j)^* \bra{\alpha'}(\rmxx)
  = 2 \int \ud\rmxx \, \sum_j \ffh^\HF \phi_{\bfC_j}(\rmxx) \delta^j_k \bra{\beta}(\rmxx)
  = 2 \int \ud\rmxx \, \ffh^\HF \phi_{\bfC_k}(\rmxx) \bra{\beta}(\rmxx)
  = 2 \int \ud\rmxx \, \sum_\alpha \bfC^\alpha_k \ffh^\HF \ket{\alpha}(\rmxx) \bra{\beta}(\rmxx)
  = 2 \sum_\alpha \bfC^\alpha_k \obraket*{\beta}{\ffh^\HF}{\alpha}
  = 2 (\bfff^\HF(\bfC) \bfC)^\beta_k$.
  By \eqnref{complex-grad} and \eqnref{fock-matrix-decomp}, we have:
  \begin{align}
    \nabla_\bfC E(\bfC) = 2 \bfff^\HF(\bfC) \bfC = 2 (\bfT + \bfV_\Ne + \bfJ(\bfP) - \bfK(\bfP)) \bfC.
    \label{eqn:grad-energy-orbital}
  \end{align}

  For the orthonormality constraint, when adopting the general complex derivative \eqnref{general-complex-deriv}, we have
  $\fracdelta{}{(\phi_l)^*} \bmclE^j_i (\braket{\phi_i}{\phi_j} - \delta^i_j)
  = \bmclE^j_i (\delta^i_l \phi_j + \delta_{lj} \phi_{i'} \delta^{i'i}) = \bmclE^j_l \phi_j + \bmclE^{i'}_l \phi_{i'}$.
  So $\partial_{(\bfC^\beta_k)^*} \bmclE^j_i \lrparen{ \braket{\phi_{\bfC_i}}{\phi_{\bfC_j}} - \delta^i_j }
  = \int \ud\rmxx \, \sum_l \fracdelta{}{(\phi_l)^*} \bmclE^j_i \braket{\phi_i}{\phi_j}(\rmxx) \partial_{(\bfC^\beta_k)^*} \phi_{\bfC_l}^*(\rmxx)
  = \int \ud\rmxx \, \sum_l \fracdelta{}{(\phi_l)^*} \bmclE^j_i \braket{\phi_i}{\phi_j}(\rmxx) \delta^l_k \bra{\beta}(\bfxx)
  = \int \ud\rmxx \, (\bmclE^j_k \phi_{\bfC_j}(\rmxx) + \bmclE^i_k \phi_{\bfC_i}(\rmxx)) \bra{\beta}(\bfxx)
  = \bmclE^j_k \bfC^\alpha_j \braket{\beta}{\alpha} + \bmclE^i_k \bfC^\alpha_i \braket{\beta}{\alpha}$, or:
  \begin{align}
    \nabla_\bfC \bmclE^j_i \braket{\phi_{\bfC_i}}{\phi_{\bfC_j}} = 2 \bfS \bfC \bmclE.
    \label{eqn:grad-orthonormality-orbital}
  \end{align}

\itemtwo From $E[\{\phi_i\}_i]$ \eqnref{energy-hf}, by plugging in the basis expansion,
  we have $E(\bfC) := E[\{\phi_{\bfC_i}\}_i] = E(\bfP = \bfP(\bfC))$ which recovers \eqnref{energy-dm-matrix}, where $\bfP(\bfC) := \bfC \bfC^\dagger$.
  Then $\partial_{(\bfC^\beta_k)^*} E(\bfC) = \sum_{\alpha,\alpha'} \partial_{(\bfP^{\alpha'}_\alpha)^*} E(\bfP = \bfP(\bfC)) \partial_{(\bfC^\beta_k)^*} (\bfP^{\alpha'}_\alpha(\bfC))^*
  = \sum_{\alpha,\alpha'} (\nabla_\bfP E(\bfP = \bfP(\bfC)))^{\alpha'}_\alpha \partial_{(\bfC^\beta_k)^*} \sum_i \bfC^\alpha_i (\bfC^{\alpha'}_i)^*
  = \sum_{\alpha,\alpha'} (\nabla_\bfP E(\bfP = \bfP(\bfC)))^{\alpha'}_\alpha \sum_i (
    \delta^{\beta\alpha} \delta_{ki} \bfC^{\alpha''}_{i''} \delta_{\alpha'\alpha''} \delta^{ii''} + \bfC^\alpha_i \delta^\beta_{\alpha'} \delta^i_k )
  = \sum_{\alpha,\alpha'} (\nabla_\bfP E(\bfP = \bfP(\bfC)))^{\alpha'}_\alpha ( \delta^{\beta\alpha} \bfC^{\alpha''}_k \delta_{\alpha'\alpha''} + \bfC^\alpha_k \delta^\beta_{\alpha'} )
  = (\nabla_\bfP E(\bfP = \bfP(\bfC)))^\beta_{\alpha''} \bfC^{\alpha''}_k + (\nabla_\bfP E(\bfP = \bfP(\bfC)))^\beta_\alpha \bfC^\alpha_k
  = 2 (\nabla_\bfP E(\bfP = \bfP(\bfC)) \bfC)^\beta_k$.
  Using \eqnref{grad-energy-dm-matrix}, we have:
  $\nabla_\bfC E(\bfC) = 2 (\bfT + \bfV_\Ne + \bfJ(\bfP) - \bfK(\bfP)) \bfC$.

  For the orthonormality constraint $\braket{\phi_{\bfC_i}}{\phi_{\bfC_j}} = \delta^i_j$, plugging in the basis expansion yields $\bfC^\dagger \bfS \bfC = \bfI$.
  Introducing Lagrange multipliers $\bmclE^j_i$, and noting $\bfS$ is Hermitian, we have
  $\nabla_\bfC \tr(\bmclE (\bfC^\dagger \bfS \bfC - \bfI)) \stackrel{\text{\eqnref{complex-grad-quadratic}}}{=} 2 \bfS \bfC \bmclE$,
  or $\nabla_\bfC \bmclE^j_i \braket{\phi_{\bfC_i}}{\phi_{\bfC_j}} = 2 \bfS \bfC \bmclE$.

Results of the two processes coincide. Also note that $\nabla_\bfP E(\bfP) = \bfff^\HF(\bfC)$.
The orbital solution satisfies:
$\bfff^\HF(\bfC^\star) \bfC^\star = \bfS \bfC^\star \bmclE^\star$ and ${\bfC^\star}^\dagger \bfS \bfC^\star = \bfI$,
which takes the form of the Roothaan equation \eqnref{roothaan}.

When using SCF (self-consistent field), \ie, iteratively solving generalized eigenvalue problems, orbital orthogonality can naturally be achieved if there is no degeneracy,
as discussed in \secref{hf-roothaan} (Paragraph ``Orbital orthonormality'') and \secref{hf-hf} (beginning).
Hence only normalization constraint is required and the Lagrange multiplier matrix is diagonal, $\bfveps = \Diag[\veps_1, \cdots, \veps_N]$.
\begin{align}
  \bfff^\HF(\bfC^\star) \bfC^\star = \bfS \bfC^\star \bfveps^\star, \quad \text{and} \quad
  {\bfC^\star}^\dagger \bfS \bfC^\star = \bfI,
  \label{eqn:roothaan-new}
\end{align}
which is the Roothaan equation \eqnref{roothaan}.

\subsection{Atomic Force Calculation} \label{sec:hf-force}

Using SCF orbital:
$\nabla_\bfR E_\bfR(\bfC^\star_\bfR) = (\nabla_\bfR E_\bfR) (\bfC^\star_\bfR) + \tr\Big( \big( \nabla_\bfC E_\bfR(\bfC^\star_\bfR) \big)^\dagger \nabla_\bfR \bfC^\star_\bfR \Big)
\stackrel{\text{\eqnsref{grad-energy-orbital,roothaan-new}}}{=} (\nabla_\bfR E_\bfR) (\bfC^\star_\bfR) + \tr\Big( 2 \big( \bfS_\bfR \bfC^\star_\bfR \bfveps^\star_\bfR \big)^\dagger \nabla_\bfR \bfC^\star_\bfR \Big)
= (\nabla_\bfR E_\bfR) (\bfC^\star_\bfR) + 2 \tr\Big( \bfveps^\star_\bfR {\bfC^\star_\bfR}^\dagger \bfS_\bfR \nabla_\bfR \bfC^\star_\bfR \Big)$.
From the constraint ${\bfC^\star_\bfR}^\dagger \bfS_\bfR \bfC^\star_\bfR = \bfI$, we have
$\nabla_\bfR {\bfC^\star_\bfR}^\dagger \bfS_\bfR \bfC^\star_\bfR + {\bfC^\star_\bfR}^\dagger \nabla_\bfR \bfS_\bfR \bfC^\star_\bfR + {\bfC^\star_\bfR}^\dagger \bfS_\bfR \nabla_\bfR \bfC^\star_\bfR = 0$,
so $2 {\bfC^\star_\bfR}^\dagger \bfS_\bfR \nabla_\bfR \bfC^\star_\bfR = -{\bfC^\star_\bfR}^\dagger \nabla_\bfR \bfS_\bfR \bfC^\star_\bfR$, and:
\begin{align}
  \nabla_\bfR E_\bfR(\bfC^\star_\bfR) ={} & (\nabla_\bfR E_\bfR) (\bfC^\star_\bfR) - \tr\Big( \bfveps^\star_\bfR {\bfC^\star_\bfR}^\dagger \nabla_\bfR \bfS_\bfR \bfC^\star_\bfR \Big)
  = (\nabla_\bfR E_\bfR) (\bfC^\star_\bfR) - \tr\Big( \bfC^\star_\bfR \bfveps^\star_\bfR {\bfC^\star_\bfR}^\dagger \nabla_\bfR \bfS_\bfR \Big) \\
  ={} & (\nabla_\bfR E_\bfR) (\bfC^\star_\bfR) - \tr\Big( \bfQ^\star_\bfR \nabla_\bfR \bfS_\bfR \Big),
  \label{eqn:hf-force-orbital}
\end{align}
where $\bfQ := \bfC \bfveps \bfC^\dagger, \bfQ^\alpha_{\alpha'} = \sum_i \bfveps_i \bfC^\alpha_i (\bfC^{\alpha'}_i)^*$
is the energy-weighted density matrix~\citep{pople1992kohn}.

Using density matrix gives the same result:
$\nabla_\bfR E_\bfR(\bfP^\star_\bfR) = (\nabla_\bfR E_\bfR) (\bfP^\star_\bfR) + \tr\Big( \big( \nabla_\bfP E_\bfR(\bfP^\star_\bfR) \big)^\dagger \nabla_\bfR \bfP^\star_\bfR \Big)
\stackrel{\text{\eqnsref{grad-energy-dm-matrix,dm-optim-eq}}}{=} (\nabla_\bfR E_\bfR) (\bfP^\star_\bfR) + \tr\Big( \big( \bfLambda^\star_\bfR \bfP^\star_\bfR \bfS_\bfR + \bfS_\bfR \bfP^\star_\bfR \bfLambda^\star_\bfR \big) \nabla_\bfR \bfP^\star_\bfR - \bfLambda^\star_\bfR \nabla_\bfR \bfP^\star_\bfR \Big)$.
From the constraint $\bfP^\star_\bfR \bfS_\bfR \bfP^\star_\bfR = \bfP^\star_\bfR$, we have $\nabla_\bfR \bfP^\star_\bfR = \nabla_\bfR \big( \bfP^\star_\bfR \bfS_\bfR \bfP^\star_\bfR \big)$, so:
\begin{align}
  \nabla_\bfR E_\bfR(\bfP^\star_\bfR) ={} & (\nabla_\bfR E_\bfR) (\bfP^\star_\bfR) - \tr\big( \bfLambda^\star_\bfR \bfP^\star_\bfR \nabla_\bfR \bfS_\bfR \bfP^\star_\bfR \big)
  = (\nabla_\bfR E_\bfR) (\bfP^\star_\bfR) - \tr\big( \bfP^\star_\bfR \bfLambda^\star_\bfR \bfP^\star_\bfR \nabla_\bfR \bfS_\bfR \big).
  \label{eqn:hf-force-dm}
\end{align}
Noting that $\bfP^\star = \bfC^\star {\bfC^\star}^\dagger$, from \eqnsref{dm-eigen,roothaan-new}, we have
$\bfff^\HF(\bfP^\star) \bfP^\star = \bfS \bfP^\star \bfLambda^\star \bfP^\star = \bfff^\HF(\bfP^\star) \bfC^\star {\bfC^\star}^\dagger = \bfS \bfC^\star \bfveps^\star {\bfC^\star}^\dagger$.
Since $\bfS$ is non-singular, this implies:
\begin{align}
  \bfP^\star \bfLambda^\star \bfP^\star = \bfC^\star \bfveps^\star {\bfC^\star}^\dagger = \bfQ^\star.
  \label{eqn:constraint-relation-orbital-dm}
\end{align}
This then equates \eqnref{hf-force-dm} to \eqnref{hf-force-orbital}.

But from derivation using SCF orbital,
$\nabla_\bfR E_\bfR(\bfC^\star_\bfR) = (\nabla_\bfR E_\bfR) (\bfC^\star_\bfR) + \tr\Big( 2 \big( \bfS_\bfR \bfC^\star_\bfR \bfveps^\star_\bfR \big)^\dagger \nabla_\bfR \bfC^\star_\bfR \Big)
= (\nabla_\bfR E_\bfR) (\bfC^\star_\bfR) + \tr\Big( \nabla_{\bfC^\star_\bfR}^\dagger \tr\big( \big( {\bfC^\star_\bfR}^\dagger \bfS_\bfR \bfC^\star_\bfR - \bfI \big) \bfveps^\star_\bfR \big) \nabla_\bfR \bfC^\star_\bfR \Big)
= (\nabla_\bfR E_\bfR) (\bfC^\star_\bfR) + \nabla_\bfR \tr\big( \big( {\bfC^\star_\bfR}^\dagger \bfS_\bfR \bfC^\star_\bfR - \bfI \big) \bfveps^\star_\bfR \big)$,
and $\diag\big( \big( {\bfC^\star_\bfR}^\dagger \bfS_\bfR \bfC^\star_\bfR - \bfI \big) \bfveps^\star_\bfR \big) = 0$ since columns of $\bfC^\star_\bfR$ are exactly $\bfS_\bfR$-normalized for any $\bfR$. Then we have:
\begin{align}
  \nabla_\bfR E_\bfR(\bfC^\star_\bfR) = (\nabla_\bfR E_\bfR) (\bfC^\star_\bfR).
  \label{eqn:hf-force-simp}
\end{align}
Similarly, from the derivation using density matrix, we can obtain the same conclusion \eqnref{hf-force-simp} above, since the idempotency $\bfP^\star_\bfR \bfS_\bfR \bfP^\star_\bfR = \bfP^\star_\bfR$ holds exactly since ${\bfC^\star_\bfR}^\dagger \bfS_\bfR \bfC^\star_\bfR = \bfI$ exactly since the $\bfS_\bfR$-orthogonality also holds exactly in general since $\bfC^\star_\bfR$ are eigenvectors of different eigenvalues in general.
This \eqnref{hf-force-simp} is also noted and acknowledged in the original paper of Pulay force~\citep{pulay1969ab} (p.199, first equation in Sec.~2, for which he cited Hellmann's~\citep{hellman1937einfuhrung} and Feynman's work~\citep{feynman1939forces}).

However, $(\nabla_\bfR E_\bfR) (\bfC^\star_\bfR)$ and $(\nabla_\bfR E_\bfR) (\bfP^\star_\bfR)$ are not the Hellmann-Feynman force (\secref{xc-approx-adiab-conn}).
Noting that $E_\bfR(\bfC) := E_\bfR(\bfP = \bfC \bfC^\dagger)$, in \eqnref{energy-dm-matrix} or equivalently \eqnref{energy-dm-matrix-flatten}, only a part of ${\bfPb^\star_\bfR}^\dagger \nabla_\bfR \bfVb_{\Ne,\bfR}$ is the Hellmann-Feynman force:
\begin{align}
  (\nabla_\bfR E_\bfR) (\bfP^\star_\bfR)
  ={} & {\bfPb^\star_\bfR}^\dagger \nabla_\bfR \bfTb_\bfR + \frac{1}{2} {\bfPb^\star_\bfR}^\dagger \nabla_\bfR (\sfbJb_\bfR - \sfbKb_\bfR) \bfPb^\star_\bfR \\
  &{} + \underbrace{ \underbrace{ {\bfPb^\star_\bfR}^\dagger \big( \obraket*{\alpha'_\bfR}{\nabla_\bfR V_{\Ne,\bfR}}{\alpha_\bfR} \big)^{(\alpha'\alpha)} }_{\text{Hellmann-Feynman force}} + {\bfPb^\star_\bfR}^\dagger \big( \obraket*{\nabla_\bfR \alpha'_\bfR}{V_{\Ne,\bfR}}{\alpha_\bfR} + \obraket*{\alpha'_\bfR}{V_{\Ne,\bfR}}{\nabla_\bfR \alpha_\bfR} \big)^{(\alpha'\alpha)} }_{{\bfPb^\star_\bfR}^\dagger \nabla_\bfR \bfVb_{\Ne,\bfR}}.
  \label{eqn:pulay-force}
\end{align}
The rest part of $(\nabla_\bfR E_\bfR) (\bfP^\star_\bfR)$ is called the wavefunction force in~\citep{pulay1969ab}.
In principle, the Hellmann-Feynman theorem can be applied and the Hellmann-Feynman force can well approximate the true force.
But under atomic basis which is incomplete, the condition that the Hellmann-Feynman theorem holds (true ground-state wavefunction) does not hold exactly, causing error in the force.
The error can be corrected by other terms in \eqnref{pulay-force}, which appear due to the dependency of the basis functions on atom coordinates.

\subsection{Correlation Missed by the Hartree-Fock Method} \label{sec:hf-correlation}

Define the \textbf{correlation energy} $E^\HF_\tnC$ as the difference between the true ground-state energy and the minimal HF energy.
It reflects the approximation gap due to the Slater-determinant form (the HF form) of wavefunction.
Since the Slater determinant is the wavefunction of non-interacting systems, this gap then arises from the effect of interaction, hence called ``correlation''.

From~\citep{koch2001chemist},
``Electron correlation is mainly caused by the \textbf{dynamical correlation} (Sinano\u{g}lu, 1964, according to~\citep{tsuneda2014density}), \ie, the instantaneous repulsion of the electrons, which is not covered by the effective HF potential. Pictorially speaking, the electrons get often too close to each other in the Hartree-Fock scheme, because \emph{the electrostatic interaction is treated in only an average manner}.''
``The second main contribution to $E^\HF_\tnC$ is the \textbf{non-dynamical} or \textbf{static correlation}. It is related to the fact that in certain circumstances the ground-state Slater determinant is not a good approximation to the true ground-state, because there are other Slater determinants with comparable energies. A typical example is provided by one of the famous laboratories of quantum chemistry, the $\rmH_2$ molecule.''

RHF deviates from the experimental truth as the inter-H distance increases, where dynamical correlation becomes a minor issue (since the Coulomb interaction is weak).
When choosing the one-electron orbitals as $\{\phi_\tngg(\rrv) \delta^\uparrow_s, \phi_\tngg(\rrv) \delta^\downarrow_s\}$
where the molecular orbital $\phi_\tngg(\rrv) := \frac{1}{\sqrt{2}} (\phi_\tnL(\rrv) + \phi_\tnR(\rrv))$ is composed of atomic orbitals of the left-H and right-H,
the Slater determinant gives equal weights to $\{\rmH^{\uparrow} \rmH^{\downarrow}, \rmH^{\downarrow} \rmH^{\uparrow}, \rmH^{\uparrow\downarrow} \rmH, \rmH^{\downarrow\uparrow} \rmH\}$
(\eqnref{slater-determ-rhf} with $N^\uparrow = N^\downarrow = 1$ gives the same value when $(s_1,s_2)$ is $(\uparrow,\downarrow)$ or $(\downarrow,\uparrow)$, and $\phi_\tngg$ assigns a spin value to either atoms with equal weights),
while the latter two states are less likely for distant H atoms.
From another perspective, another basis for the atomic orbitals combination $\phi_\tnuu(\rrv) := \frac{1}{\sqrt{2}} (\phi_\tnL(\rrv) - \phi_\tnR(\rrv))$ should be considered,
and the dominant states $\{\rmH^{\uparrow} \rmH^{\downarrow}, \rmH^{\downarrow} \rmH^{\uparrow}\}$ for distant H atoms is then representable as the summation of the Slater determinants using both molecular orbitals.

UHF ameliorates this problem a lot, since it allows the two dominant states $\{\rmH^{\uparrow} \rmH^{\downarrow}, \rmH^{\downarrow} \rmH^{\uparrow}\}$ where the electrons are unpaired.
But its accuracy also suffers from not being the total-spin eigenfunction.

%% file: dft.tex
\section{Density Functional Theory} \label{sec:dft}

\subsection{The Hohenberg-Kohn Theorems} \label{sec:dft-hk}

The two HK theorems (1964)~\citep{hohenberg1964inhomogeneous} follow the variational principle of the Schr\"{o}dinger equation.
They laid the foundation of DFT that it makes physical sense to take the ground-state density as the basic variable.
Define the system-independent operator as:
\begin{align}
  \Fh := \Th + \Wh_\ee.
\end{align}

\paragraph{The first theorem:}
\emph{The non-degenerate ground-state density $\rho_0$ of a system uniquely determines the external potential $V_{\ext[\rho_0]}(\rrv)$ (up to a constant) of the system,
consequently the Hamiltonian $\Hh_{[\rho_0]} = \Th + \Wh_\ee + \Vh_{\ext[\rho_0]}$ (\cf, \eqnref{hamiltonian}) and all of the entire the system.}
($V_\ext(\rrv)$ is restricted to a local, one-electron potential s.t. $\obraket*{\psi}{\Vh_\ext}{\psi} = \bbE_{\rho_{[\psi]}}[V_\ext]$.)
\vspace{2pt}

Otherwise, there is a different (more than a constant) $V'_\ext$ whose ground-state $\psi'_0$ yields the same density $\rho_{[\psi'_0]} = \rho_{[\psi_0]}$.
By the variational principle, $E_0 = \obraket*{\psi_0}{\Fh + \Vh_\ext}{\psi_0} < \obraket*{\psi'_0}{\Fh + \Vh_\ext}{\psi'_0}$, while $E'_0 = \obraket*{\psi'_0}{\Fh + \Vh'_\ext}{\psi'_0}$.
Subtracting them yields $E_0 - E'_0 < \obraket*{\psi'_0}{\Vh_\ext - \Vh'_\ext}{\psi'_0} = \bbE_{\rho_{[\psi'_0]}} [V_\ext - V'_\ext]$.
Symmetrically, $E'_0 - E_0 < \bbE_{\rho_{[\psi_0]}} [V'_\ext - V_\ext]$.
Adding the two inequalities yields a contradiction $0 < \bbE_{\rho_{[\psi'_0]}} [V_\ext - V'_\ext] - \bbE_{\rho_{[\psi_0]}} [V_\ext - V'_\ext] = 0$.

This theorem indicates the ground-state wavefunction $\psi_{[\rho_0]}$ of a system is a functional of the ground-state density $\rho_0$ of the system,
hence the ground-state energy: $\exists \text{ functional } E_0[\rho_0]$.
Since $E_0[\rho_0] = \obraket*{\psi_{[\rho_0]}}{\Fh}{\psi_{[\rho_0]}} + \bbE_{\rho_0} [V_{\ext[\rho_0]}]$
where the second term is an explicit functional of $\rho_0$ if we know the $V_\ext$ that yields $\rho_0$,
we know \emph{if $\rho_0$ is the ground-state density of some system, then the ``kinetic + ee'' energy of that system is uniquely determined by $\rho_0$}.
This defines the \textbf{Hohenberg-Kohn functional} $F^\HK[\rho_0]$.

By now, the $\rho$ is still required to be the ground-state density of \emph{some} system (\ie $\exists V_\ext$ s.t. this $\rho$ is the ground-state density).
Characterizing such $\rho$ is called the \textbf{V\textsubscript{ext}-representability} problem (\ie, finding the pre-image set of $V_{\ext[\rho]}$; Lieb's $\scA_N$ set).
A commonly satisfactory, much \emph{weaker} condition is \textbf{N-representability}, meaning that $\rho$ stems from an antisymmetric wavefunction of $N$ electrons (Lieb's $\scI_N$ set; for the Levy-Lieb functional).

\paragraph{The second theorem:}
\emph{The ground-state energy $E[V_\ext]$ and density $\rho_{[V_\ext]}$ of a system specified by $V_\ext(\rrv)$ are:}
\begin{align}
  (E[V_\ext], \rho_{[V_\ext]}) = \pminargmin_{\rho: \text{V\textsubscript{ext}-repr}} \lrbrace*[\big]{ E[\rho] := F^\HK[\rho] + \bbE_\rho[V_\ext] }.
  \label{eqn:hk-formulation}
\end{align}

The theorem follows the variational principle.
Note that it is inappropriate to minimize $E_0[\rho] = F^\HK[\rho] + \bbE_\rho[V_{\ext[\rho]}]$, in which $V_\ext$ also changes with $\rho$ so the minimization process is not for a fixed system.

\paragraph{The Levy constrained-search approach (1979)~\citep{levy1979universal}.}
The variational principle can be carried out in two stages:
\begin{align}
  E[V_\ext] = \min_{\psi} \obraket*[\big]{\psi}{\Fh + \Vh_\ext}{\psi}
  = \min_{\rho: \text{N-repr}} \min_{\psi: \rho_{[\psi]} = \rho} \obraket*[\big]{\psi}{\Fh + \Vh_\ext}{\psi}
  = \min_{\rho: \text{N-repr}} \lrparen*[\Big]{
    \underbrace{ \min_{\psi: \rho_{[\psi]} = \rho} \obraket*[\big]{\psi}{\Fh}{\psi} }_{=: F[\rho], \textbf{ universal functional}}
    + \bbE_{\rho}[V_\ext]
  }.
  \label{eqn:levy-formulation}
\end{align}
The \textbf{universal functional} $F[\rho] = (T + E_\ee)[\rho]$ (Lieb's $F^\Levy[\rho]$) (allows all N-representable $\rho$; Lieb's $\scI_N$) differs from $F^\HK[\rho]$ (allows only V\textsubscript{ext}-representable $\rho$; Lieb's $\scA_N$) only in the domains.
They coincide for V\textsubscript{ext}-representable $\rho$.
This formulation recovers the two HK theorems, and also lifts the non-degenerate restriction.

\paragraph{Corollaries}
\itemone For a non-interacting system described by \eqnref{hamiltonian-nonintera} whose solutions are Slater determinants (see the Explanation of A2 in \secref{hf-wavefn-uhf}), there is no $\Wh_\ee$ term.
  So the universal/HK functional in this case is \emph{a functional of density representing the kinetic energy of a non-interacting system, denoted as $T_\tnS[\rho]$}, although its explicit form in $\rho$ is unknown.
  When the non-interacting system is constructed to recover the same density as an (strictly) interacting system, $T_\tnS < T$~\citep{koch2001chemist}.

\itemtwo Note that from \eqnref{energy-decomp}, $F[\rho] = \min_{P^{12}_{1'2'}: \rho_{[P^{12}_{1'2'}]} = \rho} T[P^1_{1'}] + E_\ee[\rho_{12}]
  = \min_{P^{12}_{1'2'}: \rho_{[P^{12}_{1'2'}]} = \rho} T[P^1_{1'}] + E_\XC[\rho; h_\XC] + J[\rho]$.
  Since $J[\rho]$ is naturally a functional of $\rho$ thus fixed in the minimization,
  $(T + E_\XC)[\rho] := \min_{P^{12}_{1'2'}: \rho_{[P^{12}_{1'2'}]} = \rho} T[P^1_{1'}] + E_\XC[\rho; h_\XC]$ is also a density functional.

\paragraph{Some details.}
There are two ways to define density functionals (see Lieb's paper~\citep{lieb1983density} or \secref{dft-lieb} for more rigorous details):

\itemone $F^\HK[\rho] := F[\psi_{[\rho]}^\HK]$, $T^\HK[\rho] := T[\psi_{[\rho]}^\HK]$ (this is not Lieb's $T_\tnS^\KS$; this considers interaction),
  where $\psi_{[\rho]}^\HK$ satisfies $\psi_{[\rho]}^\HK \in \argmin_{\psi: \text{N-repr}} F[\psi] + \bbE_{\rho_{[\psi]}} [V_{\ext[\rho]}]$ and $\rho_{[\psi_{[\rho]}^\HK]} = \rho$.
  By construction, they are defined only for $\rho$ that is some potential's ground-state density (Lieb's $\scA_N$).

\itemtwo $F[\rho] := \min_{\psi: \text{N-repr}, \rho_{[\psi]} = \rho} F[\psi]$, $T_\tnS[\rho] := \min_{\psi: \text{N-repr}, \rho_{[\psi]} = \rho} T[\psi]$ (Lieb's $F^\Levy[\rho]$ and $T_\tnS^\Levy[\rho]$).
  Denote the minimizers as $\psi_{[\rho]}^{(F)}$ and $\psi_{[\rho]}^{(T)}$.
  By construction, they can be defined for all N-representable $\rho$ (Lieb's $\scI_N$), which contains $\scA_N$.

Now consider their coincidence and their eligibility for finding $E_0$ for a given $V_\ext$.

Obviously, for $\rho \in \scA_N$, $F[\rho] \le F^\HK[\rho]$ since $F[\rho]$ is the minimal achievable value by $\rho$.
If equality does not hold, then $F[\psi_{[\rho]}^{(F)}] + \bbE_{\rho_{[\psi_{[\rho]}^{(F)}]}} [V_{\ext[\rho]}]
= F[\psi_{[\rho]}^{(F)}] + \bbE_{\rho} [V_{\ext[\rho]}]
< F[\psi_{[\rho]}^\HK] + \bbE_{\rho} [V_{\ext[\rho]}]
= F[\psi_{[\rho]}^\HK] + \bbE_{\rho_{[\psi_{[\rho]}^\HK]}} [V_{\ext[\rho]}]$
which violates that $\psi_{[\rho]}^\HK$ is a minimizer.
So $F[\rho] = F^\HK[\rho]$ on $\scA_N$, and $\psi_{[\rho]}^\HK \in \argmin_{\psi: \text{N-repr}, \rho_{[\psi]} = \rho} F[\psi]$.
The eligibility is obvious by considering $F[\rho]$ in Levy's two-stage formulation, \eqnref{levy-formulation}.

Obviously, $T_\tnS[\rho] = T[\psi_{[\rho]}^{(T)}] \le T[\psi_{[\rho]}^{(F)}] = T[\psi_{[\rho]}^\HK] = T^\HK[\rho]$, and ``$=$'' holds only when $W_\ee = \const$ (non-interacting case).
As for eligibility, in a general (interacting) case, $T_\tnS[\rho] + E_\ee[\rho]
:= \min_{\psi: \rho_{[\psi]} = \rho} T[\psi] + \min_{\psi: \rho_{[\psi]} = \rho} E_\ee[\psi]
< \min_{\psi: \rho_{[\psi]} = \rho} T[\psi] + E_\ee[\psi]
= F[\rho]$, so using $T_\tnS[\rho] + E_\ee[\rho]$ is not eligible.

The results also hold when replacing $E_\ee$ with $E_\XC$, since the difference $J[\rho]$ is an explicit functional of $\rho$.

Does $T_\tnS[\rho_{[\psi_{[V]}^\KS]}] = T[\psi_{[V]}^\KS]$ where $\psi_{[V]}^\KS := \argmin_{\psi_\tnS} T[\psi_\tnS] + J[\psi_\tnS] + E_\XCb[\psi_\tnS] + \bbE_{\rho_{[\psi_\tnS]}}[V]$?
Obviously l.h.s $\le$ r.h.s by definition of $T_\tnS$.
Also, the additional terms $J$, $E_\XCb$ and $\bbE_{\rho}[V]$ are all functionals of $\rho$ (for $E_\XCb$, it is defined as $(T + E_\XC)[\rho] - T_\tnS[\rho]$),
so following a similar argument as for $F^\HK = F$ above, $\psi_{[V]}^\KS$ also minimizes $T[\psi]$ for the given density $\rho_{[\psi_{[V]}^\KS]}$.
So the kinetic energy computed from the DFT orbital solution can be taken as the label for learning $T_\tnS$.
More generally, this conclusion also holds even when the specifically chosen approximation $E_\XCb$ is not exact as long as it is a density functional (the inexactness only biases the $\rho$ solution from fitting the given system).

\subsection{The Kohn-Sham Approach} \label{sec:dft-ks}

\paragraph{Background}
The Thomas-Fermi model (1927)~\citep{thomas1927calculation, fermi1928statistische} and extensions ``fail miserably when results better than mere qualitative trends are the target''~\citep{koch2001chemist}.
The scheme approximates the kinetic energy by that of the uniform electron gas,
\footnote{Why regarded ``uniform'' while $\rho(\rrv)$ changes spacially? Because it is not regarding the density as uniform, but taking the energy density at $\rrv$ as the energy density as if the density is everywhere equal to the value $\rho(\rrv)$.}
\begin{align}
  T^\TF[\rho] := \frac{3}{10} (3 \pi^2)^\frac{2}{3} \int \ud\rrv \, \rho^{\frac{5}{3}}(\rrv).
  \label{eqn:kinfn-thomas-fermi}
\end{align}
This approximation does not meet chemical interest which is on the bonds arising from non-uniform electron density.
Calls for more accurate kinetic energy estimation.
The Kohn-Sham approach~\citep{kohn1965self} is developed in 1965.

\paragraph{Overview}
``Introduce a fictitious, \emph{non-interacting} reference system such that the major part of the kinetic energy can be computed to good accuracy.
The remainder is merged with the non-classical contributions $E_\XC$ to the electron-electron repulsion, which are also unknown but usually fairly small.
By this method, as much information as possible is computed exactly, leaving only a small part of the total energy to be determined by an approximate functional.''~\citep{koch2001chemist}

\paragraph{Description}
Finding the ground-state energy of a given system with potential $V_\ext$ can be done by varying density to minimize \eqnref{hk-formulation} or \eqnref{levy-formulation}, but the problem is that $F^\HK$ or $F$, or equivalently $(T+E_\XC)[\rho]$, is unknown.
But we know the exact wavefunction of a non-interacting system $\Hh = \Th + \Vh_\tnS$ (\eqnref{hamiltonian-nonintera}) is in the form of the Slater determinant $\psi_\tnS$ (see the Explanation of A2 in \secref{hf-wavefn-uhf}).
Choose the effective one-electron potential $V_\tnS$ such that the ground-state density of the reference system $\rho_\tnS(\rrv) = \sum_{k \in [N]} \sum_s \lrvert{\phi_k^s(\rrv)}^2$ (\eqnref{density-hf}) equals that $\rho_0(\rrv)$ of the real system of interacting electrons, $\Hh = \Th + \Vh_\Ne + \Wh_\ee$ (\eqnref{hamiltonian}).
If the energy functionals of the two systems coincide, %
the real ground-state energy can also be recovered by the reference system.
(The significance of the HK theorems here is that it tells us it suffices to use a reference system that shares the energy as a functional only of the density,
but not have to be of the wavefunction which is definitely not possible (the real one is interacting while the reference's is not; particularly the real one has correlation for antiparallel-spin electrons while the reference's does not; see \eqnref{pairdensity-hf-antiparallel}).)

To achieve this, reformulate the \emph{real, target} total energy \eqnref{energy-decomp} by leveraging Corollaries~(1) and~(2) that both $T_\tnS[\rho]$ and $(T + E_\XC)[\rho]$ are functionals of density $\rho$:
\begin{align}
  E[\rho] = \overbrace{ (T + E_\XC)[\rho] + \overbrace{E_\Ne[\rho]}^{E_\ext^{\lambda=1}[\rho]} + J[\rho]
  }^{\text{interacting, real, physical view}}
  = \overbrace{ T_\tnS[\rho] + \overbrace{ E_\Ne[\rho] + J[\rho] + \underbrace{(T + E_\XC)[\rho] - T_\tnS[\rho]}_{=: E_\XCb[\rho]} }^{E_{\ext,\tnS}[\rho] = E_\ext^{\lambda=0}[\rho]}
  }^{\text{non-interacting, reference, effective view}}.
  \label{eqn:energy-ks}
\end{align}
This $E_\XCb[\rho]$ overloads the name ``exchange-correlation energy'' in the KS scheme.
It contains not only the non-classical effects of self-interaction correction, exchange and correlation (\ie, $E_\XC$), but also a portion belonging to the kinetic energy (roughly $T^\HK[\rho] - T_\tnS[\rho]$).
By construction, it is a functional of electron density $\rho$, unlike $E_\XC$ in the general case,
so taking its variation w.r.t orbitals gives a \emph{local} effective potential, which is what needed for a non-interacting reference system.
This enables finding the ground-state energy by varying orbitals:
\begin{align}
  \min_{\rho: \text{N-repr}} E[\rho]
  \stackrel{\text{\eqnref{energy-ks}}}{=} {} &
  \min_{\rho: \text{N-repr}} \lrparen*[\Big]{ \min_{\psi_\tnS: \rho_{[\psi_\tnS]} = \rho} T[\psi_\tnS] + E_{\ext,\tnS}[\rho] }
  = \min_{\rho: \text{N-repr}} \lrparen*[\Big]{ \min_{\psi_\tnS: \rho_{[\psi_\tnS]} = \rho} T[\psi_\tnS] + E_{\ext,\tnS}[\rho_{[\psi_\tnS]}] } \\
  ={} & \min_{\psi_\tnS: \text{N-repr}} T[\psi_\tnS] + E_{\ext,\tnS}[\rho_{[\psi_\tnS]}],
\end{align}
and similarly to \eqnref{hf-eigen}, taking the variation of the objective w.r.t orbital $\phi_i$ subject to the normalization constraint gives:
\begin{gather}
  \ffh^\KS \phi_i := \big( \Th + \Vh_\tnS^\KS \big) \phi_i = \veps_i \phi_i, \quad \text{where}
  \label{eqn:ks-eigen} \\
  V_\tnS^\KS(\rrv_1) := \fracdelta{E_{\ext,\tnS}[\rho]}{\rho}(\rrv_1)
  = \underbrace{ -\sum\nolimits_{A \in [N_\tnA]} \frac{Z_A}{r_{1A}} }_{= V_\Ne(\rrv_1)}
  + \underbrace{ \int \ud\rrv_2 \, \frac{\rho(\rrv_2)}{r_{12}} }_{= V_\tnJ(\rrv_1) \text{ (see \eqnref{VJ})}}
  + \underbrace{ \fracdelta{E_\XCb[\rho]}{\rho} (\rrv_1) }_{=: V_\XCb(\rrv_1)}.
  \label{eqn:vs-ks}
\end{gather}
Note that the potential functions $V_\tnJ$, $V_\XCb$ and $V_\tnS^\KS$ are all determined by the density $\rho$; hence is the Fock operator $\ffh^\KS$.

\paragraph{Correlations and holes}
Referring to \eqnref{exc}, we define the exchange-correlation hole in the KS scheme via:
\begin{align}
  E_\XCb[\rho] =: \frac{1}{2} \int \ud\rrv_1 \ud\rrv_2 \, \frac{\rho(\rrv_1) h_\XCb(\rrv_2|\rrv_1)}{r_{12}}.
  \label{eqn:hxc-ks}
\end{align}
This is indeed the difference from the true ground-state energy.
Similar to the HF case, the exchange energy can also be handled by the exchange hole $h_\tnX^\KS := h_\tnX^\HF$ (see \eqnref{hx-spacial-hf}) using the KS-optimized orbitals.
The corresponding Coulomb hole is accordingly defined as $h_\tnCb^\KS := h_\XCb - h_\tnX^\KS$ which also accounts for the correlation in the kinetic energy.

\paragraph{Difference from the Hartree-Fock method}
Comparing the generalized eigenvalue problems \eqnref{hf-eigen} and \eqnref{ks-eigen}, the difference lies in the exchange-correlation operator:
the HF method uses the exchange operator $-\sum_{j \in [N]} \Kh^{(j)}$ which is non-local (cannot be written as the integral of the operand with a potential) and misses some correlation (\secref{hf-correlation}),
while the KS scheme uses a local exchange-correlation potential $V_\XCb(\rrv)$ (determined by $\rho$) that also takes into account the Coulomb and kinetic correlation.
``The Kohn-Sham approach is in principle exact!''~\citep{koch2001chemist}
Due to the use of the local operator, ``the KS formulation has a structure actually formally less complicated than the HF approximation''~\citep{koch2001chemist}.
The advantage of KSDFT over HF lies in performance rather than computational cost (it also solves orbitals). It is cheaper than post-HF methods \eg, CI, CCSD.

By definition, the ground-state energy can be recovered by both HF and KS with respective correlation energy:
$E_0 = E[\psi_{\tnS0}^\HF] + E_\tnC^\HF = E[\psi_{\tnS0}^\KS] + E_\tnCb^\KS$,
where $E[\psi_\tnS] = T[\psi_\tnS] + E_\Ne[\rho_{[\psi_\tnS]}] + J[\rho_{[\psi_\tnS]}] + E_\tnX^\HF[\psi_\tnS]$ is the energy under Slater determinant (see \eqnref{energy-hf}; assume A2.2 so that $E_\tnX^\HF = E_\XC^\HF$),
and since $E[\psi_{\tnS0}^\HF]$ is the minimal achievable energy by a Slater determinant, we have $E_\tnC^\HF > E_\tnCb^\KS$.
From another perspective, by construction of KS, $\rho_{\tnS0}^\KS = \rho_0$ while $\rho_{\tnS0}^\HF \ne \rho_0$.
So $h_\tnC^\HF(\rrv_2|\rrv_1) = (\rho_0(\rrv_1) - \rho_{\tnS0}^\HF(\rrv_2)) + (h_\XCb(\rrv_2|\rrv_1) - h_\tnX^\HF(\rrv_2|\rrv_1)) = \rho_0 - \rho_{\tnS0}^\HF + h_\tnCb^\KS$,
since the deviation in density also makes HF introduce error from the $T_\tnS$, $E_\Ne$ and $J$ terms.
The correlation energy $E_\tnC^\HF$ is simply ignored in HF optimization, while $E_\tnCb^\KS$ is approximated in KS together with $E_\tnX^\KS$ through $E_\XCb$.

\paragraph{Discussions}
Since the use of orbitals in KS is for approximating (the major part of) the kinetic energy, it does not require as expressive an orbital function class as HF does, whose target is indeed the true wavefunction.

Define the \textbf{non-interacting pure-state-V\textsubscript{S} representability} of a possibly interacting system as: whether the non-interacting $N$-electron ground-state that shares the same density as the interacting system can be generated by a single Slater determinant built from orbitals that are obtained as the $N$ energetically lowest lying orbitals of a simple local Kohn-Sham potential $V_\tnS$.
An interpretation could be, although the ground-state density of the real, interacting system can be recovered by that of a non-interacting reference system with some effective $V_\tnS$, the ground-state wavefunction of the reference system may need multiple determinants or virtual excited orbitals to be expressed (a single determinant with $N$ lowest orbitals may not be the only solution).
Recall \secref{energy-hf} that a single determinant in the HF scheme missed the left-right correlation in a $\rmH_2$ molecule hence overestimates the energy.
But given an exact $V_\XCb$, a single determinant is obtained as the solution to the KS reference system [Gritsenko \& Baerends, 1997] (but the KS orbitals are far different from the HF ones, and form a bad wavefunction approximator), \ie, the $\rmH_2$ system is non-interacting pure-state-V\textsubscript{S} representable.
There also exist exceptions, \eg the density of the non-degenerate $^1\Sigma_\rmgg^+$ ground state of the $\rmC_2$ molecule cannot be represented as a single determinant KS solution [Schipper, Gritsenko, \& Baerends, 1998].
That system is instead \textbf{non-interacting ensemble-V\textsubscript{S} representable} using a small number of accidentally degenerated determinants.

In principle, even for a truly open-shell system, using restricted, spin-pair-shared KS orbitals (RKS) suffices from the derivation, unless the Hamiltonian explicitly depends on spin.
However in practice, using unrestricted, spin-dependent orbitals (UKS) is required to account for the open-shell problem.
Unrestricted techniques give qualitatively correct energies but wrong densities, whereas restricted methods show the opposite behavior.
The deviation of $\lrangle*{\Svh^2}$ by UKS is in most cases considerably less significant than by UHF, as KS does not aim to approximate the wavefunction.

\subsection{Approximating the Exchange-Correlation Energy Functional \texorpdfstring{$E_\XCb[\rho]$}{} in the KS Formalism} \label{sec:dft-exc}

``Unlike in wavefunction-based methods, in density functional theory there is no systematic way towards improved approximate functionals.''~\citep{koch2001chemist}
There are constraints however, like normalization, boundary (cusp), scaling, and asymptotes.

\subsubsection{Exact expression via adiabatic connection} \label{sec:xc-approx-adiab-conn}
Developed by \citet{harris1984adiabatic}.
The central idea of the KS scheme is the density-minimizer equivalence between the real, interacting system $\Hh = \Th + \Vh_\Ne + \Wh_\ee$ (see \eqnref{hamiltonian}),
and the carefully chosen, non-interacting reference system $\Hh = \Th + \Vh_\tnS^\KS$ (see \eqnsref{hamiltonian-nonintera,ks-eigen}). %
A simple connection between them can be constructed as $\Hh_\lambda := \Th + \Vh_\lambda + \lambda \Wh_\ee$,
where $\lambda \in [0,1]$ is the \emph{coupling strength parameter},
and $\Vh_\lambda: \obraket*{\psi}{\Vh_\lambda}{\psi} = \bbE_{\rho_{[\psi]}(\rrv)}[V_\lambda(\rrv)] = E_\ext^\lambda[\rho_{[\psi]}]$ is the external potential operator that makes the ground-state of $\Hh_\lambda$ recover the real ground-state density $\rho_0$.
By construction, $E_{\lambda=0} = E_{\lambda=1}$ (see \eqnref{energy-ks}), and
$V_{\lambda=0} = V_\tnS^\KS$, $V_{\lambda=1} = V_\Ne$ (all local).
Since the density %
is kept the same along the process, it is called \emph{adiabatic}.
Although the one-electron density is constantly $\rho_0$, the pair density and thus the hole depend on $\lambda$.

\textbf{Hellmann–Feynman theorem}~\citep{hellman1937einfuhrung,feynman1939forces}:
\emph{$\fracdiff{E_\lambda}{\lambda} = \obraket*[\big]{\psi_\lambda}{\fracdiff{\Hh_\lambda}{\lambda}}{\psi_\lambda}$, where $(\psi_\lambda, E_\lambda)$ is an eigenfunction-eigenvalue pair of the Hamiltonian $\Hh_\lambda$, and $\psi_\lambda$ is normalized}
(\href{https://handwiki.org/wiki/Physics:Hellmann%E2%80%93Feynman_theorem}{handwiki}).
To see this, note $\obraket*{\psi_\lambda}{\Hh_\lambda}{\psi_\lambda} = E_\lambda \braket*{\psi_\lambda}{\psi_\lambda} = E_\lambda$, so
$\fracdiff{E_\lambda}{\lambda} = \obraket*[\big]{\psi_\lambda}{\fracdiff{\Hh_\lambda}{\lambda}}{\psi_\lambda}
+ \obraket*[\big]{\fracdiff{\psi_\lambda}{\lambda}}{\Hh_\lambda}{\psi_\lambda} + \obraket*[\big]{\psi_\lambda}{\Hh_\lambda}{\fracdiff{\psi_\lambda}{\lambda}}$.
Noting $\Hh_\lambda$ is Hermite and $E_\lambda$ is real, we know that
$\fracdiff{E_\lambda}{\lambda} = \obraket*[\big]{\psi_\lambda}{\fracdiff{\Hh_\lambda}{\lambda}}{\psi_\lambda} + E_\lambda \braket*[\big]{\fracdiff{\psi_\lambda}{\lambda}}{\psi_\lambda} + E_\lambda \braket*[\big]{\psi_\lambda}{\fracdiff{\psi_\lambda}{\lambda}}
= \obraket*[\big]{\psi_\lambda}{\fracdiff{\Hh_\lambda}{\lambda}}{\psi_\lambda} + E_\lambda \fracdiff{\braket{\psi_\lambda}{\psi_\lambda}}{\lambda}$.
If $\psi_\lambda$ keeps normalized as $\lambda$ varies, then the second term is zero.
Alternatively, the variational principle indicates $\fracdelta{}{\psi} E_\lambda[\psi=\psi_\lambda] = \mu \fracdelta{\braket{\psi_\lambda}{\psi_\lambda}}{\psi}$, where $E_\lambda[\psi] := \frac{\obraket*{\psi}{\Hh_\lambda}{\psi}}{\braket*{\psi}{\psi}}$, and the second term comes from the constraint $\braket{\psi}{\psi} = 1$ during optimization.
So $\fracdiff{E_\lambda}{\lambda} = \fracdiff{E_\lambda[\psi=\psi_\lambda]}{\lambda}
= \fracpartial{E_\lambda[\psi=\psi_\lambda]}{\lambda} + \int \ud \bfxx \, \fracdelta{E_\lambda[\psi=\psi_\lambda]}{\psi} \fracdiff{\psi_\lambda}{\lambda}
= \fracpartial{E_\lambda[\psi=\psi_\lambda]}{\lambda} - \mu \int \ud \bfxx \, \fracdelta{\braket{\psi_\lambda}{\psi_\lambda}}{\psi} \fracdiff{\psi_\lambda}{\lambda}
= \obraket*[\big]{\psi_\lambda}{\fracdiff{\Hh_\lambda}{\lambda}}{\psi_\lambda} - \mu \fracdiff{\braket{\psi_\lambda}{\psi_\lambda}}{\lambda}$,
which is again $\obraket*[\big]{\psi_\lambda}{\fracdiff{\Hh_\lambda}{\lambda}}{\psi_\lambda}$ if $\psi_\lambda$ keeps normalized as $\lambda$ varies.

The theorem applies to the ground state $\psi_\lambda$ of $\Hh_\lambda := \Th + \Vh_\lambda + \lambda \Wh_\ee$ above of course, so $\fracdiff{E_\lambda}{\lambda}
= \obraket*[\big]{\psi_\lambda}{\fracdiff{\Vh_\lambda}{\lambda} + \Wh_\ee}{\psi_\lambda}
= \bbE_{\rho_0(\rrv)} \big[ \fracdiff{V_\lambda(\rrv)}{\lambda} \big] + \frac{1}{2} \bbE_{\rho_\lambda(\rrv_1,\rrv_2)}[1/r_{12}]
= \bbE_{\rho_0(\rrv)} \big[ \fracdiff{V_\lambda(\rrv)}{\lambda} \big] + J[\rho_0]
+ \frac{1}{2} \int \ud\rrv_1 \ud\rrv_2 \, \frac{\rho_0(\rrv_1) h_\XC^\lambda(\rrv_2|\rrv_1)}{r_{12}}$.
Consequently,
\begin{align}
  0 & {}= E_{\lambda=1} - E_{\lambda=0} = \int_0^1 \ud\lambda \, \fracdiff{E_\lambda}{\lambda}
  = \int_0^1 \ud\lambda \, \bbE_{\rho_0(\rrv)} \lrbrack{ \fracdiff{V_\lambda(\rrv)}{\lambda} }
  + J[\rho_0] + \frac{1}{2} \int_0^1 \ud\lambda \int \ud\rrv_1 \ud\rrv_2 \, \frac{\rho_0(\rrv_1) h_\XC^\lambda(\rrv_2|\rrv_1)}{r_{12}} \\
  & {}= \bbE_{\rho_0(\rrv)} [\overbrace{V_\Ne(\rrv)}^{V_{\lambda=1}}] - \bbE_{\rho_0(\rrv)} [\overbrace{V_\tnS^\KS(\rrv)}^{V_{\lambda=0}}]
  + \bbE_{\rho_0(\rrv)} [V_\tnJ(\rrv)] + \frac{1}{2} \int \ud\rrv_1 \ud\rrv_2 \, \frac{\rho_0(\rrv_1) \int_0^1 \ud\lambda \, h_\XC^\lambda(\rrv_2|\rrv_1)}{r_{12}} \\
  & {} \stackrel{\text{\eqnref{vs-ks}}}{=}
  -\bbE_{\rho_0(\rrv)} [V_{\XCb[\rho_0]}(\rrv)] + \bbE_{\rho_0(\rrv_1)} \lrbrack*[\Big]{ \frac{1}{2} \int \ud\rrv_2 \, \frac{\int_0^1 \ud\lambda \, h_\XC^\lambda(\rrv_2|\rrv_1)}{r_{12}} } \\
  & \stackrel[\text{from \eqnsref{vs-ks,hxc-ks}}]{\text{assume $V_{\XCb[\rho_0]}(\rrv_1) = \frac{1}{2} \int \ud\rrv_2 \, \frac{h_\XCb(\rrv_2|\rrv_1)}{r_{12}}$}}{\Longrightarrow}
  h_\XCb(\rrv_2|\rrv_1) = \int_0^1 \ud\lambda \, h_\XC^\lambda(\rrv_2|\rrv_1).
  \label{eqn:hole-ks-adiabatic}
\end{align}

The theorem is also useful for computing the atomic force (derivative w.r.t coordinates):
\begin{align}
  \nabla_{\rrv_A} E_0 = Z_A \bbE_{\rho_0(\rmxx)} \lrbrack*[\bigg]{ \frac{\rrv - \rrv_A}{\lrVert*{\rrv - \rrv_A}^3} } - Z_A \sum_{B \ne A} Z_B \frac{\rrv_B - \rrv_A}{\lrVert{\rrv_B - \rrv_A}^3}.
  \label{eqn:hellmann-feynman-force}
\end{align}
This is called the Hellmann-Feynman force.
Under a complete basis, and exact optimization, the condition for the theorem holds, \ie the wavefunction is an eigenstate and normalized.
But under an atomic basis which is incomplete, error appears due to inaccurate approximation to the ground state.
A more accurate estimation of the force is by directly taking the gradient of the total energy w.r.t atom coordinates, which gives other terms in addition to the Hellmann-Feynman force; see \eqnsref{hf-force-simp,pulay-force}.
The other terms are called the Pulay force~\citep{pulay1969ab}.
When using plane-wave basis, it is easier to approach completeness than atomic basis, and moreover, it is naturally independent of atom coordinates, hence the other terms (Pulay force) in \eqnref{pulay-force} vanishes.
This makes the Hellmann-Feynman force still a good approximation.

One may also think of applying the Hellmann-Feynman theorem to $\nabla_{\rrv_A} \obraket*{\psi^\star_{\tnS,\bfR}}{\ffh^\star_\bfR}{\psi^\star_{\tnS,\bfR}}$, where $\ffh^\star_\bfR$ is the converged Fock operator in a DFT (\eqnref{ks-eigen}) or HF calculation (\eqnref{hf-eigen}), and $\psi^\star_{\tnS,\bfR}$ is the corresponding Slater determinant of converged orbitals.
Since the orbitals are eigenstates of $\ffh^\star_\bfR$ and orthonormal, the theorem still holds.
But the problem is that $\obraket*{\psi^\star_{\tnS,\bfR}}{\ffh^\star_\bfR}{\psi^\star_{\tnS,\bfR}}$ is not the total energy; see \eqnref{energy-hf-eigen}.
Moreover, $\ffh^\star_\bfR$ in turn also depends on the current state $\psi^\star_{\tnS,\bfR}$ (see \eqnref{energy-hf-JKopr} for the Hartree-Fock case, where $\Jh$ and $\Kh$ depend on orbitals thus on $\bfR$ for a non-orthonormal, conformation-dependent basis), which then introduces additional dependency on $\bfR$.

\subsubsection{Direct approximations}

\paragraph{Local Density Approximation (LDA)}
``Uniform electron gas is the only system for which we know the form of the exchange and correlation energy functionals exactly or at least to very high accuracy.''~\citep{koch2001chemist}
Its $E_\XCb[\rho]$ is already used in~\citep{kohn1965self}.
The model assumes $E_\XCb^\LDA[\rho] = \bbE_{\rho(\rrv)}[V_{\XCb[\rho]}^\LDA(\rrv)] = \bbE_\rho[V_{\tnX[\rho]}^\LDA] + \bbE_\rho[V_{\tnCb[\rho]}^\LDA]$.

The first term handles two parallel-spin electrons, which are set apart by the exchange/Fermi hole.
For uniform electron gas, the hole $h_\tnX^\LDA(\rrv_2|\rrv_1)$ is spherically symmetric w.r.t $\rrv_2$ and is constantly $-\rho(\rrv_1)$ (due to Property~(b)) within a so-called \textbf{Wigner-Seitz} radius $r_\WS$ and is zero outside.
Property~(n) $-1 = \int \rrv_2 \, h_\tnX^\LDA(\rrv_2|\rrv_1) = - \rho(\rrv_1) \frac{4}{3} \pi r_\WS^3$ gives $r_\WS(\rrv_1) = \sqrt[3]{\frac{3}{4 \pi}} \rho^{-\frac{1}{3}}(\rrv_1)$.
Standard electrostatics shows the potential is inversely proportional to the radius, so
\begin{align}
  V_{\tnX[\rho]}^\LDA(\rrv) \propto -\rho^{\frac{1}{3}}(\rrv), \text{ and }
  E_\tnX^\LDA[\rho] = C_\tnX \textstyle\int \ud\rrv \, \rho(\rrv)^{\frac{4}{3}}
  \label{eqn:excfn-lda}
\end{align}
depends only on the \emph{local} values of the density.
This functional is frequently called \textbf{Slater exchange}.
Bloch (1929) and Dirac (1930) derived this and made the Thomas-Fermi-Dirac model,
\begin{align}
  E^\TFD[\rho,V] = T^\TF[\rho] + E_\Ne[\rho,V] + J[\rho] + E_\tnX^\LDA[\rho]
  \label{eqn:engfn-thomas-fermi-dirac}
\end{align}
(among the first DFT methods; kinetic correction and Coulomb correlation not considered).
Slater (1951) also derived this and developed the Hartree-Fock-Slater method that uses it with $C_\tnX \leftarrow \alpha C_\tnX$ ($\alpha = 2/3, 1$, \etc) in the HF method.

The second term does not have an explicit form.
Various approximations based on QMC by CA80, interpolation by VWN80, and by PW92.
LDA is sometimes specified as SVWN.
\textbf{Local Spin-Density approximation} (LSD): treat $\rho^\uparrow(\rrv)$ and $\rho^\downarrow(\rrv)$ separately.

LDA is good for equilibrium structure, harmonic frequencies and charge moments, but is not for energetical details (\eg, bond energies).
It is due to its notorious overbinding tendency.

\paragraph{Generalized Gradient Approximation (GGA)}
Supplement the density $\rho(\rrv)$ with the gradient information $\nabla \rho(\rrv)$ (or the dimensionless \textbf{reduced density gradient} or local inhomogeneity parameter $\varrho(\rrv) := \frac{\lrVert{\nabla \rho(\rrv)}}{\rho^{4/3}(\rrv)}$) to account for the non-homogeneity of the true electron density.
Gradient expansion approximation (GEA): $E_\XCb^\GEA[\rho^\uparrow, \rho^\downarrow] = \bbE_{\rho} [V_{\XCb[\rho^\uparrow, \rho^\downarrow]}^\GEA] + \sum_{s,s'} \bbE_{\rho} [C_\XCb^{s,s'} \varrho^s \varrho^{s'}] + \dots$.
GGA: truncate $V_\XCb^\GEA$, $C_\XCb^{s,s'}$, \etc to satisfy Properties~(b) and~(n).
``Non-local functional'' is a misleading and sloppy synonym.

Often, reformulate $E_\tnX^\GGA[\rho] = E_\tnX^\LDA[\rho] - \sum_s \int \ud\rrv \, f(\varrho^s(\rrv)) \rho^s(\rrv)^{4/3}$,
where $f$ is designed by B88 and variants FT97, PW91, CAM(A) and CAM(B) (1993), and as a rational function by B86, P86, LG93, PBE96.
GGA correlation functionals have even more complicated analytical forms and cannot be understood by simple physically motivated reasonings.
Examples: P86, PW91, and LYP (based on an accurate correlated theory by Colle \etal'75) (none considered non-dynamical effects).
Common combinations: BP86, BLYP, BPW91.

Meta-GGA (Perdew \etal, 1999): incorporates higher-order derivatives of density and (non-interacting) kinetic energy density.
LAP (Proynov \etal, 1994) involves Laplacian of density.
Filatov \& Thiel (1998).
Schmider \& Becke (1998b) extended their B97.
VSXC (van Voorhis \& Scuseria, 1998).
\textbf{Empirical density functionals} (EDF1; Adamson, Gill, Pople, 1998): heavy use of parameterization.

\subsubsection{Hybrid functionals}
Since self-interaction and exchange correlation are exactly handled in HF (\secref{hf-correlation}) via
$E_\tnX^\HF[\psi_\tnS] = -\frac{1}{2} \sum_{k,l \in [N]} \obraket*{\phi_k}{\Kh^{(l)}}{\phi_k}$ (see \eqnsref{exc-hf,JK-opr}; with A2.2 s.t. $h_\XC^\HF = h_\tnX^\HF$ \eqnref{hx-spacial-hf}),
and the orbitals (one-electron wavefunctions) $\{\phi_k\}_k$ are also available in KS (solved for computing $T_\tnS$),
we can leverage this exact yet tractable expression for the exchange part, and only need to find an approximation for $E_\tnCb[\rho]$ the (Coulomb) correlation part and the kinetic residual.
But this is not easy: the total hole is local but the exact exchange hole is global, so the chosen approximate correlation hole must also be global and has to carefully cancel the exchange hole at distance.

The adiabatic connection is a finer description: $E_\XCb = \int \ud\lambda \, E_\XC^\lambda$ (see \eqnref{hole-ks-adiabatic}),
and we know $\lambda = 0$ yields an effective non-interacting system so $E_\XC^{\lambda=0} = E_\tnX^\HF$ can be exactly computed,
while $\lambda = 1$ recovers the original system for whose $E_\XC^{\lambda=1}$ we have introduced a set of approximations.
Functionals hybridizing these two parts are called \textbf{DFT/HF hybrid functionals}, or \textbf{ACM (adiabatic connection method) functionals}.
Half-and-half combination (Becke, 1993a) $E_\XCb^\textnormal{HH} := \frac{1}{2} E_\tnX^\HF + \frac{1}{2} E_\XCb^\LDA$.
B3 (Becke, 1993b) $E_\XCb^\textnormal{B3} := \cdots$.
B3LYP (Stephens \etal, 1994) takes the same spirit and parameters as B3 (fitted on the G2 dataset), and is ``currently the most popular hybrid functional'':
$E_\XCb^\textnormal{B3LYP} := 0.20 E_\tnX^\HF + (1-0.20) E_\tnX^\textnormal{LSD} + 0.72 E_\tnX^\textnormal{B88} + 0.81 E_\tnCb^\textnormal{LYP} + (1-0.81) E_\tnCb^\textnormal{LSD}$.
B1 (Becke, 1996) $E_\XCb^\textnormal{B1} := E_\XCb^\textnormal{B1B95} + 0.28 (E_\tnX^\HF - E_\XCb^\textnormal{B1B95})$ uses one empirical parameter to achieve a similar performance, at the cost of involving kinetic energy density.
B97 (Becke, 1997) and B98 (Schmider \& Becke, 1998, fitted on the extended G2 dataset vs G2) and HCTH (and HCTH/120, HCTH/147, Boese \etal, 2000; use 4th-order vs 2nd-order)
separate the spins and introduce an elaborate power-series fitting procedure (many parameters).
Theoretical grounded fraction of $E_\tnX^\HF$ in $E_\XCb$ is between $0.20$ and $0.25$ by Perdew \etal (1996) and Burke \etal (1997).
Their PBE1PBE (or called PBE0): $E_\XCb^\textnormal{PBE} + 0.25 (E_\tnX^\HF - E_\tnX^\textnormal{PBE})$.

\subsubsection{Constraints on the functional}

\paragraph{Self-interaction}
For a one-electron system, there is no electron-electron interaction, and no need of correcting kinetic energy, $E_\XCb = E_\XC$.
So $E_\XCb[\rho] = -J[\rho]$ for $N = 1$.
None of the currently used functionals satisfies this.
Self-interaction corrected (SIC) functionals (PZ81) improves the results for atoms, but is worse on molecules, and the orbitals then do not share the same potential.

\paragraph{Asymptotic behavior}
$V_\XCb(\rrv) := \fracdelta{E_\XCb}{\rho}[\rho(\rrv)] \sim -\frac{1}{r} + E_\tnI + \veps_\textnormal{max}, r \to \infty$ (\eg, Tozer \& Handy, 1998),
where $\veps_\textnormal{max}$ is the highest occupied KS orbital energy (see \eqnref{hf-eigen}).
Crucial for properties that also rely on the virtual orbitals, \eg, response to electromagnetic fields, polarizability, excitation energy (Rydberg states).
None of the current popular choices satisfies this.
Their continuity w.r.t $N$ also violates the \emph{derivative discontinuity} (at integer $N$) in DFT (Perdew \etal, 1982),
which implies the potential should not vanish asymptotically.

Improvements:
LB94 potential (van Leeuwen \& Baerends, 1994) $\sim -1/r$ but violates derivative discontinuity and other issues.
HCTH(AC) (asymptotically corrected) (Tozer \& Handy, 1998) improves mentioned property prediction.
Hybrid functionals also ameliorate this problem, as $V_\tnX^\HF \sim -1/r$ and is discontinuous at integer $N$.

\subsection{Lieb's Paper (1983)} \label{sec:dft-lieb}

Ref: \citet{lieb1983density}.

\paragraph{Setups}
\begin{itemize}
  \item Let $Q$ be the number of spin values ($Q=2$ for electrons).
  \item Kinetic energy is taken as $T[\psi] = \sum_{i \in [N]} \int \ud\bfxx \, \lrvert{\nabla_i \psi(\bfxx)}^2$. ($\frac{1}{2}$ is missing, possibly due to closed-shell systems are considered)
    Given symmetry or antisymmetry, $T[\psi] = N \int \ud\bfxx \, \lrvert{\nabla_1 \psi(\bfxx)}^2$.
  \item $\rmDeltab^n := \Set{ \{\lambda_i\}_{i=1}^n }{ \sum_{i=1}^n \lambda_i = 1, \lambda_i \ge 0, \forall i \in [n] }$.
  \item $\Gamma_{[\psi]}(\bfxx; \bfxx') := \ket{\psi(\bfxx)} \bra{\psi(\bfxx')} = \psi(\bfxx) \psi^*(\bfxx')$ ($\Gamma$ admitting this form is called achievable by a pure state).

    For an $N$-particle density matrix $\Gamma$,
    $P_{[\Gamma]}(\rrv; \rrv') := N \sum_{s} \int \ud\rmxx_2 \cdots \ud\rmxx_N \, \Gamma((\rrv, s), \rmxx_2, \cdots, \rmxx_N; (\rrv', s), \rmxx_2, \cdots, \rmxx_N)$.

    $\rho_{[P]}(\rrv) := P(\rrv; \rrv)$.

    $\rho_{[\psi]}(\rrv) := (\rho_{[\cdot]} \circ P_{[\cdot]} \circ \Gamma_{[\psi]}) (\rrv) = N \sum_{s} \int \ud\rmxx_2 \cdots \ud\rmxx_N \, \lrvert{\psi((\rrv, s), \rmxx_2, \cdots, \rmxx_N)}^2$.
\end{itemize}

\paragraph{Variable spaces (\tabref{rho-spaces})}

\begin{itemize}
  \item Sobolev space $\bbW^{k,p}(\bbR^n) := \{\text{functions on $\bbR^n$ whose first $k$ weak derivatives are in $\bbL^p(\bbR^n)$}\}$ for $k \in \bbN$ and $1 \le p < \infty$
    with norm $\lrVert{f}_{\bbW^{k,p}} := ( \sum_{\bfii: \lrvert{\bfii} \le k} \lrVert{\partial_{\bfii} f}_{\bbL^p}^p )^{1/p}$,
    where $\bfii := (i_1, \cdots, i_n) \in \bbN^n$, $\lrvert{\bfii} := \sum_{j \in [n]} i_j$, and $\partial_{\bfii} f := \frac{\partial^{\lrvert{\bfii}} f}{\partial x_1^{i_1} \cdots \partial x_n^{i_n}}$.
    They are Banach spaces.
    $\bbH^k(\bbR^n) := \bbW^{k,2}(\bbR^n)$ with inner product $\lrangle{f, g}_{\bbH^k} := \sum_{\bfii: \lrvert{\bfii} \le k} \lrangle{\partial_\bfii f, \partial_\bfii g}_{\bbL^2}$ are Hilbert spaces.
    Note $\bbL^p(\bbR^n) = \bbW^{0,p}(\bbR^n)$.

    Sobolev embedding theorem: if $\frac{1}{p_1} - \frac{k_1}{n} = \frac{1}{p_2} - \frac{k_2}{n}$ and $k_1 > k_2$, $1 \le p_1 < p_2 < \infty$ and $p_1 < n$, then $\bbW^{k_1,p_1}(\bbR^n) \subseteq \bbW^{k_2,p_2}(\bbR^n)$, and the embedding is continuous.
    Particularly, $\bbH^1(\bbR^3) = \bbW^{1,2}(\bbR^3) \subseteq \bbW^{0,6}(\bbR^3) = \bbL^6(\bbR^3)$.

\end{itemize}

\begin{table}[h]
  \centering
  \setlength{\tabcolsep}{2pt}
  \renewcommand{\arraystretch}{1.5}
  \caption{Definitions and properties of spaces of $\psi$, $\rho$, $\Gamma$, $P$, and $V$.}
  \label{tab:rho-spaces}
  \begin{tabular}{p{7.0cm}cp{9.4cm}}
    \toprule
    Spaces of $\psi$ ($\supseteq$ order), $\rho$ ($\subseteq$ order), $\Gamma$, $P$, $V$ & Convex & Remarks
    \\ \midrule
    $\psi \in \bbL^2(\bbR^{3N}) = \bbH^0(\bbR^{3N})$
    & \cmark
    & Need redefining $T[\psi] := \bbE_{\lrvert{\psit(\bmkkv)}^2} \lrVert*{\bmkkv}^2$, where $\psit$ (exists and in $\bbL^2$) is the Fourier transform of $\psi \in \bbL^2$.\footnote{
      Under the transformation $\psit(\bmkkv) = (2 \pi)^{-\frac{3N}{2}} \int \ud\bmrrv \, \psi(\bmrrv) \exp(-\ii \bmkkv \cdot \bmrrv)$,
      the original function is $\psi(\bmrrv) = (2 \pi)^{-\frac{3N}{2}} \int \ud\bmkkv \, \psit(\bmkkv) \exp(\ii \bmkkv \cdot \bmrrv)$.
      So $\bmnabla \psi(\bmrrv) = \ii (2 \pi)^{-\frac{3N}{2}} \int \ud\bmkkv \, \psit(\bmkkv) \bmkkv \exp(\ii \bmkkv \cdot \bmrrv)$,
      and $\bmnabla^2 \psi(\bmrrv) = -(2 \pi)^{-\frac{3N}{2}} \int \ud\bmkkv \, \psit(\bmkkv) \lrVert*{\bmkkv}^2 \exp(\ii \bmkkv \cdot \bmrrv)$.
      So $T[\psi] = -\int \ud\bmrrv \, \psi^* \bmnabla^2 \psi
      = (2 \pi)^{-3N} \int \ud\bmrrv \int \ud\bmkkv' \, \psit^*(\bmkkv') \int \ud\bmkkv \, \psit(\bmkkv) \lrVert*{\bmkkv}^2 \exp(\ii (\bmkkv - \bmkkv') \cdot \bmrrv)
      \stackrel{\text{\ftnref{fourier-transform}}}{=} \int \ud\bmkkv' \, \psit^*(\bmkkv') \int \ud\bmkkv \, \psit(\bmkkv) \lrVert*{\bmkkv}^2 \delta(\bmkkv - \bmkkv')
      = \int \ud\bmkkv \, \lrvert*{\psit(\bmkkv)}^2 \lrVert*{\bmkkv}^2$,
      or $T[\psi] = \int \ud\bmrrv \, \lrvert{\bmnabla \psi}^2
      = (-\ui) \ii (2 \pi)^{-3N} \int \ud\bmrrv \int \ud\bmkkv' \, \psit^*(\bmkkv') \int \ud\bmkkv \, \psit(\bmkkv) \bmkkv' \cdot \bmkkv \exp(\ii (\bmkkv - \bmkkv') \cdot \bmrrv)
      \stackrel{\text{\ftnref{fourier-transform}}}{=} \int \ud\bmkkv' \, \psit^*(\bmkkv') \int \ud\bmkkv \, \psit(\bmkkv) \bmkkv' \cdot \bmkkv \delta(\bmkkv - \bmkkv')
      = \int \ud\bmkkv \, \lrvert*{\psit(\bmkkv)}^2 \lrVert*{\bmkkv}^2$.
    }
    \\
    $\bbH^1(\bbR^{3N})$
    & \cmark
    & To make the standard $T[\psi]$ well-defined. \newline
      $\rho^{1/2}_{[\bbH^1(\bbR^{3N})]} \subseteq \bbH^1(\bbR^3)$ (Thm.~1.1) and $\rho^{1/2}_{[\cdot]}$ is continuous between them (Thm.~1.3).
    \\
    $\scW_N := \{ \psi \in \bbH^1(\bbR^{3N}) \mid \braket{\psi}{\psi} = 1 \}$
    \footnote{Originally, $\scW_N := \{ \psi \mid \lrVert{\psi} = 1, T[\psi] < \infty \}$.}
    & \cmark
    & $\rho_{[\scW_N]} \subseteq \scI_N$ (Thm.~1.1 + normalize).
      Hilbert space.
    \\ \midrule
    $\scA_N := \{ \rho \mid \rho \text{ comes from a ground state}\}$
    & \xmark &
    $\exists \rho \in \scI_N \setminus \scA_N$ if $N > Q$ (Thm.~3.4(ii)).\footnote{
      Such a $\rho$ is a convex combination of $\rho$'s in $\scA_N$ (hence $\scA_N$ is non-convex).
      It exists since there are $V$ with a degenerate ground state. (But can such $V$ achievable by a molecule?)
      ``Even for $N=1$, there are densities which never vanish but do not come from any $V$, even if allowing density matrices.''
    }
    \\
    $\scI_N := \{\rho \mid \rho^{1/2} \in \bbH^1(\bbR^3), \rho \ge 0, \int \ud\rrv \, \rho = N\}$
    & \cmark
    & Achievable by pure state in $\scW_N$ (Thm.~1.2).
    \\
    $\scR_N := \{\rho \mid \rho \in \bbL^3(\bbR^3), \rho \ge 0, \int \ud\rrv \, \rho = N\}$
    & \cmark
    & $\scI_N \subseteq \scR_N$ by Sobolev emb. \newline
      For lin. extension of $\scI_N$ ($(\bbH^1)^2$ is not linear using $\bbH^1$ structures).
    \\
    $\bbL^3 \cap \bbL^1$
    & \cmark
    & $\scR_N \subseteq \bbL^3 \cap \bbL^1$ (Sobolev emb.) $\subseteq \bbL^p$ for $1 \le p \le 3$. \newline
      Seemingly the linear extension of $\scI_N$.
    \\ \midrule
    $N$-particle density matrix $\Gamma(\bfxx; \bfxx')$: a kernel with $\Gamma \ge 0$, $\tr(\Gamma) = 1$ and proper symmetry.
    & \cmark
    & Universal expr. $\Gamma(\bfxx; \bfxx') = \sum_{i=1}^\infty \lambda_i \psi_i(\bfxx) \psi_i^*(\bfxx')$, where $\{\lambda_i\}_i \in \rmDeltab^\infty$ and $\{\psi_i\}_i$ are orthonormal.
    \\ \midrule
    admissible kernel $P(\rrv; \rrv')$: $\tr(P) = N$, and $0 \le P \le Q$ for fermions or $0 \le P \le N$ for bosons.
    & \cmark
    & Achievable by density matrix. \newline
      Achievable by pure state for bosons but not for fermions (Thm.~2.2).
    \\ \midrule
    $V \in \bbL^\frac{3}{2} + \bbL^\infty$
    & \cmark
    & $\bbL^\frac{3}{2} + \bbL^\infty = (\bbL^3 \cap \bbL^1)^*$ but $\bbL^3 \cap \bbL^1 \subset (\bbL^\frac{3}{2} + \bbL^\infty)^*$ {\footnotesize{(topo}. dual)}. \newline
    To make $\int \ud \rrv \, \rho V$ well-defined for $\rho \in \bbL^3 \cap \bbL^1$. \newline
    Banach sp. with norm $\lrVert{V} = \inf_{g \in \bbL^\frac{3}{2}, h \in \bbL^\infty, g + h = V} \lrVert{g}_\frac{3}{2} + \lrVert{h}_\infty$.
    \\
    $\scV_N := \{V \mid \Fh + \Vh \text{ has a ground state}\}$
    & -
    & $\scV_N \subseteq \bbL^\frac{3}{2} + \bbL^\infty$.
    \\ \bottomrule
  \end{tabular}
\end{table}

\paragraph{Universal functionals (\tabref{defs-univfns})}

\begin{itemize}
  \item For a one-particle potential $V(\rrv)$ (as $V_\ext$),
    denote $\Fh := \Th + \Wh_\ee$, so that $\obraket*{\psi}{\Fh}{\psi} = T[\psi] + E_\ee[\psi]$ is system-independent.
  \item Ground-state energy:
    $E[V] := \inf_{\psi \in \scW_N} \obraket*{\psi}{\Fh + \Vh}{\psi}
    = \inf_{\psi \in \scW_N} \obraket*{\psi}{\Fh}{\psi} + \int \ud\rrv \, \rho_{[\psi]} V$,
    for all $V \in \bbL^\frac{3}{2} + \bbL^\infty$ so that $\int \ud\rrv \, \rho_{[\psi]} V$ is well-defined.

    Any minimizing $\psi \in \scW_N$ is called a \emph{ground state}, which would satisfy $(\Fh + \Vh) \psi = E[V] \psi$ in the distributional sense.

    (Thm.~3.1) $E[V]$ is concave, monotone decreasing ($V_1 \le V_2$ if it holds for all $\rrv$), continuous, and finite.

  \item (Thm.~3.2, first HK theorem) Ground states $\psi_1$, $\psi_2$ of $V_1$, $V_2 \ne V_1 + \const$ lead to $\rho_1 \ne \rho_2$.
    So $\rho_1 = \rho_2 \Longrightarrow V_1 = V_2 + \const$, and $\rho$ determines a unique $V$ up to an additive constant.

  \item A functional $f[\rho]$ is (weakly) \emph{lower semi-continuous} (l.s.c) if $\rho_n$ (weakly) converging to $\rho$ implies $f[\rho] \le \liminf f[\rho_n]$.

    Weak l.s.c is equivalent to that $\{\rho \mid f[\rho] \le \lambda\}$ is (weakly) closed for all real $\lambda$.

    weak l.s.c $\Longrightarrow$ l.s.c. \qquad l.s.c and convexity $\Longrightarrow$ weak l.s.c.

    On $\bbR^n$, finite convex functions are continuous.
    But on infinite-dimensional spaces, even l.s.c is unnecessarily implied.

  \item The \emph{convex envelope} of $f$ on $\bbA \subseteq \bbX$ is $\CE_{[f]} [\rho] := \sup \{ g[\rho] \mid g$ is weakly l.s.c and convex on $\bbX$, $g < f$ on $\bbA \}$.

    $\CE_{[f]}$ is convex, weakly l.s.c, and $\CE_{[f]} \le f$ on $\bbA$.

    $\CE_{[f]} = (f^\tncc)^\tncc$, the double conjugate of $f$.

    $(f^\tncc)^\tncc = f$ $\stackrel{\text{Fenchel–Moreau ``[29]''}}{\Longleftrightarrow}$ $f$ is properly convex and l.s.c.

    Roughly, things starts with $F^\Levy$ on $\scI_N$.
    $E = (F^\Levy)^\tncc$ is the Legendre transform / Fenchel dual / convex conjugate of $F^\Levy$ (sign of the inner product is negated so $E$ is concave),
    which is defined on $\bbL^\frac{3}{2} + \bbL^\infty = (\bbL^3 \cap \bbL^1)^*$ the topo. dual of the linear extension of $\scI_N$.
    $F^\Lieb = E^\tncc$ is defined as the Legendre transform of $E$, which is defined on $\bbL^3 \cap \bbL^1 \subset (\bbL^\frac{3}{2} + \bbL^\infty)^*$ and is convex.
    This means $F^\Lieb = ((F^\Levy)^\tncc)^\tncc = \CE_{[F^\Levy]}$, and $\exists \rho \in \scI_N$ s.t. $F^\Lieb[\rho] = ((F^\Levy)^\tncc)^\tncc[\rho] < F^\Levy[\rho]$ since $F^\Levy$ is not convex.

  \item A \emph{tangent functional} TF of $f$ at $\rho_0 \in \bbA \subseteq \bbX$ where $\bbX$ is a Banach space, is a linear functional $\ell$ on $\bbX$ s.t. $f[\rho] \ge f[\rho_0] - \ell[\rho - \rho_0]$ on $\bbA$.
    $f$ has at least one TF if $f$ is convex and $f[\rho_0] < \infty$ (Hahn-Banach).

    All continuous linear functionals on $\bbL^3 \cap \bbL^1$ takes the form $\ell[\rho] = \int \ud\rrv \, \rho V$ for some $V \in \bbL^\frac{3}{2} + \bbL^\infty$.
    So a continuous TF on $\bbL^3 \cap \bbL^1$ (as $\bbX$) is represented by $V$.

    (Thm.~3.10) Let $\rho_0 \in \scI_N$. Then the followings are equivalent:
    \itemi $\rho_0 \in \scA_N$.
    \itemii $F^\Levy[\rho_0] = F^\Lieb[\rho_0]$ and $F^\Lieb$ has a continuous TF at $\rho_0$.
    \itemiii $F^\Levy$ has a continuous TF at $\rho_0$.
    \itemiv $E[V] = F^\Levy[\rho_0] + \int \ud\rrv \, \rho_0 V$ with some $V$.
    Moreover, \itemiii and \itemiv hold with the same set of $V$;
    the $V$ in \itemiv is in $\scV_N$ whose ground-state density is $\rho_0$;
    $F^\Lieb$ and $F^\Levy$ have the same continuous TF at $\rho_0$ which is unique up to a constant.

    Roughly, for $\rho_0 \in \scA_N$, ``$\fracdelta{F}{\rho}[\rho_0]$'' $= V$, the potential that admits $\rho_0$ as the ground-state density of $\Fh + \Vh$.

    (Thm.~3.11) Let $\rho_0 \in \scI_N$ and $V \in \bbL^\frac{3}{2} + \bbL^\infty$.
    Then, $F^\Lieb$ has continuous TF $V$ at $\rho_0$ $\Longleftrightarrow$ $E[V] = F^\Lieb[\rho_0] + \int \ud\rrv \, \rho_0 V$.

  \item (Thm.~4.1) Let $(\rho)_{\lrbbrack{N}}$ denote a $\rho \in \bbL^3 \cap \bbL^1$ s.t. $\int \ud\rrv \, \rho = N$, and $E_N[V]$ denote the dependence on $N$ through $\scW_N$. Then,
    $F^\Lieb$ is jointly convex in $N$ and $\rho$: $\frac{1}{2} \lrparen{ F^\Lieb[(\rho_1)_{\lrbbrack{N+1}}] + F^\Lieb[(\rho_2)_{\lrbbrack{N-1}}] } \ge F^\Lieb[(\frac{1}{2} \rho_1 + \frac{1}{2} \rho_2)_{\lrbbrack{N}}]$ \\
    $\Longleftrightarrow$
    $E$ is convex in $N$: $\frac{1}{2} (E_{N+1}[V] + E_{N-1}[V]) \ge E_N[V], \forall V$ (\ie, the ionization energy decreases as \#electrons increases, \eg Coulomb systems (conjecture)).
\end{itemize}

\begin{table}[h]
  \centering
  \setlength{\tabcolsep}{2pt}
  \renewcommand{\arraystretch}{1.5}
  \caption{Definitions and properties of universal functional versions.}
  \label{tab:defs-univfns}
  \begin{tabular}{p{5.4cm}@{\hspace{8pt}}ccp{4.1cm}@{\hspace{8pt}}p{5.3cm}}
    \toprule
    Universal Functional & Domain & Convex\footnote{``If $F$ is to be used in a variational principle, it is clearly desirable that $F$ be a convex functional.''}
    & Variational prin.: $E[V] \!=$ & Remarks
    \\ \midrule
    HK functional: \newline
    $F^\HK[\rho] := E[V] - \int \ud\rrv \, \rho V$,
    where $V$ is determined from $\rho$ by the first HK thm. (the const. cancels).
    & $\scA_N$ & -
    & $\min_{\rho \in \scA_N} F^\HK[\rho] + \int \ud\rrv \, \rho V$, only for $V \in \scV_N$.
    & Seemingly $F^\HK[\rho] = \inf_{\psi \in \scW_N: \rho_{[\psi]} = \rho} \obraket*{\psi}{\Fh}{\psi}$, only for $\rho \in \scA_N$.
    \\
    Levy's functional~\citep{levy1979universal}: \footnote{\label{ftn:inf-min}
      Original notation in the paper is ``$\Ft$''.
      That ``min'' can replace ``inf'' in defining $F^\Levy$ and $F^\DM$ is proven by Thm.~3.3 and Cor.~4.5(ii), respectively.
      It is implied to also hold for $T_\tnS^\Levy$ and $T_\tnS^\DM$.
    } \newline
    $F^\Levy[\rho] := \min_{\psi \in \scW_N: \rho_{[\psi]} = \rho} \obraket*{\psi}{\Fh}{\psi}$.
    & $\scI_N$ & \parbox[t]{1.4cm}{\xmark {\footnotesize{ ($N \!>\! Q$,}} \\ {\footnotesize{Th.3.4(i))}}}
    & $\inf_{\rho \in \scI_N} F^\Levy[\rho] + \int \ud\rrv \, \rho V$.\footnote{
      Resembles the Legendre transform of $F^\Levy$.
    }
    & $F^\Levy = F^\HK$ on $\scA_N$. \newline
      $F^\Levy < +\infty$ on $\scI_N$ {\footnotesize{(Th.1.2)}}.
    \\
    Lieb's functional: \footnote{\label{ftn:legendre-resemble}
      Original notation in the paper is ``$F$''.
      Resembles the Legendre transform of $E[V]$ under $\bbL^2$ inner product, though $E[V]$ is concave on $\bbL^\frac{3}{2} + \bbL^\infty$.
    } \newline
    $F^\Lieb[\rho] := \sup\limits_{V \in \bbL^\frac{3}{2} + \bbL^\infty} E[V] - \int \ud\rrv \, \rho V$.
    & \parbox[t]{1.2cm}{$\scR_N$, \footnote{
        Originally $\scI_N$.
        Can be taken as $\scR_N$ or even $\bbL^3 \cap \bbL^1$ if $F^\Lieb$ is allowed to give $\infty$, since $F^\Lieb[\rho] = +\infty$ outside $\scI_N$ (Thm.~3.8).
        Variant of $F^\Lieb$ is possible to make finite value on a dense subset of nonnegative functions on $\bbL^3 \cap \bbL^1$.
    } \\ $\bbL^3 \cap \bbL^1$}
    & \cmark
    & $\inf_{\rho \in \bbL^3 \cap \bbL^1} F^\Lieb[\rho] + \int \ud\rrv \, \rho V$ \newline
      $= \inf_{\rho \in \scI_N} F^\Lieb[\rho] + \int \ud\rrv \, \rho V$ {\footnotesize{(Th.3.5)}}.
    & $F^\Lieb \!=\! F^\Levy$ on $\scA_N$, $F^\Lieb \!\le\! F^\Levy$ on $\scI_N$. \newline
      $\exists \rho \in \scI_N$ s.t. $F^\Lieb[\rho] < F^\Levy[\rho]$. %
      \newline
      $F^\Lieb \!<\! +\infty \Leftrightarrow \rho \!\in\! \scI_N$ {\footnotesize{(Th.1.2, 3.8)}}. \newline
      $F^\Lieb$ is weakly l.s.c {\footnotesize{(Th.3.6)}}. \newline
      $F^\Lieb = \CE_{[F^\Levy]}$ on $\bbL^3 \cap \bbL^1$ {\footnotesize{(Th.3.7)}}.
    \\
    Density matrix functional: \footref{ftn:inf-min} \newline
    $F^\DM[\rho] := \min_{\Gamma: \rho_{[\Gamma]} = \rho} \tr(\Fh \Gamma)$ \newline
    $= \min\limits_{\substack{\{\lambda_i\}_{i=1}^\infty \in \rmDeltab^\infty, \\ \{\psi_i\}_{i=1}^\infty \text{ orthonormal}, \\ \sum_{i=1}^\infty \lambda_i \rho_{[\psi_i]} = \rho}}
    \sum\limits_{i=1}^\infty \lambda_i \obraket*{\psi_i}{\Fh}{\psi_i}$.
    & \parbox[t]{1.2cm}{$\scI_N$ ($= +\infty$ outside)}
    & \cmark
    & $\inf_{\rho \in \scI_N} F^\DM[\rho] + \int \ud\rrv \, \rho V$.
    & $F^\DM = F^\Lieb$ on $\scI_N$ {\footnotesize{(Th.4.3)}}. \newline
      $\stackrel{\scriptstyle\text{(Th.4.2)}}{=} \!
      \inf_{\!\!\!\! \substack{K \in \bbN^*, \\ \{\lambda_i\}_i^K \in \rmDeltab^K, \\ \{\rho_i\}_i^K \in \scI_N, \\ \sum_i^K \lambda_i \rho_i = \rho}}
      \!\! \sum_i^K \!\! \lambda_i F^\Levy[\rho_i]$. \newline
      $F^\DM$ is weakly l.s.c {\footnotesize{(Cor.4.5(i))}}.
    \\ \bottomrule
  \end{tabular}
\end{table}

\paragraph{Kinetic energy functional (\tabref{defs-kinfns})}

\begin{itemize}
  \item $E_\tnS[V] := \inf_{\psi \in \scW_N} \obraket*{\psi}{\Th}{\psi} + \int \ud\rrv \, \rho_{[\psi]} V$, $\forall V \in \bbL^\frac{3}{2} + \bbL^\infty$.
  \item ``All the previous theorems carry over to these kinetic energy functionals.'' (p.265, l.-6).
  \item $\scA_N$ is the collection of $\rho$ that comes from the ground state of $\Fh + \Vh$ for some one-body potential $V \in \scV_N$ that allows a ground state. Correspondingly, define $\scA'_N$ as the collection of $\rho$ that comes from the ground state of $\Th + \Vh$.
  \item (Thm.~3.1') $E_\tnS[V]$ is concave, monotone decreasing ($V_1 \le V_2$ if it holds for all $\rrv$), continuous, and finite.
  \item (Thm.~3.2', first HK theorem) $\rho \in \scA'_N$ determines $V$ uniquely up to an additive constant s.t. $\rho$ is the ground-state density of $\Th + \Vh$.
  \item (Thm.~3.10') Roughly, for $\rho_0 \in \scA'_N$, ``$\fracdelta{T_\tnS}{\rho}[\rho_0]$'' $= V$, the potential that admits $\rho_0$ as the ground-state density of $\Th + \Vh$.
  \item (Thm.~4.1, remark in p.262, l.-8) $T_\tnS^\Lieb[\rho]$ is jointly convex in $N$ and $\rho$, since $E_{\tnS,N}[V]$ is convex in $N$.
  \item (Thm.~4.8) $\exists \rho \in \scA'_N$ s.t. $T_{\det} > T_\tnS^\Levy$ (this is the last statement in \tabref{defs-kinfns}).
    This means ``not every ground state of $\Th + \Vh$ for any $\Vh$ is a determinant
    when degeneracy is present'' (p.266, l.8)
    (but the ground state is a linear combination of determinants, p.267, l.-5),
    since $\rho \in \scA'_N$ means there is a one-body potential $\Vh$ such that $\rho$ is the ground-state density of $\Th + \Vh$, and the potential energy is determined only by $\rho \in \scA'_N$ thus fixed while the kinetic energy is lower if allowing more than determinantal wavefunctions.
    This means $\scA'_N \supset \scA''_N$ (p.269, l.2), where $\scA''_N$ denotes (p.269, l.3) the collection of $\rho$ that comes from a \emph{determinantal} ground state.
  \item (Thm.~4.6) The ground state of $\Th + \Vh$ is indeed a determinant when $\Vh$ incurs a \emph{nondegenerate} ground state of $\Th + \Vh$.
    Denote $\scA'''_N$ (p.269, l.4) as the collection of $\rho$ that comes from nondegenerate ground states (which must be determinantal by the theorem).
    Moreover, there may exist a ground state of $\Th + \Vh$ which is a determinant but is degenerate (Thm.~4.8 Remark(ii)).
    This means $\scA''_N \supset \scA'''_N$ (p.269, l.4).
  \item (Thm.~4.7) In the constrained search $\min_{\psi \in \scW_N: \rho_{[\psi]} = \rho} \obraket*{\psi}{\Th}{\psi}$ for any $\rho \in \scI_N$, there exists a minimizing $\psi$ that is a determinant (see also \ftnref{thm4.7}).
    But this does not mean any $\rho \in \scI_N$ (not even for any $\rho \in \scA'_N$) comes from a determinantal ground state:
    the ``argmin'' / minimizing determinantal $\psi$ may not be \emph{a} ground state (if $\rho \in \scI_N \setminus \scA'_N$, and may not be \emph{the} ground state of $\Th + \Vh$ if $\rho \in \scA'_N$ where $\Vh$ is the unique potential determined by $\rho$ through the first HK theorem).
\end{itemize}

\begin{table}[h]
  \centering
  \setlength{\tabcolsep}{2pt}
  \renewcommand{\arraystretch}{1.5}
  \caption{Definitions and properties of kinetic functional versions. Theorem numbers with prime denote the carried-over theorems from the universal functionals to the kinetic functionals.}
  \label{tab:defs-kinfns}
  \begin{tabular}{p{5.7cm}@{\hspace{8pt}}ccp{4.1cm}@{\hspace{8pt}}p{5.2cm}}
    \toprule
    Kinetic Functional & Domain & Convex & Variational prin.: $E_\tnS[V] \!=$ & Remarks
    \\ \midrule
    $T_\tnS^\KS[\rho] := E_\tnS[V] - \int \ud\rrv \, \rho V$,
    where $V$ is determined from $\rho$ by the first HK thm. (the const. cancels).
    & $\scA'_N$ & -
    & $\min_{\rho \in \scA_N} T_\tnS^\KS[\rho] + \int \ud\rrv \, \rho V$, only for $V \in \scV_N$.
    & Seemingly $T_\tnS^\KS[\rho] = \inf_{\psi \in \scW_N: \rho_{[\psi]} = \rho} \obraket*{\psi}{\Th}{\psi}$, only for $\rho \in \scA'_N$.
    \\
    $T_\tnS^\Levy[\rho] := \min_{\psi \in \scW_N: \rho_{[\psi]} = \rho} \obraket*{\psi}{\Th}{\psi}$. \footref{ftn:inf-min}
    & $\scI_N$ & \parbox[t]{1.4cm}{\xmark {\footnotesize{ ($N \!>\! Q$,}} \\ {\footnotesize{Th.3.4(i)')}}}
    & $\inf_{\rho \in \scI_N} T_\tnS^\Levy[\rho] + \int \ud\rrv \, \rho V$.\footnote{
      Resembles the Legendre transform of $T_\tnS^\Levy$.
    }
    & $T_\tnS^\Levy = T_\tnS^\KS$ on $\scA'_N$. \newline
      $T_\tnS^\Levy < +\infty$ on $\scI_N$ {\footnotesize{(Th.1.2')}}.
    \\
    $T_\tnS^\Lieb[\rho] := \sup\limits_{V \in \bbL^\frac{3}{2} + \bbL^\infty} E_\tnS[V] - \int \ud\rrv \, \rho V$.\footref{ftn:legendre-resemble}
    & \parbox[t]{1.2cm}{$\scR_N$, \footnote{
        Originally $\scI_N$.
        Can be taken as $\scR_N$ or even $\bbL^3 \cap \bbL^1$ if $T_\tnS^\Lieb$ is allowed to give $\infty$, since $T_\tnS^\Lieb[\rho] = +\infty$ outside $\scI_N$ (Thm.~3.8').
    } \\ $\bbL^3 \cap \bbL^1$}
    & \cmark
    & $\inf_{\rho \in \bbL^3 \cap \bbL^1} T_\tnS^\Lieb[\rho] + \int \ud\rrv \, \rho V$ \newline
      $= \inf_{\rho \in \scI_N} T_\tnS^\Lieb[\rho] + \int \ud\rrv \, \rho V$ {\footnotesize{(Th.3.5')}}.
    & $T_\tnS^\Lieb \!=\! T_\tnS^\Levy$ on $\scA'_N$, $T_\tnS^\Lieb \!\le\! T_\tnS^\Levy$ on $\scI_N$. \newline
      $\exists \rho \in \scI_N$ s.t. $T_\tnS^\Lieb[\rho] < T_\tnS^\Levy[\rho]$. %
      \newline
      $T_\tnS^\Lieb \!<\! +\infty \Leftrightarrow \rho \!\in\! \scI_N$ {\footnotesize{(Th.1.2', 3.8')}}. \newline
      $T_\tnS^\Lieb$ is weakly l.s.c {\footnotesize{(Th.3.6')}}. \newline
      $T_\tnS^\Lieb = \CE_{[T_\tnS^\Levy]}$ on $\bbL^3 \cap \bbL^1$ {\footnotesize{(Th.3.7')}}.
    \\
    $T_{\det}[\rho] := \min\limits_{\substack{\psi \in \scW_N: \rho_{[\psi]} = \rho, \\ \psi \text{ is a determinant}}} \obraket*{\psi}{\Th}{\psi}$.
    \footnote{\label{ftn:thm4.7}That ``min'' can replace ``inf'' is proven by Thm.~4.7.}
    & $\scI_N$ & -
    & $\inf_{\rho \in \scI_N} T_{\det}[\rho] + \int \ud\rrv \, \rho V$ {\footnotesize{(Th.4.9)}}
    & $T_{\det} \ge T_\tnS^\Levy$ on $\scI_N$. \newline
      $T_{\det} = T_\tnS^\Levy$ for $\rho \in \scA''_N$ which comes from a determinantal ground state (p.269, l.3, l.8). \newline %
      $\exists \rho \in \scA'_N$ s.t. $T_{\det} > T_\tnS^\Levy$ {\footnotesize{(Th.4.8)}}.
    \\ \bottomrule
  \end{tabular}
\end{table}

\paragraph{Bounds (\tabref{functional-bounds})}

\begin{itemize}
  \item Define $T^\vW[\rho] := \int \ud\rrv \, \lrVert{\nabla \rho^{1/2}(\rrv)}^2 = \frac{1}{4} \int \ud\rrv \, \lrVert{\nabla \rho(\rrv)}^2 / \rho(\rrv)$, which is convex.
  \item $\Et_\XC[\rho] := \inf_{\Gamma: \rho_{[\Gamma]} = \rho} E_\XC[\Gamma]$ is convex on $\scI_N$, so a lower bound exists.
    But it is ``extremely complicated, \eg it is nonlocal''.
    The nonlocality has something to do with induced dipolar forces or van der Waals forces ``[25]''.
  \item For a molecular Coulomb system $\{(Z_A, \rrv_A)\}_{A \in [N_\tnA]}$, under the scaling $Z_A \asn \lambda Z_A$ and $N \asn \lambda N$, $\lim_{\lambda \to \infty} E^\TF_0[V] / E[V] = 1$ (``[14]'').
    It also holds for $E^\TFD$ and $E^\TFW$ (``[25]'').
\end{itemize}

\paragraph{2-RDM}
``It is very difficult to decide when a given two-body density matrix is, in fact, the reduction of an admissible $N$-body density matrix $\Gamma$. This is called the N-representability problem.
\ldots It is possible, however, to find some necessary conditions and some sufficient conditions for N-representability, henceforth bounds of $E[V]$.''

\begin{table}[h]
  \centering
  \setlength{\tabcolsep}{2pt}
  \caption{Functional bounds.}
  \label{tab:functional-bounds}
  \begin{tabular}{l@{}p{7.4cm}p{9.1cm}}
    \toprule
    & \hspace{2.0cm} Lower bound & \hspace{2.0cm} Upper bound
    \\ \midrule
    $T[\psi]$
    & \vspace{-10pt}
    \begin{itemize}
      \item $T[\psi] \ge T^\vW[\rho_{[\psi]}]$ {\footnotesize{(Cauchy-Schwartz)}} and
        $T[\psi] \ge 3 \big(\frac{\pi}{2}\big)^\frac{4}{3} \lrVert{\rho_{[\psi]}}_3$ {\footnotesize{(Sobolev)}}
        on $\bbH^1(\bbR^{3N})$ {\footnotesize{(Th.1.1)}}.
      \item $T[\psi] \ge \frac{3}{5 Q^\frac{2}{3}} \big(\frac{3\pi}{2}\big)^\frac{2}{3} \int \ud\rrv \, \rho_{[\psi]}^\frac{5}{3}$ {\footnotesize{(for fermions in 3-dim. ``Lieb\&Thirring [21]'')}}.
        It is conjectured to hold after multiplying $(4\pi)^\frac{2}{3}$, which gives $T^\TF[\rho]$ for $Q = 2$ (up to the $\frac{1}{2}$ factor from \eqnref{kinfn-thomas-fermi}).
    \end{itemize}
    & \vspace{-10pt}
    \begin{itemize}
      \item No upper bound in terms of $\rho$.
      \item $\forall \rho \in \scI_N$, $\exists \psi^\star$ s.t. $\rho_{[\psi^\star]} = \rho$, and: \newline
        for bosons, $T[\psi^\star] = T^\vW[\rho]$;\footnote{Originally ``$\le$'', which means ``$=$'' when considering Thm.~1.1.}
        for fermions, $\psi^\star$ is a determinant and $T[\psi^\star] \le (4\pi)^2 N^2 T^\vW[\rho]$ {\footnotesize{(Th.1.2)}}.
      \item For fermions, $\forall \rho \in \scI_N$, $\exists \psi^\star$ s.t. $\rho_{[\psi^\star]} = \rho$, $\psi^\star$ is a determinant, and
        $T[\psi^\star] \le \frac{C_d}{Q^\frac{2}{d}} \int \ud\rrv \, \rho^\frac{d+2}{d} + T^\vW[\rho]$ {\footnotesize{(``March\&Young~[17]''; $C_1 = \frac{\pi^2}{3}$; found wrong proof for $d > 1$)}}.
    \end{itemize}
    \\
    $F^\Lieb[\rho]$
    & \vspace{-10pt}
    \begin{itemize}
      \item $F^\Lieb[\rho] \ge T^\vW[\rho]$ on $\scI_N$ {\footnotesize{(Th.3.8)}}.
      \item $\forall V \in \bbL^\frac{3}{2} + \bbL^\infty$, $\exists C$ s.t. $F^\Lieb[\rho] \ge \int \ud\rrv \, \rho V + C$ on $\bbL^3 \cap \bbL^1$ {\footnotesize{(Th.3.12)}}.
    \end{itemize}
    & \vspace{-10pt}
    \begin{itemize}
      \item $F^\Lieb[\rho] \le (4 \pi)^2 N^2 T^\vW[\rho] + J[\rho]$ on $\scI_N$ {\footnotesize{(also for $F^\Levy$; Th.3.9)}}.
    \end{itemize}
    \\
    $T_\tnS^\Lieb[\rho]$
    & \vspace{-10pt}
    \begin{itemize}
      \item $T_\tnS^\Lieb[\rho] \ge T^\vW[\rho]$ on $\scI_N$ {\footnotesize{(Th.3.8')}}.
      \item (Not sure if Th.3.12' holds for $T_\tnS^\Lieb$.)
    \end{itemize}
    & \vspace{-10pt}
    \begin{itemize}
      \item $T_\tnS^\Lieb[\rho] \le (4 \pi)^2 N^2 T^\vW[\rho]$ on $\scI_N$ {\footnotesize{(also for $T_\tnS^\Levy$; Th.3.9')}}.
    \end{itemize}
    \\
    $J[\rho]$
    & \vspace{-10pt}
    \begin{itemize}
      \item -
    \end{itemize}
    & \vspace{-10pt}
    \begin{itemize}
      \item $J[\rho] \le \const \lrVert{\rho}_{6/5}^2 \le \const \sqrt{\lrVert{\rho}_1^3 \lrVert{\rho}_3}$.
      \item $J[\rho] \le \const N^{3/2} T^\vW[\rho]^{1/2} \le \const (N + N^2 T^\vW[\rho])$.
    \end{itemize}
    \\
    $E_\XC[\Gamma]$
    & \vspace{-10pt}
    \begin{itemize}
      \item $E_\XC[\Gamma] \ge -1.68 \int \ud\rrv \, \rho_{[\Gamma]}^\frac{4}{3}$ for any $Q$ and both fermions and bosons {\footnotesize{(``[24]''; \cf X-LDA \eqnref{excfn-lda})}}. \newline
        For $E_\ee[\Gamma]$, $J[\rho] - C \int \ud\rrv \, \rho^\frac{4}{3}$ is not convex; not even positive.
    \end{itemize}
    & \vspace{-10pt}
    \begin{itemize}
      \item No upper bound in terms of $\rho$ {\footnotesize{(``[24]'')}}.
      \item $E_\XC[\Gamma] < 0$ for any pure, determinantal state.
    \end{itemize}
    \\
    $E[V]$
    & \vspace{-10pt}
    \begin{itemize}
      \item $E[V] \ge E^\TFD[\rho_0,V] := E^\TF[\rho_0,V] - C \int \ud\rrv \, \rho_0^\frac{4}{3}$,
        where $E^\TF[\rho,V] := T^\TF[\rho] + J[\rho] + \int \ud\rrv \, \rho V$ {\footnotesize{($\TF$-Dirac; \cf \eqnref{engfn-thomas-fermi-dirac})}},
        and $\rho_0$ is the true ground-state density.
    \end{itemize}
    & \vspace{-10pt}
    \begin{itemize}
      \item $E[V] \le E^\TFW[\rho,V] := \frac{C_d}{Q^\frac{2}{d}} \int \ud\rrv \, \rho^\frac{d+2}{d} + T^\vW[\rho] + J[\rho] + \int \ud\rrv \, \rho V,
        \, \forall \rho \in \scI_N$ {\footnotesize{($\TF$-Weizs\"acker)}}.
      \item $E[V] \le \tr(P(-\nabla^2 + V)) + \frac{1}{2} \int \ud\rrv \ud\rrv' \, \big( P(\rrv; \rrv) P(\rrv'; \rrv') - \lrvert{P(\rrv; \rrv')}^2 \big) \frac{1}{\lrVert{\rrv - \rrv'}}$ for any admissible kernel $P$.
        The bound-minimizing $P$ comes from a pure, determinantal state.
    \end{itemize}
    \\ \bottomrule
  \end{tabular}
\end{table}

\subsection{Random Notes}

\paragraph{Scaling rule of functionals}

Source: [\href{https://dft.uci.edu/theses/takeyce.pdf}{Ref. thesis}].
Also in Lieb's~\citep{lieb1983density} proof of Thm.~3.4 for $E_\ext$.

Let $\gamma_\# \rho$ denote the electron density after the shrinking the space by a scale $\gamma$. The coordinates after the transformation is $\rrv' = \rrv / \gamma$.
The density is transformed as: $\gamma_\# \rho (\rrv') = \gamma^3 \rho(\gamma \rrv')$.

\begin{itemize}
  \item $T_\tnS[\gamma_\# \rho] = \gamma^2 T_\tnS[\rho]$.
  \item $J[\gamma_\# \rho] = \gamma J[\rho]$.
  \item $E_\tnX[\gamma_\# \rho] = \gamma E_\tnX[\rho]$.
  \item $E_\tnC[\gamma_\# \rho] > \gamma E_\tnC[\rho]$ for $\gamma > 1$.
\end{itemize}

\paragraph{Chemical potential calculation}
For a (effective) non-interacting system with one-electron potential $V$, no Hartree term nor XC term arises (or, the kinetic energy can be replaced by the non-interacting kinetic energy functional $T_\tnS$), so solving for the ground-state density amounts to:
\begin{align}
  \min_\rho T_\tnS[\rho] + \int \rho(\rrv) V(\rrv) \dd \rrv - \mu \lrparen*[\Big]{ \int \rho(\rrv) \dd \rrv - N },
\end{align}
where the Lagrange multiplier $\mu$ is the chemical potential to make the optimized $\rho$ normalized. Taking the variation yields:
\begin{align}
  \fracdelta{T_\tnS[\rho]}{\rho(\rrv)} + V(\rrv) = \mu.
\end{align}
Since $T_\tnS[\rho] = \min_{\{\phi_i\}_i: \text{ orthonormal}, \rho_{[\{\phi_i\}_i]} = \rho} T[\{\phi_i\}_i]$,
which is $T_\tnS[\rho] = T[\{\phi_{[\rho]i}\}_i]$ in terms of the optimal orbitals $\{\phi_{[\rho]i}\}_i$. So:
\begin{align}
  \fracdelta{T_\tnS[\rho]}{\rho(\rrv)} = \fracdelta{}{\rho(\rrv)} T[\{\phi_{[\rho]i}\}_i]
  = \sum_i \int \fracdelta{}{\phi_i(\rrv')} T[\{\phi_{[\rho]i}\}_i] \fracdelta{\phi_{[\rho]i}(\rrv')}{\rho(\rrv)} \dd \rrv'
  = 2 \sum_i \int \Th \phi_{[\rho]i}(\rrv') \fracdelta{\phi_{[\rho]i}(\rrv')}{\rho(\rrv)} \dd \rrv'.
\end{align}
The optimal orbitals recovers the density, $\sum_i \lrvert{\phi_{[\rho]i}(\rrv)}^2 = \rho(\rrv)$, whose variation gives:
\begin{align}
  2 \sum_i \phi_{[\rho]i}(\rrv') \fracdelta{\phi_{[\rho]i}(\rrv')}{\rho(\rrv)} = \delta(\rrv-\rrv').
\end{align}
So:
\begin{align}
  \mu ={} & \fracdelta{T_\tnS[\rho]}{\rho(\rrv)} + \int V(\rrv') \delta(\rrv-\rrv') \dd \rrv'
  = 2 \sum_i \int \Th \phi_{[\rho]i}(\rrv') \fracdelta{\phi_{[\rho]i}(\rrv')}{\rho(\rrv)} \dd \rrv' + \int V(\rrv') 2 \sum_i \phi_{[\rho]i}(\rrv') \fracdelta{\phi_{[\rho]i}(\rrv')}{\rho(\rrv)} \dd \rrv' \\
  ={} & 2 \sum_i \int \lrparen{ \Th \phi_{[\rho]i}(\rrv') + V(\rrv') \phi_{[\rho]i}(\rrv') } \fracdelta{\phi_{[\rho]i}(\rrv')}{\rho(\rrv)} \dd \rrv'
\end{align}

In another more direct way, using $N$ orbitals,
\begin{align}
  \min_{\{\phi_i\}_i} T[\{\phi_i\}_i] + \int \rho_{[\{\phi_i\}_i]}(\rrv) V(\rrv) \dd \rrv - \sum_i \veps_i \lrparen*[\Big]{ \int \lrvert{\phi_i(\rrv)}^2 \dd \rrv - 1 },
\end{align}
where the Lagrange multipliers for orthonormal constraints are omitted (satisfied when non-degenerate).
Taking the variation w.r.t $\phi_i$ yields:
\begin{align}
  \Th \phi_i + V \phi_i = \veps_i \phi_i,
\end{align}
and the optimal $\rho(\rrv) = \sum_i \lrvert{\phi_i(\rrv)}^2$.
These orbitals also give the right KEDF value: $T_\tnS[\rho_{[\{\phi_i\}_i]}] = T[\{\phi_i\}_i]$, so:
\begin{align}
  \mu = 2 \sum_i \int \lrparen{ \Th \phi_i(\rrv') + V(\rrv') \phi_i(\rrv') } \fracdelta{\phi_{[\rho]i}(\rrv')}{\rho(\rrv)} \dd \rrv'
  = 2 \sum_i \int \veps_i \phi_i(\rrv') \fracdelta{\phi_{[\rho]i}(\rrv')}{\rho(\rrv)} \dd \rrv'
  = \sum_i \veps_i \int \fracdelta{\lrvert{\phi_{[\rho]i}(\rrv')}^2}{\rho(\rrv)} \dd \rrv'
  \stackrel{\text{?}}{=} \frac{\sum_i \veps_i}{N}.
\end{align}

To make the density normalized, the projected variation $\fracdelta{T_\tnS[\rho]}{\rho(\rrv)} + V(\rrv) - \mu$, should integrate to zero. So:
\begin{align}
  \mu = \frac{\int \fracdelta{T_\tnS[\rho]}{\rho(\rrv)} + V(\rrv) \dd \rrv}{\int \ud \rrv}.
\end{align}